\documentclass[prd,letterpaper,twocolumn,preprintnumbers,superscriptaddress,nofootinbib]{revtex4}
\usepackage[letterpaper, hdivide={1.91cm,,1.165cm}, vdivide={1.83cm,,2.4cm}]{geometry}

\usepackage{slashed}
\usepackage{amsmath,amssymb}
\usepackage{graphicx}
\usepackage{units}
\usepackage{bbold}
\usepackage{xcolor}
\usepackage{dsfont}
\usepackage[hyperfootnotes=false,colorlinks,citecolor=blue]{hyperref}

\usepackage{slashed}
\usepackage{array}
\usepackage{hhline}
\usepackage{float}

\usepackage{comment}
\usepackage[normalem]{ulem}

\def \vec#1{{\boldsymbol{#1}}}

\def \L {\mathcal{L}} 
\newcommand{\hc}{\ensuremath{\text{h.c.}}}
\newcommand{\BR}{\ensuremath{\text{BR}}}

\newcommand{\dd}{{\rm d}}
\newcommand{\ii}{{\rm i}}

\begin{document}

\title{Lepton flavor violation by three units}

\author{Julian Heeck}
\email[Email: ]{heeck@virginia.edu}
\thanks{ORCID: \href{https://orcid.org/0000-0003-2653-5962}{0000-0003-2653-5962}.}
\affiliation{Department of Physics, University of Virginia,
Charlottesville, Virginia 22904-4714, USA}

\author{Mikheil Sokhashvili}
\email[Email: ]{ms2guc@virginia.edu}
\thanks{ORCID: \href{https://orcid.org/0000-0003-0844-7563}{0000-0003-0844-7563}.}
\affiliation{Department of Physics, University of Virginia,
Charlottesville, Virginia 22904-4714, USA}

\author{Anil Thapa}
\email[Email: ]{a.thapa@colostate.edu}
\thanks{ORCID: \href{https://orcid.org/0000-0003-4471-2336}{0000-0003-4471-2336}.}
\affiliation{Department of Physics, University of Virginia,
Charlottesville, Virginia 22904-4714, USA}
\affiliation{Physics Department, Colorado State University, Fort Collins, CO 80523, USA}

\hypersetup{
pdftitle={Lepton flavor violation by three units},   
pdfauthor={Julian Heeck, Mikheil Sokhashvili, Anil Thapa}
}

\begin{abstract}
The conservation of lepton flavor is a prediction of the Standard Model and is still an excellent approximate symmetry despite our observation of neutrino oscillations. Lepton flavor violation by one or two units have been discussed for decades, with several dedicated experiments exploring the vast model landscape but no discoveries so far. Here, we explore operators and processes that violate at least one lepton flavor by three units and identify testable signatures. In the Standard Model effective field theory, such operators already arise at mass dimension 7 and can be tested through their contributions to Michel parameters in leptonic decays. True neutrinoless charged-lepton flavor violation arises at mass dimension 10 and can realistically only be seen in the tau decay channels $\tau \to eee\bar{\mu}\bar{\mu}$ or $\tau \to \mu\mu\mu\bar{e}\bar{e}$, for example in Belle II. Testable rates for these tau decays require light new particles and subsequently predict an avalanche of remarkably clean but so-far unconstrained collider signatures.
\end{abstract}

\maketitle

\section{Introduction}

The three individual lepton flavors -- electron, muon, and tau number -- are conserved quantum numbers in the Standard Model (SM) of particle physics that forbid  otherwise seemingly allowed decays such as $\mu^-\to e^- \gamma$~\cite{Pontecorvo:1967fh}. No short-range process violating this symmetry $U(1)_{L_e}\times U(1)_{L_\mu}\times U(1)_{L_\tau}$ has ever been observed~\cite{Calibbi:2017uvl,Davidson:2022jai}, in perfect agreement with the SM, but through \textit{neutrino oscillations}, $\nu_\alpha\to\nu_\beta$, we have learned that these symmetries are at least broken in the neutrino sector. This is a spectacular observation that forces us to extend the SM by new particles to account for non-zero neutrino masses, but by itself does \textit{not} imply testable rates for \textit{charged} lepton flavor violation, since these are typically suppressed by the tiny neutrino masses~\cite{Petcov:1976ff,Bilenky:1977du,Marciano:1977wx,Lee:1977qz,Lee:1977tib}. $U(1)_{L_e}\times U(1)_{L_\mu}\times U(1)_{L_\tau}$, or any of its subgroups, then remain excellent \textit{approximate} symmetries in the charged-lepton sector, which could have an entirely different flavor breaking pattern than in the neutrino sector~\cite{Heeck:2016xwg,Hambye:2017qix}. Lepton flavor violation (LFV) by \textit{one} unit, as in $\mu\to e\gamma$, might then be suppressed compared to LFV by \textit{two} units, such as the muonium conversion $\mu^+ e^-\to \mu^- e^+$, as recently emphasized more generally in Ref.~\cite{Heeck:2024uiz}. Despite the more complicated experimental signatures, it is important to search for LFV by \textit{two} units in order not to miss new physics due to theoretical bias.

Here, we take this argument one step further and investigate whether LFV by \textit{three} units could be detectable. While LFV by one or two units arise already at mass dimension $d=5$ (together with total lepton number $L$ violation) or $d=6$ (without $L$ violation) in the Standard Model Effective Field Theory (SMEFT)~\cite{Isidori:2023pyp}, LFV by \textit{three} units first arises at $d=7$ and is thus generically more suppressed. Nevertheless, we show below that these operators can lead to potentially observable modifications of muon and tau decays, parametrized by familiar Michel parameters. \textit{Charged} LFV by three units first arises at $d=10$ and requires UV-complete models with new light particles for testable rates, but could then lead to clean tau decays such as  $\tau\to eee\bar{\mu}\bar{\mu}$ and  $\tau\to \mu\mu\mu\bar{e}\bar{e}$~\cite{Heeck:2016xwg}, potentially observable at Belle~II.

The rest of this article is organized as follows: 
Sec.~\ref{sec:dim7} is devoted to $d=7$ LFV operators, the lowest mass dimension in which LFV by three units occurs; 
in Sec.~\ref{sec:dim9} we briefly sketch LFV by three units in $d=9$. 
Sec.~\ref{sec:dim10} has a long discussion of LFV at $d=10$, the lowest mass dimension with \textit{charged} LFV by three units, including a detailed discussion of one UV completion. 
We conclude in Sec.~\ref{sec:conclusions}.
App.~\ref{sec:UV_dim7},~\ref{sec:UV_dim9}, and~\ref{sec:UV_dim10} list UV completions for the $d=7$, $d=9$, and $d=10$ operators, respectively.
App.~\ref{sec:LEPlimits} provides a detailed analysis of LEP limits from $e^+e^-\to \mu^+\mu^-$, relevant for the UV completion in Sec.~\ref{sec:dim10}.

\section{Dimension 7}
\label{sec:dim7}

The lowest SMEFT operators violating lepton flavor by three units arise at dimension 7, and are of the simple form
\begin{align}
   \frac{1}{\Lambda^3} \bar{\mu}_R L_e L_e L_e H =  \frac{1}{\Lambda^3} (\mu_R^c L_e^i) (L_e^j L_e^k) H^l \epsilon_{ij}\epsilon_{kl}\,,
   \label{eq:d7operator}
\end{align}
or flavor permutations thereof, where the brackets indicate Lorentz singlets and $i,j,k,l$ are $SU(2)_L$ indices. The SM fields and their gauge quantum numbers are collected in Tab.~\ref{tab:fields}.  This operator violates $\Delta L_e = 3 = -3\Delta L_\mu$; if only one such operator is added to the SM Lagrangian, the SMEFT scale $\Lambda$ can be chosen to be real without loss of generality. Tree-level UV completions for this operator are listed in App.~\ref{sec:UV_dim7}. This operator can be enhanced compared to lower-$d$ LFV using the anomaly-free flavor symmetry $U(1)_{L_e -4 L_\tau +3 L_\mu}$~\cite{Heeck:2016xwg,Hambye:2017qix}, although that symmetry needs to be broken in the neutrino sector to comply with neutrino-oscillation data. Notice that the operator also violates \textit{total} lepton number $L$ by two units, but will \textit{not} lead to any kind of Majorana neutrino masses due to the flavor symmetry. Stated differently, one cannot close loops to turn this operator into a neutrino propagator.

\begin{table}[tb]
\centering
    \renewcommand{\arraystretch}{1.2} % Increase row height
\begin{tabular}{c c c c} 
 \hline
 Field & Chirality & Generations & $SU(3)_C\times SU(2)_L\times U(1)_Y$   \\ [0.5ex] 
 \hline\hline
 $Q$ & left & 3 & $\left(\vec{3},\vec{2},\tfrac16\right)$\\
 $u_R$ & right & 3 & $\left(\vec{3},\vec{1},\tfrac23\right)$\\
 $d_R$ & right & 3 & $\left(\vec{3},\vec{1},-\tfrac13\right)$\\
 $L$ & left & 3 & $\left(\vec{1},\vec{2},-\tfrac12\right)$\\
 $l_R$ & right & 3 & $\left(\vec{1},\vec{1},-1\right)$\\
 $H$ & scalar & 1 & $\left(\vec{1},\vec{2},\tfrac12\right)$\\
 \hline
\end{tabular}
\caption{SM fields and quantum numbers; hypercharge is related to electric charge via $Q = Y + T_3$.}
\label{tab:fields}
\end{table}

While the operator violates lepton flavor by three units, it does not generate \textit{charged} LFV; after electroweak symmetry breaking, the operator becomes $\langle H\rangle \,\mu_R^c e_L \nu_e \nu_e/\Lambda^3$ and thus contains two neutrinos. The best experimental signature is likely the induced $\mu\to e\nu_e\nu_e$ decay. Due to the different neutrino flavors, the new operator does not interfere with the SM muon decay contribution and thus \textit{shortens} the muon lifetime. More importantly, the electron distribution will be different from the SM Michel spectrum~\cite{Michel:1949qe}. In Ref.~\cite{Langacker:1988cm}, it has been shown that even LFV and $\Delta L$ operators inducing $\mu\to e \nu\nu$ can be mapped onto the usual Michel parameters, but our case is different still since we have two identical neutrinos in the final state. The explicit calculation below shows that our operator can nevertheless be captured in the form of Michel parameters.

\begin{figure}[b]
    \centering
    \includegraphics[width=0.37\textwidth]{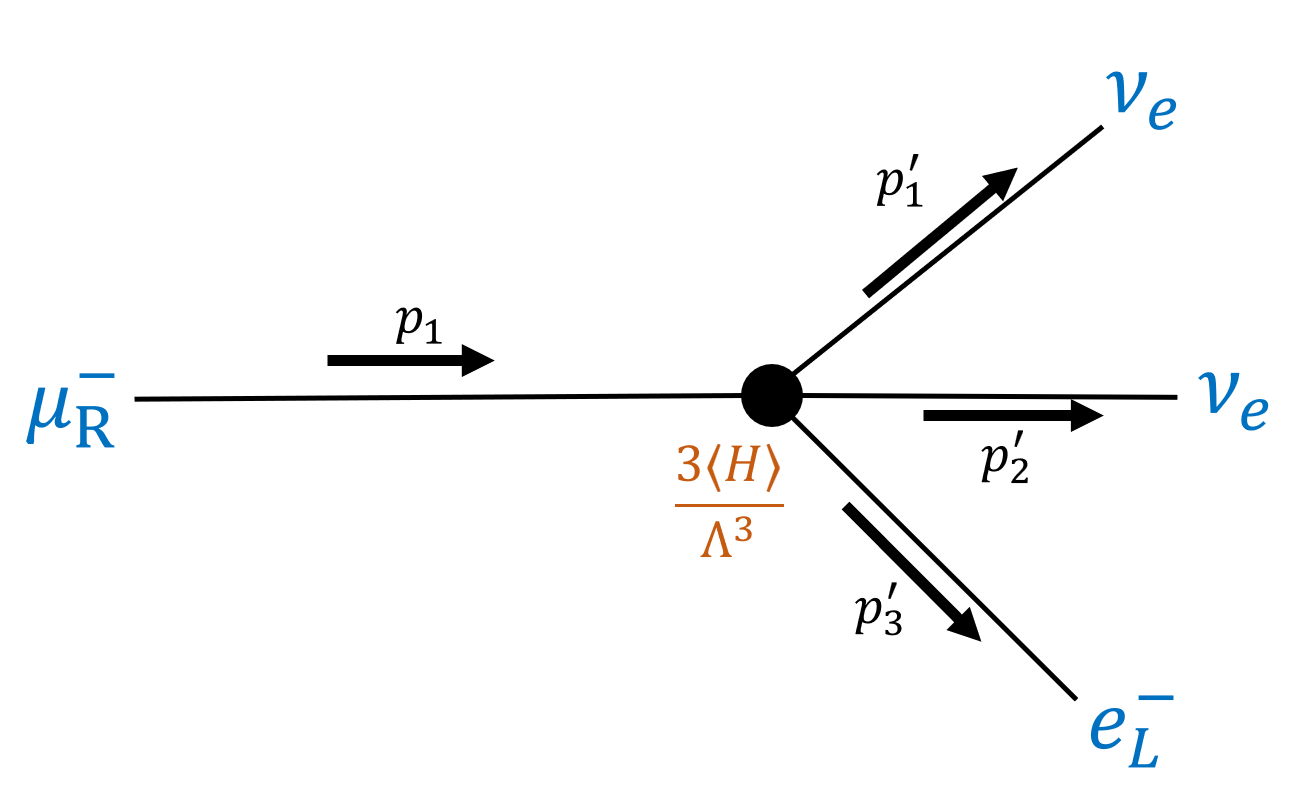}
    \caption{Muon decay $\mu\to e\nu_e\nu_e$ through the $\Delta L_e= 3=-3\Delta L_\mu $ LFV operator from Eq.~\eqref{eq:d7operator}.
   }
    \label{fig:muon_decay}
\end{figure}

\begin{figure}[tb]
    \centering
    \includegraphics[width=0.47\textwidth]{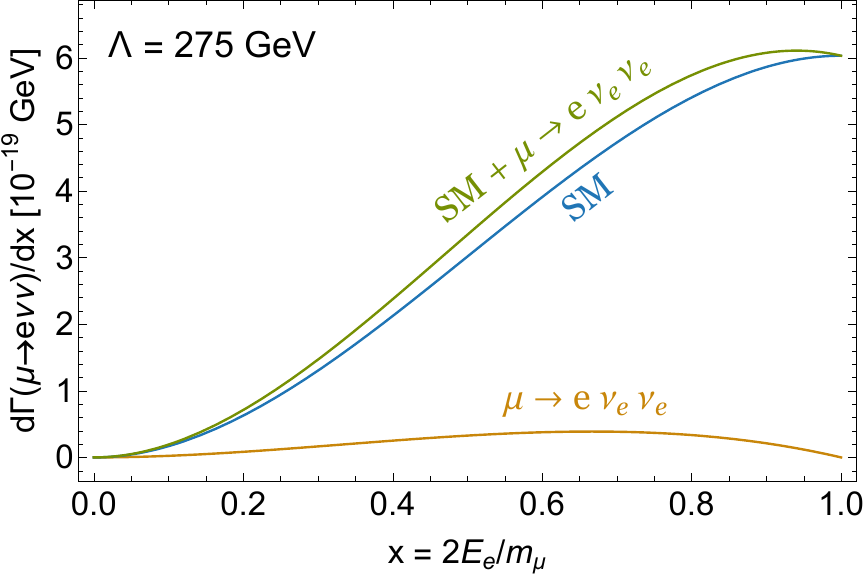}
    \caption{Electron energy distribution in $\mu\to e \nu\nu$ from both SM and our LFV-by-three-units operator~\eqref{eq:d7operator} with $\Lambda = \unit[275]{GeV}$.}
    \label{fig:muon_decay_rate}
\end{figure}

The amplitude $\mathcal{M}$ for polarized muon decay $\mu\to e\nu_e\nu_e$ via Eq.~\eqref{eq:d7operator} [Feynman diagram in Fig.~\ref{fig:muon_decay}] can be written as
\begin{align}
   \mathcal{M} = \frac{3v}{\sqrt2 \Lambda^3} 
    \left [ \overline{\Sigma_1 v_1} P_R v_1'  \overline{\Sigma_3' v_3'} P_R v_2' 
    - \overline{\Sigma_1 v_1} P_R v_2'  \overline{\Sigma_3' v_3'} P_R v_1' \right],
\end{align}
where we defined the Higgs' vacuum expectation value $v\simeq \unit[246]{GeV}$ and introduced the spin projection operators $\Sigma _1$ and $\Sigma _3'$ for the muon and the electron, respectively:
\begin{align}
    \Sigma _1 \equiv \frac{1 + \gamma^5  \slashed{s}_1}{2}\,, \quad 
    \Sigma _3' \equiv \frac{1 + \gamma^5  \slashed{s}_3'}{2} \,.
\end{align}
Considering neutrinos and  electron to be massless, we find
\begin{align}
    \langle|\mathcal{M}|^2\rangle 
    = \frac{9v^2}{2\Lambda^6} 
    (p_1' \cdot p_2') (p_3' \cdot [p_1 + m_\mu s_1])\,,
\end{align}
notably independent of the electron polarization, just like in the SM case~\cite{Rubbia:2022hry}. This means we have to multiply $\langle|\mathcal{M}|^2\rangle$ by a factor of two, as we would have gotten the same answer had we not introduced the spin projection of the electron in the first place; but to account for the identical neutrinos, we also divide $\langle|\mathcal{M}|^2\rangle$ by a factor of two, so nothing changes. 
The electron four momentum dotted with the muon spin direction vector, $s_1 = (0,\vec{s})$, is $p_3' \cdot s_1 = P_\mu \cos(\theta) E_3'$, with $\theta$ being the angle between electron momentum and muon spin, and muon polarization $P_\mu$. Further defining $x \equiv 2 E_3'/m_\mu$, we eventually find the double differential rate for the electron energy and emission direction as
\begin{align}
    \frac{\dd\Gamma}{\dd x\, \dd \cos\theta} 
    = \frac{9\, m_\mu^5\, v^2\,}{4096\, \pi^3\, \Lambda^6}\left[1 + P_\mu \cos(\theta)\right] (1 - x)\, x^2 \,.
\end{align}
For comparison, the SM decay $\mu\to e \nu_\mu \bar{\nu}_e$ takes the form
\begin{align}
    \frac{\dd\Gamma_\text{SM}}{\dd x \,\dd \cos\theta} 
    =\frac{G_F^2 m_\mu^5 }{192\, \pi^3} 
    \left[3 - 2x + P_\mu \cos(\theta) (2x - 1)\right]x^2 \,.
    \label{eq:our_muon_decay_rate}
\end{align}
The final electron-energy distribution is illustrated in Fig.~\ref{fig:muon_decay_rate}; the LFV contribution enhances the number of electrons with energies $E_e\sim m_\mu/3$.

Our additional contribution to $\mu\to e\nu\nu$ can be mapped onto the following Michel parameters, assuming $\Lambda \gg v$:
\begin{align}
\begin{split}
    &\rho = \frac{3}{4} - \frac{27}{512 \sqrt{2} G_F^3 \Lambda^6} + \mathcal{O}\left(\frac{1}{\Lambda^{12}}\right),\\
    &\delta =  \frac{3}{4} - \frac{81}{512 \sqrt{2} G_F^3 \Lambda^6} + \mathcal{O}\left(\frac{1}{\Lambda^{12}}\right),\\
    &\xi = 1 + \frac{9}{64 \sqrt{2} G_F^3 \Lambda^6} + \mathcal{O}\left(\frac{1}{\Lambda^{12}}\right).
    \label{eq:MichelParam}
\end{split}
\end{align}
Swapping $\mu_R$ for $\tau_R$ in Eq.~\eqref{eq:d7operator} similarly gives $\tau\to e\nu_e\nu_e$, and the third possible flavor combination gives $\tau\to \mu\nu_\mu\nu_\mu$, all of which result in the same Michel parameters in the limit of massless final-state leptons.\footnote{Three other operators arise from flavor exchange \textit{within} each operator, e.g.~$\bar{e}_R L_\mu L_\mu L_\mu H$, which now violates $\Delta L_\mu = 3 = -3\Delta L_e$ and gives rise to $\mu^-\to e^-\bar{\nu}_\mu\bar{\nu}_\mu$ with   $P_\mu\to -P_\mu$ in Eq.~\eqref{eq:our_muon_decay_rate}.
}

\begin{table}[tb]
    \centering
    \renewcommand{\arraystretch}{1.4} % Increase row height
    \begin{tabular}{|c|c|c|}
    \hline
        $\mu \rightarrow e\nu\nu$ & $\tau \rightarrow \mu\nu\nu$ & $\tau \rightarrow e\nu\nu$ \\
        \hline\hline
        $\rho = 0.74979 \pm 0.00026$ & $\rho = 0.763 \pm 0.020$ & $\rho = 0.747 \pm 0.010$ \\
        \hline
        $\delta = 0.75047 \pm 0.00034$ & $\delta \, \xi = 0.778 \pm 0.037$ & $\delta \, \xi = 0.734 \pm 0.028$ \\
        \hline
        $\xi P_\mu = 1.0009^{+ 0.0016}_{- 0.0007}$ & $\xi = 1.030 \pm 0.059$ & $\xi = 0.994 \pm 0.040$ \\
        \hline
    \end{tabular}
    \caption{Experimental Michel-parameter measurements from Refs.~\cite{ParticleDataGroup:2024cfk,CLEO:1997obd,ALEPH:2001gaj,TWIST:2011jfx}. Each column lists the measured values of $\rho$, $\delta$ (or $\delta \, \xi$), and $\xi$ for the respective process indicated in the header. 
    }
    \label{tab:limitsMichel}
\end{table}

Experimental limits on the three sets of relevant Michel parameters are listed in Tab.~\ref{tab:limitsMichel}; we construct a simple $\chi^2$ function to fit $\Lambda$ to all three Michel parameters, using $\Delta \chi^2 = 3.84$ to get 95\% CL. limits via
\begin{align}
    \chi^2(\Lambda) \leq \chi^2_{\text{min}} + \Delta \chi^2\,.
\end{align} 
The actual minimum $\chi^2_{\text{min}}$ is achieved for negative $\Lambda^6$, so we use the smallest physically allowed value, $\chi^2_{\text{min}} = \chi^2(\infty)$, although this gives essentially the same limits, collected in Tab.~\ref{tab:limitsDim7}.
Despite coming from $d=7$ operators and relatively unclean signatures, these are relevant SMEFT limits, illustrating that we are already sensitive to LFV by three units. 
Future precision measurements of muon and tauon decays~\cite{Aiba:2021bxe,Shimizu:2017uce,Bodrov:2021hfe} could improve these limits further. The outgoing undetectable neutrinos make it impossible to prove lepton flavor or number violation in these decays, but the correlation of the Michel parameters shown in Eq.~\eqref{eq:MichelParam} would be a strong indication for our operators.

\begin{table}[tb]
    \centering
    \renewcommand{\arraystretch}{1.3} % Increase row height
    \setlength{\tabcolsep}{28pt} % Increase horizontal spacing
    \begin{tabular}{|c|c|}
    \hline
        Process & 95\% C.L.~lower limit on $\Lambda$ \\
        \hline \hline 
        $\mu \rightarrow e\nu\nu$ & $0.74$ TeV \\
        \hline
        $\tau \rightarrow \mu\nu\nu$ & $0.32$ TeV  \\
        \hline
        $\tau \rightarrow e\nu\nu$ & $0.33$ TeV  \\
        \hline
    \end{tabular}
    \caption{95\% C.L.~limits on the magnitudes of the Wilson coefficients using Eq.~\eqref{eq:MichelParam} and experimental values from \cite{ParticleDataGroup:2024cfk}.}
    \label{tab:limitsDim7}
\end{table}

\section{Dimension 9}
\label{sec:dim9}

The next instances of LFV by three units in the SMEFT arise at $d=9$. Of the non-derivative operators, one type, $Qu_Ru_RL_eL_eL_e$, violates baryon number~\cite{Hambye:2017qix, Fonseca:2018ehk}, 
not under consideration here, while the others again violate $\Delta L = 2$ and still do not induce Majorana neutrino masses on account of the mismatch in flavor violation. 
The operators involving Higgs fields,
\begin{align}
   \frac{1}{\Lambda^5}  \bar{\ell}_R L_e L_e L_e H |H|^2\,, &&
    \frac{1}{\Lambda^5} \bar{L}_\alpha e_R L_e L_e H H H \,,
\end{align}
can be experimentally tested through Michel parameters, just like the $d=7$ case above. Purely leptonic operators are of the form $\bar{\ell}_R\bar{\ell}_R L L L L$ and thus involve four charged leptons and two neutrinos. For example, the flavor combination
\begin{align}
   \frac{1}{\Lambda^5} \bar{e}_R \bar{e}_R L_\mu L_\mu L_\mu L_\tau\,,
   \label{eq:leptonic_d9}
\end{align}
can be enhanced compared to lower-$d$ LFV using the anomaly-free flavor symmetry $U(1)_{2 L_e + 3 L_\mu -5 L_\tau}$ and induces two-neutrino processes such as $\tau \to e e \bar{\mu} \bar{\nu}_\mu \bar{\nu}_\mu$; without taus, the only flavor combination is $\bar{\mu}_R \bar{e}_R L_e L_e L_e L_e$, which induces $\mu \to e e\bar{e}\nu_e\nu_e$. All of these decays mimic SM decays~\cite{Flores-Tlalpa:2015vga} and thus require precision measurements, basically Michel parameter analyses of $\ell \to \ell\ell\ell\nu\nu$. While this can be attempted at Mu3e and Belle II, we do not expect any constraints on $\Lambda$ that would push it above the electroweak scale, so the SMEFT applicability is questionable.

Finally, the last set of operators involves quarks and comes in the forms $\bar{Q} u_R \bar{\ell}_R L L L$, $\bar{d}_R Q \bar{\ell}_R L L L$, $\bar{d}_R u_R \bar{\ell}_R \ell_R L L $, and $\bar{d}_R u_R\bar{L} L L L $, which give rise to processes involving only \emph{one} neutrino, and thus moderately more visible than the other $d=9$ operators. For example, the flavor combination
\begin{align}
   \frac{1}{\Lambda^5} \bar{d}_R u_R \bar{\mu}_R e_R L_e L_e \,,
   \label{eq:hadronic_d9}
\end{align}
yields meson decays such as $\pi^+\to \mu^- e^+ e^+ \bar{\nu}_e$, constrained to branching ratios below $10^{-6}$~\cite{ParticleDataGroup:2024cfk,Baranov:1991uj}. Kaon or $B$-meson decays into three charged leptons and one neutrino can also occur, as well as tau decays into two charged leptons, a meson, and a neutrino.

Within the SMEFT, all these operators are too suppressed to generate testable rates, esp.~considering that the experimental signatures are not as spectacular as other LFV decays, although the meson decays such as $\pi^+\to \mu^- e^+ e^+ \bar{\nu}_e$ are relatively clean and show flavor violation even without neutrino identification. Still, testable rates require $\Lambda$ far below the electroweak scale, forcing us to study UV completions of these operators with light neutral particles -- potentially light enough to be emitted on-shell in tauon or meson decays -- to confidently claim that this is a testable signature. We provide UV completions of the operators~\eqref{eq:leptonic_d9} and~\eqref{eq:hadronic_d9} in App.~\ref{sec:UV_dim9}  but will refrain from a detailed study of these $d=9$ operators and instead move on to better-visible $d=10$ operators.

\section{Dimension 10}
\label{sec:dim10}

At mass dimension $d=10$ in the SMEFT, we finally find operators that violate lepton flavor by three units without also violating baryon or total lepton number. These can give rise to fully visible neutrinoless \textit{charged} LFV. Non-derivative operators all contain one Higgs doublet $H$ and come in three different types, organized by the number of lepton doublets:
\begin{align}
   \bar{L}\bar{L}\bar{L} L  L l_R   H\,, &&
   \bar{L}\bar{L} L \bar{l}_R l_R l_R  H\,, &&
   \bar{L} \bar{l}_R\bar{l}_R l_R l_R l_R H\,.
\end{align}
In terms of their flavor structure, these operators can give LFV by three units  in the two varieties
\begin{align}
    \Delta L_\alpha &= 3\,, \Delta L_\beta=-2\,, \Delta L_\gamma=-1\,,\\
    \Delta L_\alpha &= 3\,, \Delta L_\beta=-3\,, \Delta L_\gamma=0\,, 
\end{align}
with $\alpha,\beta,\gamma$ being three different lepton flavors. Only the first variety can lead to \textit{neutrinoless} lepton decays, namely
\begin{align}
    \tau &\to eee\bar{\mu}\bar{\mu} \,, \quad \text{which conserves }\ \ U(1)_{L_e +4L_\mu - 5 L_\tau}\,,\\
    \tau &\to \mu\mu\mu\bar{e}\bar{e} \,, \quad \text{which conserves }\ \ U(1)_{L_\mu +4L_e - 5 L_\tau}\,.
\end{align}
These two decays are experimentally clean, so Belle II should be able to probe these down to branching ratios of order $10^{-8}$, and from the  theoretical perspective we can easily make either of these two decays dominant over all other LFV by imposing the appropriate $U(1)$ flavor symmetry; for details see the argument in Ref.~\cite{Heeck:2016xwg} or~\cite{Hambye:2017qix}. 
However, testable rates would require $\Lambda \sim \unit{GeV}$ for the SMEFT scale, of questionable validity given that mediator particles with such low masses would have been observed at collider experiments long ago.
This makes it necessary to study full UV completions of these LFV operators with light particles to investigate if testable LFV rates are compatible with indirect constraints. We will perform such an analysis below for one particular UV completion to highlight the challenges.

\subsection{UV completion for $\tau \to eee\bar{\mu}\bar{\mu}$}
\label{sec:UVstudy_dim10}

\begin{table}[tb]
    \centering
    \renewcommand{\arraystretch}{1.3}
    \setlength{\tabcolsep}{0.15cm}
    \begin{tabular}{|>{\centering\arraybackslash}p{3.1cm}|>{\centering\arraybackslash}p{5.2cm}|}
    
        \hline
        Generic Operator
        & Operator(s) With Flavor Indices\\
        \hline \hline
        $\bar{L} \, \bar{L} \, \bar{L} \, l \, L \, L \, H$
        & $\bar{L}_\tau \, \bar{L}_\mu \, \bar{L}_\mu \, l_e \, L_e \, L_e \, H$ \\
        & $\bar{L}_e \, \bar{L}_e \, \bar{L}_e \, l_\mu \, L_\mu \, L_\tau \, H$ \\
        & $\bar{L}_e \, \bar{L}_e \, \bar{L}_e \, l_\tau \, L_\mu \, L_\mu \, H$ \\
        \hline
        $\bar{L} \, \bar{L} \, \bar{l} \, l \, l \, L \, H$
        & $\bar{L}_\tau \, \bar{L}_\mu \, \bar{l}_\mu \, l_e \, l_e \, L_e \, H$ \\
        & $\bar{L}_\mu \, \bar{L}_\mu \, \bar{l}_\tau \, l_e \, l_e \, L_e \, H$ \\
        & $\bar{L}_e \, \bar{L}_e \, \bar{l}_e \, l_\mu \, l_\mu \, L_\tau \, H$ \\
        & $\bar{L}_e \, \bar{L}_e \, \bar{l}_e \, l_\mu \, l_\tau \, L_\mu \, H$ \\
        \hline
        $\bar{L} \, \bar{l} \, \bar{l} \, l \, l \, l \, H$
        & $\bar{L}_e \, \bar{l}_e \, \bar{l}_e \, l_\mu \, l_\mu \, l_\tau \, H$ \\
        \hline
    \end{tabular}
    \caption{Non-derivative $d=10$ operators  for $\tau \to eee\bar{\mu}\bar{\mu}$.}
    \label{tab:dim10LFVoperators}
\end{table}

\begin{figure}[tb]
    \centering
    \includegraphics[width=0.4\textwidth]{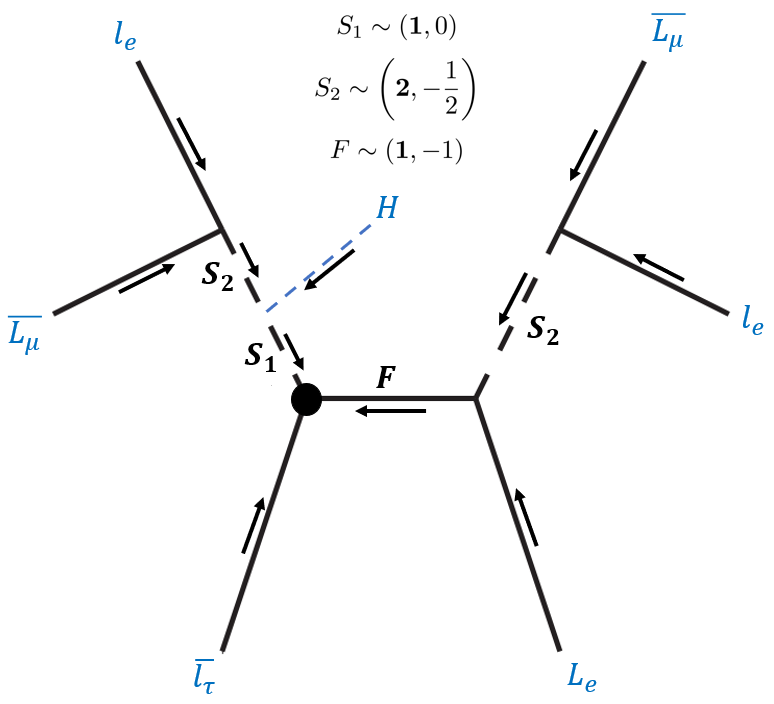}
    \caption{
        UV completion of the operator $\bar{L}_\mu \, \bar{L}_\mu \, \bar{l}_\tau \, l_e \, l_e \, L_e \, H$ of interest in Sec.~\ref{sec:UVstudy_dim10}. Arrows indicate the direction of momentum flow;  $SU(2)_L \times U(1)_Y$ quantum numbers of the new particles are also displayed.
    }
    \label{fig:uv_completion_full_diagram}
\end{figure}

We will study the decay channel $\tau \to eee\bar{\mu}\bar{\mu}$; switching electron and muon makes little difference. The relevant $d=10$ operators are listed in Tab.~\ref{tab:dim10LFVoperators} and their
UV completions compiled in App.~\ref{sec:UV_dim10}. We will focus on the operator $ \bar{L}_\mu  \bar{L}_\mu  \bar{\tau}_R  e_R  e_R  L_e  H$ and its UV completion by the diagram shown in Fig.~\ref{fig:uv_completion_full_diagram} via two new scalars $S_{1,2}$ and one fermion $F$; their charges under the SM gauge group and the conserved $U(1)_{L_e + 4 L_\mu - 5 L_\tau}$ are given in Tab.~\ref{tab:charge}.

\begin{table}[tb]
    \centering
    \renewcommand{\arraystretch}{1.2} % Increase row height
    \setlength{\tabcolsep}{8pt} % Increase horizontal spacing
    \begin{tabular}{|c|c|c|c|c|}
    \hline
         & $SU(3)_c$ & $SU(2)_L$ & $U(1)_Y$ & $U(1)_{L_e + 4 L_\mu - 5 L_\tau}$ \\
         \hline \hline
        $L_e$ & $\vec{1}$ & $\vec{2}$  & $-\tfrac12$ & $\phantom{-}1$ \\
        $e_R$ & $\vec{1}$ & $\vec{1}$  & $-1$ & $\phantom{-}1$  \\
        $L_\mu$ & $\vec{1}$ & $\vec{2}$  & $-\tfrac12$ & $\phantom{-}4$  \\
        $\mu_R$ & $\vec{1}$ & $\vec{1}$  & $-1$ & $\phantom{-}4$  \\
        $L_\tau$ & $\vec{1}$ & $\vec{2}$  & $-\tfrac12$ & $-5$  \\
        $\tau_R$ & $\vec{1}$ & $\vec{1}$  & $-1$ & $-5$  \\
        $F_{L,R}$ & $\vec{1}$ & $\vec{1}$  & $-1$ & $-2$  \\
        \hline
        $H$ & $\vec{1}$ & $\vec{2}$ & $\phantom{-}\tfrac12$ & $\phantom{-}0$  \\
        $S_1$ & $\vec{1}$ & $\vec{1}$ & $\phantom{-}0$ & $-3$  \\
        $S_2$ & $\vec{1}$ & $\vec{2}$ & $-\tfrac12$ & $-3$  \\
        \hline
    \end{tabular}
    \caption{Charge assignments for our UV completion in Sec.~\ref{sec:UVstudy_dim10}, $F$ and $S_{1,2}$ being new particles.
    }
    \label{tab:charge}
\end{table}
The  relevant couplings  of the new particles take the form
\begin{align}
    -{\cal L}\   &\supset \   Y_1 \bar{F}_L S_1 e_R + \tilde{Y}_1 \bar{F}_L S_1^\dagger \tau_R + Y_2 \bar{L}_\mu S_2^\dagger e_R +\tilde{Y}_2 \bar{L}_e S_2^\dagger F_R \notag \\
    & \quad + m_F \bar{F}_L F_R  + \mu \, S_2 H S_1^\dagger+ \hc  ,
\end{align}
with the imposed $U(1)_{L_e + 4 L_\mu - 5 L_\tau}$ forbidding many other terms that would induce LFV by one or two units. $\mu$~and $m_F$ are parameters with dimensions of mass, positive without loss of generality; the $Y$ are dimensionless Yukawa couplings. Together, these couplings give rise to the desired $\tau \to eee\bar{\mu}\bar{\mu}$ via Fig.~\ref{fig:uv_completion_full_diagram},  but of course more than that, as the new particles induce various non-LFV processes.
$F$ and $S_2$ in particular carry SM gauge charges and should thus be heavier than $m_Z/2$ to avoid egregious contributions to $Z$ decays. This is true even for the neutral component of $S_2$ because it remains a complex scalar carrying $U(1)_{L_e + 4 L_\mu - 5 L_\tau}$ charge, i.e.~is not split into two real scalars as in other two-Higgs-doublet models~\cite{Branco:2011iw}. 
$S_1$ on the other hand is an SM \textit{singlet} and could be much lighter, which is the key to enhance $\tau \to eee\bar{\mu}\bar{\mu}$. In fact, the biggest rate enhancement is achieved when $S_1$ is lighter than the tauon, splitting the five-body decay into the four-body decay $\tau \to ee \bar{\mu} S_1$ followed by near-instantaneous $S_1\to e\bar{\mu}$.

To calculate the relevant rates, we note that $\mu \, S_2 H S_1^\dagger$ mixes $S_1$ and  $S_2^0$ after electroweak symmetry breaking, with mixing angle
\begin{align}
    \theta = \frac{1}{2}\tan^{-1}\left(\frac{\sqrt{2} \mu v}{m_{S_2^0}^2-m_{S_1}^2}\right) \simeq \frac{\mu v}{\sqrt2 m_{S_2^0}^2} \,,
\end{align}
working in the small-angle limit for simplicity.
The neutral mass eigenstates $s_{1,2}$,
\begin{align}
    s_{1} &\equiv \cos\theta\ S_1 + \sin\theta\ S_2^0 \,,\\
    s_{2} &\equiv -\sin\theta\ S_1 + \cos\theta\ S_2^0\,,
\end{align}
are the relevant scalars for our tau decay, while the charged component of $S_2$ is assumed to be heavy-enough to be ignored for the time being. 
$F$ and $s_2$ are heavy enough to be integrated out, yielding the four-body tau decay rate
\begin{align}
    \Gamma ( \tau \rightarrow e e \bar{\mu} s_1 ) =
    \frac{2 |\tilde{Y}_2 \tilde{Y}_1 Y_2|^2}{m_{s_2}^4 m_F^2} \frac{m_\tau^7}{2949120 \pi^5} f\left(\frac{m_{s_1}^2}{m_\tau^2}\right)
    \label{eq:four-body_tau_decay}
\end{align}
with function
\begin{align}
\begin{split}
   f\left(x\right) &=  1 + \frac{155 }{4 }x + 20 x^2 - 55x^3 - 5 x^4 + \frac{1}{4 }x^5 \\
    &\quad + 15 x
    \left(1 + 4 x + 2x^2\right)
    \log x \,.
\end{split}
\end{align}
The scalar $s_1$ subsequently decays with rate
\begin{align}
    \Gamma(s_1\to e\bar{\mu}) \simeq \frac{m_{s_1} |Y_2|^2 \theta^2}{16\pi}
    \simeq \frac{\mu^2 |Y_2|^2 v^2  m_{s_1}}{32 \pi m_{s_2}^4} \,.
\end{align}
$s_1$ carries a typical momentum of $m_\tau/4$  in the tau rest frame, which gives a boosted decay length of 
\begin{align}
    L_\text{avg}(s_1) = \frac{m_\tau/4}{m_{s_1} \Gamma(s_1\to e\bar{\mu})} \,.
\end{align}
In most experiments, the tau rest frame is not identical to the lab frame, but we neglect this distinction here for simplicity.
The goal is to maximize $  \Gamma ( \tau \rightarrow e e \bar{\mu} s_1 )$ while keeping $ L_\text{avg} (s_1)$ short enough  so that $s_1$ decays \textit{inside} the detector. For this, we need to study limits on the relevant masses and couplings. Since the tau decay rate of Eq.~\eqref{eq:four-body_tau_decay} scales with $m_{s_2}^{-4}$, it is desirable to make the neutral $S_2$ component as light as possible.

\subsection{Limits}

$s_2\sim S_2^0$ has the $Y_2$ coupling to electrons and muons, which in particular induces an $\bar{e}e\bar{\mu}\mu$ four-lepton operator that conserves lepton flavor and number, but is still subject to experimental constraints:
\begin{itemize}

 \item It contributes to the anomalous magnetic moments of muon and electron, $\Delta a_{\mu,e}$, which in the limit  $m_\ell \ll m_{S_2^0}$ read as \cite{Lavoura:2003xp}
    \begin{equation}
        \Delta a_\ell = \frac{2 m_\ell^2 |Y_2|^2}{96 \pi^2 m_{S_2^0}^2 } \, ,
    \end{equation}
    the factor of 2 arising from the degenerate real scalar and pseudo-scalar inside the complex $S_2^0$.
    Comparing this with the experimental values $\Delta a_e = (-87 \pm 36)\times 10^{-14}$ \cite{Hanneke:2008tm,Aoyama:2017uqe,Parker:2018vye} and $\Delta a_\mu = (2.49 \pm 0.48)\times 10^{-9}$ \cite{Muong-2:2023cdq,Muong-2:2024hpx} gives the limits
    \begin{align}
        |Y_2| < 4.1 \left( \frac{m_{S_2^0}}{100\ {\rm GeV}} \right), \hspace{5mm} {\rm from\ }\Delta a_e \,,\\
        |Y_2| < 1.45 \left( \frac{m_{S_2^0}}{100\ {\rm GeV}} \right), \hspace{5mm} {\rm from\ }\Delta a_\mu\,.
    \end{align}
    Here, we took the $5\sigma$ C.L.~range of the measured values to derive these limits, indicating that the maximum allowed values for a given mediator mass beyond which new physics contributions would exceed the experimentally measured values.
    Given the current mismatch between experimental measurement~\cite{Muong-2:2023cdq,Muong-2:2024hpx} and SM prediction~\cite{Aoyama:2020ynm,Colangelo:2022jxc} for $\Delta a_\mu$, the limit should be taken with a grain of salt. In our case, $\Delta a_\ell$ gives weaker limits than LEP anyway, so the details are less important.
    
    \item Integrating out $S_2^0$ gives the four-fermion operator
    \begin{align}
        {\cal L}_{\rm eff} =& -\frac{|Y_{2}|^2}{2 m_{S_2^0}^2}   (\bar{e}_R \gamma^\alpha e_R) (\bar{\mu}_L \gamma_\alpha \mu_L)   
    \end{align}  
    after performing a Fierz transformation. 
Contact interactions such as this one modify $e^+ e^- \to \mu^+ \mu^-$  and have been constrained by LEP~\cite{ALEPH:2013dgf}. Specifically, the above interaction corresponds to the $\Lambda^-_{LR,\mu\mu}$ parameter~\cite{Eichten:1983hw}, which is the least-constrained of all $ee\mu\mu$ contact interactions by far, with Ref.~\cite{ALEPH:2013dgf} quoting $\unit[2.2]{TeV}$. In App.~\ref{sec:LEPlimits}, we explain in detail that this is due to an approximate degeneracy in the SM$+\Lambda^-_{LR,\mu\mu}$ cross section in the narrow energy region used in the LEP fit, and that the inclusion of low-$\sqrt{s}$ data from TRISTAN~\cite{VENUS:1997bjf} lifts the degeneracy and enforces the much stronger limit $\Lambda^-_{LR,\mu\mu} > \unit[8]{TeV}$, or
\begin{align}
    m_{S_2^0}/|Y_2| > \unit[1.6]{TeV}\,.
\end{align}
In any case, this limit is only valid for $S_2$ masses far above LEP energies, not exactly the region of interest for us. Instead, we calculate the full cross section $e^+ e^- \to \mu^+ \mu^-$ and fit it to the same cross-section and forward-backward asymmetry data that underlies the contact-interaction limits to find the exclusion curve in Fig.~\ref{fig:mu-e-scalar_limits}, which reduces to the $\Lambda^-_{LR,\mu\mu}$ limit for large masses but flattens out to $Y_2  \simeq 0.1$ for light $S_2^0$. Together with the $Z$-decay limit, $m_{S_2^0}> m_Z/2$, this is stronger than the magnetic-moment constraints and even future MUonE~\cite{MUonE:2016hru} sensitivity to deviations in $\mu e\to \mu e$ scattering from such scalars~\cite{Dev:2020drf}.

      \begin{figure}[tb]
        \centering
\includegraphics[width=0.45\textwidth]{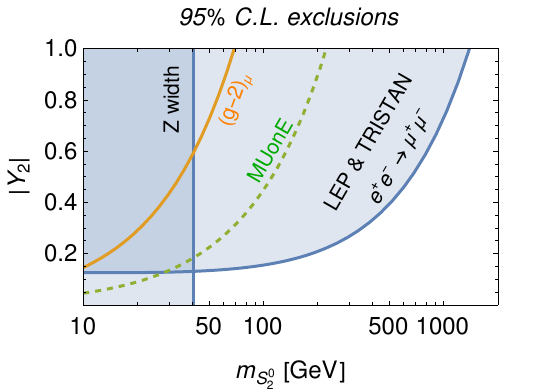}
        \caption{Limits on the complex neutral scalar $S_2^0$ with $\mu_L$--$e_R$ coupling $Y_2$, see text for details.}
        \label{fig:mu-e-scalar_limits}
    \end{figure}

\item $f\bar{f}\to Z^*\to (S_2^0\to \mu^+ e^-) (\bar{S}_2^0 \to \mu^- e^+)$ could provide excellent bounds for $m_{S_2^0} < \unit[100]{GeV}$ (LEP-II) or  $m_{S_2^0} < \mathcal{O}(\unit{TeV})$ (LHC), although we are not aware of such searches.\footnote{Aside from different lepton flavors, similar signatures have been proposed in Ref.~\cite{Iguro:2019sly,Afik:2022vpm}.} Given the clean signature and large cross section for this process, we will conservatively assume that this would have been observed; evading such limits is possible for $m_{S_2^0} > m_h+m_{S_1}$, since then the decay $S_2^0\to h S_1$ becomes kinematically allowed and can dominate over $S_2^0\to \mu^+ e^-$ if $\mu \gg \sqrt2 m_{S_2^0} Y_2$ (or $\theta \gg Y_2 v/m_{S_2^0}$), which is doable even in the small-$\theta$ regime. This will severely weaken LEP and LHC constraints on $S_2^0$, likely opening up parameter space around $Y_2\sim 0.1$, $m_{S_2^0}\gtrsim m_h$.
Again, dedicated searches for this kind of neutral scalar could drastically improve limits, so we encourage our experimental colleagues to look for this clean signature. 
Notice that this mass hierarchy also eliminates $h\to s_1 s_2$.

\end{itemize}

The charged component of $S_2$ decays almost exclusively via $S_2^-\to e^- \bar{\nu}_\mu$ and does not couple to quarks, unlike the typical charged scalars under scrutiny at LHC~\cite{ATLAS:2024itc}. It does mimic a left-handed selectron $\tilde e_L$ in the  massless-neutralino limit in existing searches for supersymmetric particles~\cite{ATLAS:2019lff,CMS:2020bfa}. 
The ATLAS analysis of the LHC Run-2 data~\cite{ATLAS:2019lff} provides a lower bound of $m_{S_2^+} > \unit[550]{GeV}$ at 95\% C.L.
This does not directly impact the tau decay of interest to us, since the diagram shown in Fig.~\ref{fig:uv_completion_full_diagram} does not involve $S_2^-$, but a large mass splitting within the $S_2$ doublet will contribute to the oblique parameters, notably $T$~\cite{Peskin:1990zt,Peskin:1991sw}. This effect arises at loop level and potentially interferes with other new physics in the electroweak sector, so we will not take this limit into account here.

The light complex singlet $s_1\sim S_1$ inherits a coupling to the $Z$ boson from the mixing with $S_2^0$, which makes it available in $Z$ decays:
\begin{align}
    \BR (Z\to s_1\bar{s}_1)\simeq \frac{3 \theta^4 }{86} \,.
\end{align}
 The total $Z$ width agrees very well with the SM prediction~\cite{ParticleDataGroup:2024cfk}, which
can be translated into a $2\sigma$ upper bound of $2\times 10^{-3}$ on any non-SM $Z$ branching ratio, enforcing $|\theta| < 0.5$, easily satisfied in the small-mixing-angle regime. A dedicated search for $Z\to s_1\bar{s}_1\to \mu^+\mu^-e^+e^-$, typically with displaced $s_1$ vertex, could vastly improve this limit.

Lastly, the new charged fermion $F$ was implicitly taken to be heavier than $S_2^0$ above, because that hierarchy maximizes the tau decay of Eq.~\eqref{eq:four-body_tau_decay}. Notice that $F$ does not mix with the SM leptons on account of the $U(1)_{L_e + 4 L_\mu - 5 L_\tau}$ symmetry. It can decay into $e S_1$ through the $Y_1$ coupling, which does not feed into $\tau \to ee \bar{\mu} s_1$, or into $\tau \bar{S}_1$ via $\tilde Y_1$, or into $L_e S_2$ via $\tilde Y_2$. The decays into $S_1$ will lead to displaced-vertex signatures, the decay $F^-\to e^- (S_2^0\to \mu^+ e^-)$ will be prompt. We are not aware of searches for such final states, although they should be fantastically clean.
Since $S_1$ will decay displaced, it might appear as missing energy, rendering the signature similar to right-handed stau searches, $pp\to \tilde \tau_R \tilde \tau_R \to \tau\tau+\slashed{E}$. Recent ATLAS limits exclude stau masses below \unit[350]{GeV}, assuming 100\% decays into $\tau+\slashed{E}$~\cite{ATLAS:2024fub}. The limits on our fermion will likely be stronger, since the Drell-Yan cross section into fermions is roughly a factor 4 larger than into scalars, albeit with a more complicated relationship near threshold.
Once again we encourage dedicated searches for this clean setup, esp.~$pp\to F^-F^+ $ followed by the decay chain $F^-\to e^- (S_2^0\to \mu^+ e^-)$. For now, we assume that order-one couplings of $F$ with $m_F\sim\unit[350]{GeV}$ are currently allowed.

\subsection{Discussion}

\begin{figure}[h]
        \centering
        \includegraphics[width=0.45\textwidth]{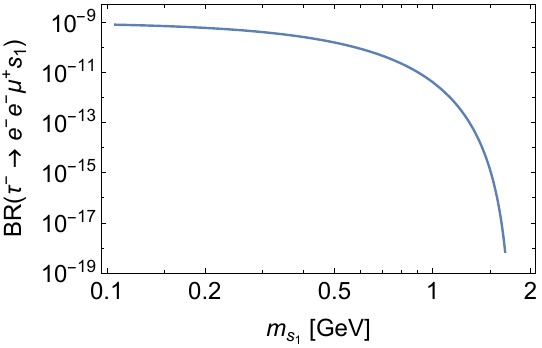}
        \caption{Branching ratio $\BR ( \tau \rightarrow e e \bar{\mu} s_1 )$ as function of $m_{s_1}$ for the benchmark values from Eq.~\eqref{eq:benchmark}. Together with the near-prompt decay $s_1\to e\bar{\mu}$ (see Fig.~\ref{fig:decayLength}), this gives the LFV-by-three-units decay $\tau \to eee\bar{\mu}\bar{\mu}$.}
        \label{fig:branchingRatio}
\end{figure}

\begin{figure}[h]
        \centering
        \includegraphics[width=0.45\textwidth]{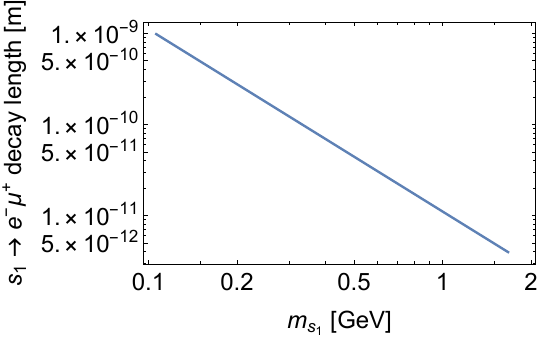}
        \caption{ Estimated decay length of $s_1$ following the tau decay $\tau \rightarrow e e \bar{\mu} s_1$ as a function of $m_{s_1}$, using the benchmark values from Eq.~\eqref{eq:benchmark}. }
        \label{fig:decayLength}
\end{figure}

We will take the following benchmark point, which satisfies current collider constraints:
\begin{align}
\begin{split}
 m_{s_2}&=\unit[140]{GeV}\,, \quad m_F=\unit[350]{GeV} \,, \quad \theta = 0.2 \,, \\
 Y_2 &= 0.1\,, \qquad\qquad \tilde Y_1 = \tilde Y_2 = 3\,,
\end{split}
\label{eq:benchmark}
\end{align}
with $m_{s_1}$ varying from $m_\mu + m_e$ (to allow for $s_1\to e\bar{\mu}$) to $m_\tau - m_\mu - 2 m_e$ (to enable $\tau \to ee\bar{\mu} s_1$. The resulting branching ratio for the LFV-by-three-units tauon decay is shown in Fig.~\ref{fig:branchingRatio}.
For very light $s_1$, near the threshold for it to decay into $\mu e$, the tau branching ratio can be of order $10^{-9}$, potentially in reach of Belle II. Dedicated collider searches for the new particles could probe the same region of parameter space. Since $s_1$ is very light here, its decay length (in the tau rest frame) is fairly long, see Fig.~\ref{fig:decayLength}, although still below a nanometer, rendering the tau decay $\tau \to eee\bar{\mu}\bar{\mu}$ effectively prompt.

Let this suffice as an example for the challenges involved in generating a large $\tau \to eee\bar{\mu}\bar{\mu}$ rate; other operators and UV completions might fare better or worse, our intend is not to promote one particular model but rather to emphasize that LFV-by-three-units signatures are not entirely hopeless but can be probed at colliders and $\tau$ factories.

\section{Conclusions}
\label{sec:conclusions}

Searches for lepton flavor violation are as old as the discovery of the muon and have steadily improved ever since, but have yet to discover signs of it. This is all the more surprising since neutrino oscillations have proven beyond doubt that lepton flavor is violated in nature. It appears as though lepton flavor is at least an excellent \textit{approximate} symmetry, neutrino masses aside. This makes possible different flavor breaking patterns in the neutrino and charged lepton sector, which could give rise to rather unusual signatures, for example flavor violation by \textit{three} units. In this article, we have shown that it is indeed theoretically possible to heavily suppress flavor violation by one or two units, making three units the dominant channel. In the SMEFT, such operators arise first at dimension 7, but come together with total lepton number violation, forcing neutrinos into any and all such processes. Still, $\ell_\alpha \to \ell_\beta \nu_\beta \nu_\beta$ provide testable signals of such operators, conveniently parametrized through Michel parameters. \textit{Charged} lepton flavor violation by three units first arises at mass dimension 10, with golden modes $\tau \to eee\bar{\mu}\bar{\mu}$ or $\tau \to \mu\mu\mu\bar{e}\bar{e}$, but with absurdly suppressed rates within the SMEFT framework. We provide tree-level UV completions for all operators of interest in this study, and investigate one example in which a new particle is lighter than the tau, enabling the parametrically faster decay $\tau\to ee\bar{\mu} (s\to e\bar{\mu})$. In this model, tau decay rates in reach of Belle II seem possible, but the same model can also be investigated at the LHC in a variety of clean signatures ranging from $pp\to Z^*\to S_2^0 \bar{S}_2^0 \to \mu^+ e^- \mu^- e^+$ to $pp\to \gamma^*/Z^*\to F^- F^+\to e^- S_2^0 e^+ S_2^0\to e^- \mu^+ e^- e^+ \mu^- e^-$. We encourage our experimental colleagues to have a look at these novel signatures, which should be useful beyond our model example and might reveal particles overlooked so far.

%%%%%%%%%%%%%%%%%%%%%%%%%

\section*{Acknowledgements}
JH thanks Martin Gruenewald and Dimitri Bourilkov for their patient correspondence regarding Ref.~\cite{ALEPH:2013dgf}. We've made excessive use of Renato Fonseca's Mathematica packages \texttt{Sym2Int}~\cite{Fonseca:2017lem} and \texttt{GroupMath}~\cite{Fonseca:2020vke}.
This work was supported by the U.S. Department of Energy under Grant No. DE-SC0007974.

\appendix

\section{UV completions for \texorpdfstring{$d=7$}{d=7} operators}
\label{sec:UV_dim7}

Here, we list all tree-level UV completions for the $d=7$ operators $\bar{l} \, L \, L \, L \, H $ of interest in Sec.~\ref{sec:dim7}. Notice that we list \emph{all} UV completions, even though not all of them give the LFV-by-three-units decays we are interested in, because they can come with unwelcome  antisymmetric couplings: the scalar $S\sim \left(\vec{1} , \vec{1} , 1\right)$ couples to $\bar{L}^c L$, which is \textit{antisymmetric} in flavor space, meaning it does not contain the desired  $L_eL_e$ coupling for Eq.~\eqref{eq:d7operator}. For this, we need to use the triplet scalar $S\sim \left(\vec{1} , \vec{3} , 1\right)$, which has a \textit{symmetric} coupling to $\bar{L}^c L$.

This Appendix consists of two sections, each starting with a Feynman diagram of a given tree-level topology. Diagrams are followed by several tables containing information about gauge numbers of mediating particles. Each row contains a set of gauge quantum numbers in $SU(3)_C\times SU(2)_L\times U(1)_Y$ space corresponding to each mediating particle. Table captions display which layout of external particles allows for the rows of mediators listed in the respective tables.

\subsection{First Topology}
\begin{figure}[H]
    \centering
    \includegraphics[width=0.35\textwidth]{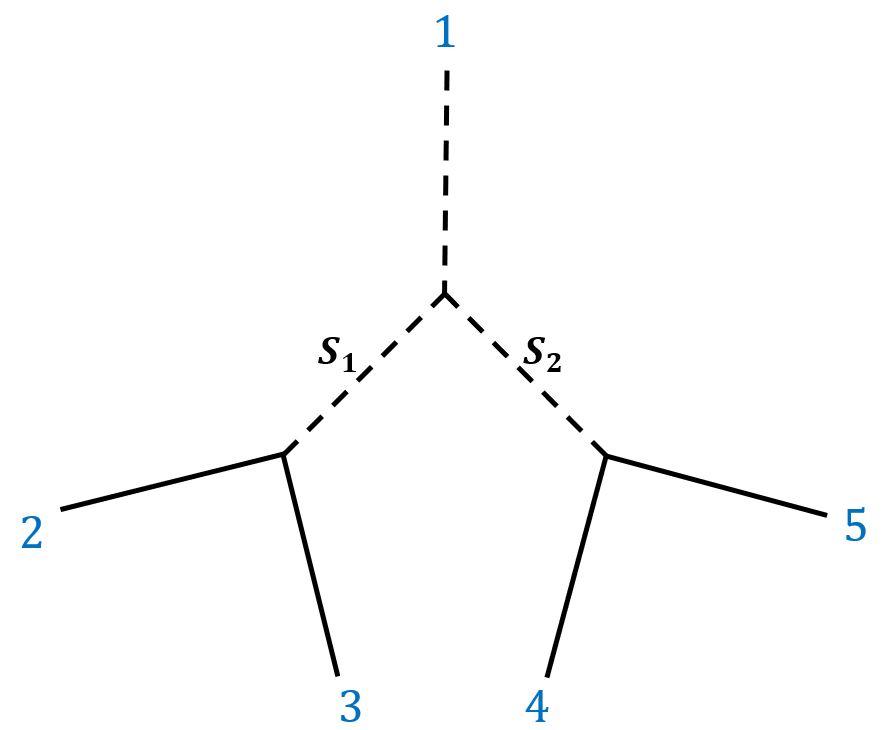}
    \caption{Feynman diagram for Tab.~\ref{tab:d7_first_first}.}
\end{figure}

\begin{table}[H]
\centering
\renewcommand{\arraystretch}{1.4}
\setlength{\tabcolsep}{0.15cm} % Adjust the column separation to make it narrower
\begin{tabular}{|>{\centering\arraybackslash}p{1.5cm}|>{\centering\arraybackslash}p{1.5cm}|}
\hline
\textbf{$\vec{S_1}$} & \textbf{$\vec{S_2}$} \\
\hhline{|==|}
$\left(\vec{1} , \vec{2} , \frac{1}{2}\right)$ & $\left(\vec{1} , \vec{3} , 1\right)$ \\
\hline
$\left(\vec{1} , \vec{2} , \frac{1}{2}\right)$ & $\left(\vec{1} , \vec{1} , 1\right)$ \\
\hline\end{tabular}
\caption{$1 \rightarrow H$, $2 \rightarrow \overline{l}$, $3 \rightarrow L$, $4 \rightarrow L$, $5 \rightarrow L$}
\label{tab:d7_first_first}
\end{table}

\subsection{Second Topology}

\begin{figure}[H]
    \centering
    \includegraphics[width=0.35\textwidth]{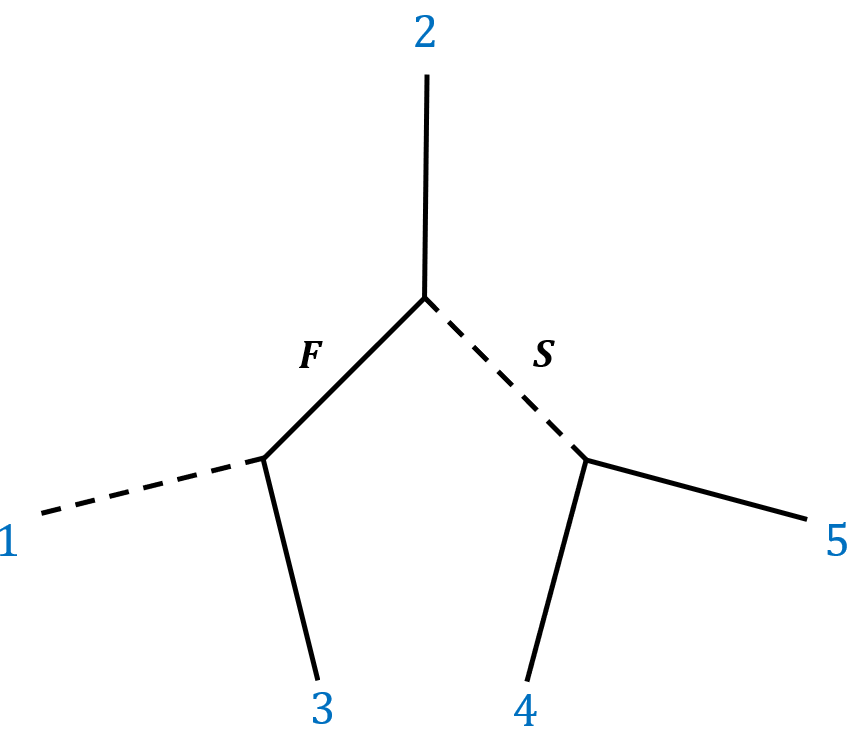}
    \caption{Feynman diagram for  Tabs.~\ref{tab:d7_second_first}--\ref{tab:d7_second_last}.}
\end{figure}

\begin{table}[H]
\centering
\renewcommand{\arraystretch}{1.4}
\setlength{\tabcolsep}{0.15cm} % Adjust the column separation to make it narrower
\begin{tabular}{|>{\centering\arraybackslash}p{1.5cm}|>{\centering\arraybackslash}p{1.5cm}|}
\hline
\textbf{$\vec{F}$} & \textbf{$\vec{S}$} \\
\hhline{|==|}
$\left(\vec{1} , \vec{3} , 0\right)$ & $\left(\vec{1} , \vec{3} , 1\right)$ \\
\hline
$\left(\vec{1} , \vec{1} , 0\right)$ & $\left(\vec{1} , \vec{1} , 1\right)$ \\
\hline\end{tabular}
\caption{$1 \rightarrow H$, $2 \rightarrow \overline{l}$, $3 \rightarrow L$, $4 \rightarrow L$, $5 \rightarrow L$}
\label{tab:d7_second_first}
\end{table}

\begin{table}[H]
\centering
\renewcommand{\arraystretch}{1.4}
\setlength{\tabcolsep}{0.15cm} % Adjust the column separation to make it narrower
\begin{tabular}{|>{\centering\arraybackslash}p{1.5cm}|>{\centering\arraybackslash}p{1.5cm}|}
\hline
\textbf{$\vec{F}$} & \textbf{$\vec{S}$} \\
\hhline{|==|}
$\left(\vec{1} , \vec{2} , \frac{3}{2}\right)$ & $\left(\vec{1} , \vec{3} , 1\right)$ \\
\hline
$\left(\vec{1} , \vec{2} , \frac{3}{2}\right)$ & $\left(\vec{1} , \vec{1} , 1\right)$ \\
\hline\end{tabular}
\caption{$1 \rightarrow H$, $2 \rightarrow L$, $3 \rightarrow \overline{l}$, $4 \rightarrow L$, $5 \rightarrow L$}
\end{table}

\begin{table}[H]
\centering
\renewcommand{\arraystretch}{1.4}
\setlength{\tabcolsep}{0.15cm} % Adjust the column separation to make it narrower
\begin{tabular}{|>{\centering\arraybackslash}p{1.5cm}|>{\centering\arraybackslash}p{1.5cm}|}
\hline
\textbf{$\vec{F}$} & \textbf{$\vec{S}$} \\
\hhline{|==|}
$\left(\vec{1} , \vec{3} , 0\right)$ & $\left(\vec{1} , \vec{2} , -\frac{1}{2}\right)$ \\
\hline
$\left(\vec{1} , \vec{1} , 0\right)$ & $\left(\vec{1} , \vec{2} , -\frac{1}{2}\right)$ \\
\hline\end{tabular}
\caption{$1 \rightarrow H$, $2 \rightarrow L$, $3 \rightarrow L$, $4 \rightarrow \overline{l}$, $5 \rightarrow L$}
\label{tab:d7_second_last}
\end{table}

\section{UV completions for \texorpdfstring{$d=9$}{d=9}  operators}
\label{sec:UV_dim9}

Here, we list tree-level UV completions for the $d=9$ operators of interest in Sec.~\ref{sec:dim9}. Notation and organization follow App.~\ref{sec:UV_dim7}. 

\subsection{First Topology}
\begin{figure}[H]
    \centering
    \includegraphics[width=0.35\textwidth]{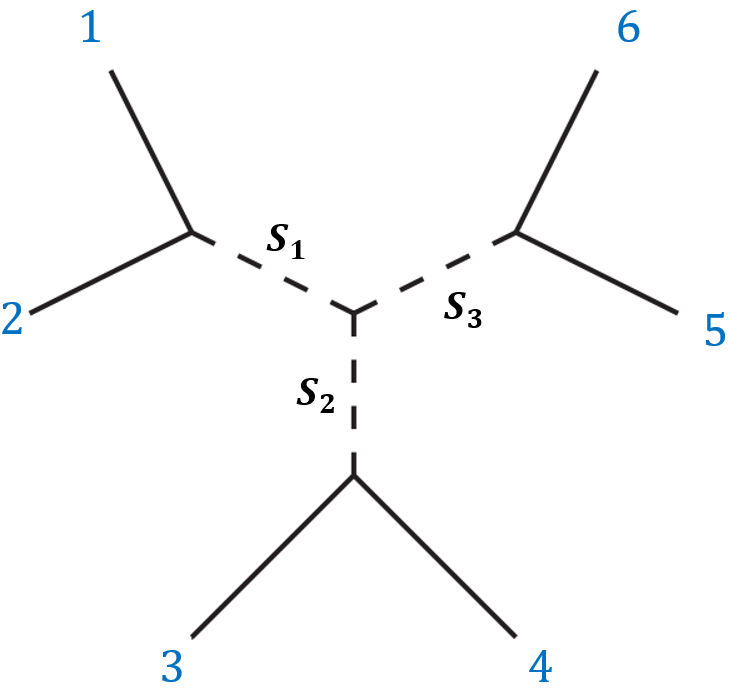}
    \caption{Feynman diagram for the given topology. The corresponding tables are Tab.~\ref{tab:d9_first_first} - Tab.~\ref{tab:d9_first_last}.}
\end{figure}

\subsubsection{$\overline{l}\, \overline{l}\, L\, L\, L\, L$}

\begin{table}[H]
\centering
\renewcommand{\arraystretch}{1.4}
\setlength{\tabcolsep}{0.15cm} % Adjust the column separation to make it narrower
\begin{tabular}{|>{\centering\arraybackslash}p{1.5cm}|>{\centering\arraybackslash}p{1.5cm}|>{\centering\arraybackslash}p{1.5cm}|}
\hline
\textbf{$\vec{S_1}$} & \textbf{$\vec{S_2}$} & \textbf{$\vec{S_3}$} \\
\hhline{|===|}
$\left(\vec{1} , \vec{1} , 2\right)$ & $\left(\vec{1} , \vec{3} , -1\right)$ & $\left(\vec{1} , \vec{3} , -1\right)$ \\
\hline
$\left(\vec{1} , \vec{1} , 2\right)$ & $\left(\vec{1} , \vec{1} , -1\right)$ & $\left(\vec{1} , \vec{1} , -1\right)$ \\
\hline\end{tabular}
\caption{$1 \rightarrow \overline{l}$, $2 \rightarrow \overline{l}$, $3 \rightarrow L$, $4 \rightarrow L$, $5 \rightarrow L$, $6 \rightarrow L$}
\label{tab:d9_first_first}
\end{table}

\begin{table}[H]
\centering
\renewcommand{\arraystretch}{1.4}
\setlength{\tabcolsep}{0.15cm} % Adjust the column separation to make it narrower
\begin{tabular}{|>{\centering\arraybackslash}p{1.5cm}|>{\centering\arraybackslash}p{1.5cm}|>{\centering\arraybackslash}p{1.5cm}|}
\hline
\textbf{$\vec{S_1}$} & \textbf{$\vec{S_2}$} & \textbf{$\vec{S_3}$} \\
\hhline{|===|}
$\left(\vec{1} , \vec{2} , \frac{1}{2}\right)$ & $\left(\vec{1} , \vec{2} , \frac{1}{2}\right)$ & $\left(\vec{1} , \vec{3} , -1\right)$ \\
\hline
$\left(\vec{1} , \vec{2} , \frac{1}{2}\right)$ & $\left(\vec{1} , \vec{2} , \frac{1}{2}\right)$ & $\left(\vec{1} , \vec{1} , -1\right)$ \\
\hline\end{tabular}
\caption{$1 \rightarrow \overline{l}$, $2 \rightarrow L$, $3 \rightarrow \overline{l}$, $4 \rightarrow L$, $5 \rightarrow L$, $6 \rightarrow L$\\
}
\end{table}

\subsubsection{$\overline{d}\, \overline{l}\, u\, l\, L\, L$}

\begin{table}[H]
\centering
\renewcommand{\arraystretch}{1.4}
\setlength{\tabcolsep}{0.15cm} % Adjust the column separation to make it narrower
\begin{tabular}{|>{\centering\arraybackslash}p{1.5cm}|>{\centering\arraybackslash}p{1.5cm}|>{\centering\arraybackslash}p{1.5cm}|}
\hline
\textbf{$\vec{S_1}$} & \textbf{$\vec{S_2}$} & \textbf{$\vec{S_3}$} \\
\hhline{|===|}
$\left(\vec{3} , \vec{1} , -\frac{1}{3}\right)$ & $\left(\vec{\overline{3}} , \vec{1} , \frac{4}{3}\right)$ & $\left(\vec{1} , \vec{1} , -1\right)$ \\
\hline\end{tabular}
\caption{$1 \rightarrow l$, $2 \rightarrow u$, $3 \rightarrow \overline{l}$, $4 \rightarrow \overline{d}$, $5 \rightarrow L$, $6 \rightarrow L$}
\end{table}

\begin{table}[H]
\centering
\renewcommand{\arraystretch}{1.4}
\setlength{\tabcolsep}{0.15cm} % Adjust the column separation to make it narrower
\begin{tabular}{|>{\centering\arraybackslash}p{1.5cm}|>{\centering\arraybackslash}p{1.5cm}|>{\centering\arraybackslash}p{1.5cm}|}
\hline
\textbf{$\vec{S_1}$} & \textbf{$\vec{S_2}$} & \textbf{$\vec{S_3}$} \\
\hhline{|===|}
$\left(\vec{3} , \vec{1} , -\frac{1}{3}\right)$ & $\left(\vec{1} , \vec{2} , \frac{1}{2}\right)$ & $\left(\vec{\overline{3}} , \vec{2} , -\frac{1}{6}\right)$ \\
\hline\end{tabular}
\caption{$1 \rightarrow l$, $2 \rightarrow u$, $3 \rightarrow \overline{l}$, $4 \rightarrow L$, $5 \rightarrow \overline{d}$, $6 \rightarrow L$}
\label{tab:d9_first_last}
\end{table}

\subsection{Second Topology}
\begin{figure}[H]
    \centering
    \includegraphics[width=0.35\textwidth]{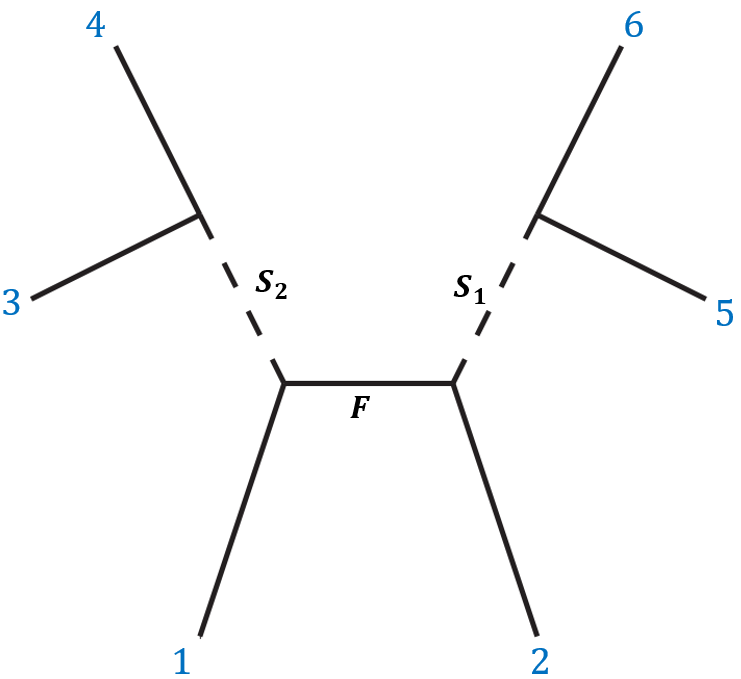}
    \caption{Feynman diagram for the given topology. The corresponding tables are Tab.~\ref{tab:d9_second_first} - Tab.~\ref{tab:d9_second_last}.}
\end{figure}

\subsubsection{$\bar{l} \, \bar{l} \, L \, L \, L \, L$}

\begin{table}[H]
\centering
\renewcommand{\arraystretch}{1.4}
\setlength{\tabcolsep}{0.15cm} % Adjust the column separation to make it narrower
\begin{tabular}{|>{\centering\arraybackslash}p{1.5cm}|>{\centering\arraybackslash}p{1.5cm}|>{\centering\arraybackslash}p{1.5cm}|}
\hline
\textbf{$\vec{S_1}$} & \textbf{$\vec{F}$} & \textbf{$\vec{S_2}$} \\
\hhline{|===|}
$\left(\vec{1} , \vec{3} , -1\right)$ & $\left(\vec{1} , \vec{3} , 0\right)$ & $\left(\vec{1} , \vec{3} , 1\right)$ \\
\hline
$\left(\vec{1} , \vec{1} , -1\right)$ & $\left(\vec{1} , \vec{1} , 0\right)$ & $\left(\vec{1} , \vec{1} , 1\right)$ \\
\hline\end{tabular}
\caption{$1 \rightarrow \overline{l}$, $2 \rightarrow \overline{l}$, $3 \rightarrow L$, $4 \rightarrow L$, $5 \rightarrow L$, $6 \rightarrow L$}
\label{tab:d9_second_first}
\end{table}

\begin{table}[H]
\centering
\renewcommand{\arraystretch}{1.4}
\setlength{\tabcolsep}{0.15cm} % Adjust the column separation to make it narrower
\begin{tabular}{|>{\centering\arraybackslash}p{1.5cm}|>{\centering\arraybackslash}p{1.5cm}|>{\centering\arraybackslash}p{1.5cm}|}
\hline
\textbf{$\vec{S_1}$} & \textbf{$\vec{F}$} & \textbf{$\vec{S_2}$} \\
\hhline{|===|}
$\left(\vec{1} , \vec{3} , -1\right)$ & $\left(\vec{1} , \vec{2} , -\frac{3}{2}\right)$ & $\left(\vec{1} , \vec{2} , -\frac{1}{2}\right)$ \\
\hline
$\left(\vec{1} , \vec{1} , -1\right)$ & $\left(\vec{1} , \vec{2} , -\frac{3}{2}\right)$ & $\left(\vec{1} , \vec{2} , -\frac{1}{2}\right)$ \\
\hline\end{tabular}
\caption{$1 \rightarrow \overline{l}$, $2 \rightarrow L$, $3 \rightarrow \overline{l}$, $4 \rightarrow L$, $5 \rightarrow L$, $6 \rightarrow L$}
\end{table}

\begin{table}[H]
\centering
\renewcommand{\arraystretch}{1.4}
\setlength{\tabcolsep}{0.15cm} % Adjust the column separation to make it narrower
\begin{tabular}{|>{\centering\arraybackslash}p{1.5cm}|>{\centering\arraybackslash}p{1.5cm}|>{\centering\arraybackslash}p{1.5cm}|}
\hline
\textbf{$\vec{S_1}$} & \textbf{$\vec{F}$} & \textbf{$\vec{S_2}$} \\
\hhline{|===|}
$\left(\vec{1} , \vec{2} , \frac{1}{2}\right)$ & $\left(\vec{1} , \vec{3} , 0\right)$ & $\left(\vec{1} , \vec{3} , 1\right)$ \\
\hline
$\left(\vec{1} , \vec{2} , \frac{1}{2}\right)$ & $\left(\vec{1} , \vec{1} , 0\right)$ & $\left(\vec{1} , \vec{1} , 1\right)$ \\
\hline\end{tabular}
\caption{$1 \rightarrow \overline{l}$, $2 \rightarrow L$, $3 \rightarrow L$, $4 \rightarrow L$, $5 \rightarrow \overline{l}$, $6 \rightarrow L$}
\end{table}

\begin{table}[H]
\centering
\renewcommand{\arraystretch}{1.4}
\setlength{\tabcolsep}{0.15cm} % Adjust the column separation to make it narrower
\begin{tabular}{|>{\centering\arraybackslash}p{1.5cm}|>{\centering\arraybackslash}p{1.5cm}|>{\centering\arraybackslash}p{1.5cm}|}
\hline
\textbf{$\vec{S_1}$} & \textbf{$\vec{F}$} & \textbf{$\vec{S_2}$} \\
\hhline{|===|}
$\left(\vec{1} , \vec{3} , -1\right)$ & $\left(\vec{1} , \vec{2} , -\frac{3}{2}\right)$ & $\left(\vec{1} , \vec{1} , -2\right)$ \\
\hline
$\left(\vec{1} , \vec{1} , -1\right)$ & $\left(\vec{1} , \vec{2} , -\frac{3}{2}\right)$ & $\left(\vec{1} , \vec{1} , -2\right)$ \\
\hline\end{tabular}
\caption{$1 \rightarrow L$, $2 \rightarrow L$, $3 \rightarrow \overline{l}$, $4 \rightarrow \overline{l}$, $5 \rightarrow L$, $6 \rightarrow L$}
\end{table}

\begin{table}[H]
\centering
\renewcommand{\arraystretch}{1.4}
\setlength{\tabcolsep}{0.15cm} % Adjust the column separation to make it narrower
\begin{tabular}{|>{\centering\arraybackslash}p{1.5cm}|>{\centering\arraybackslash}p{1.5cm}|>{\centering\arraybackslash}p{1.5cm}|}
\hline
\textbf{$\vec{S_1}$} & \textbf{$\vec{F}$} & \textbf{$\vec{S_2}$} \\
\hhline{|===|}
$\left(\vec{1} , \vec{2} , \frac{1}{2}\right)$ & $\left(\vec{1} , \vec{3} , 0\right)$ & $\left(\vec{1} , \vec{2} , -\frac{1}{2}\right)$ \\
\hline
$\left(\vec{1} , \vec{2} , \frac{1}{2}\right)$ & $\left(\vec{1} , \vec{1} , 0\right)$ & $\left(\vec{1} , \vec{2} , -\frac{1}{2}\right)$ \\
\hline\end{tabular}
\caption{$1 \rightarrow L$, $2 \rightarrow L$, $3 \rightarrow \overline{l}$, $4 \rightarrow L$, $5 \rightarrow \overline{l}$, $6 \rightarrow L$}
\end{table}

\subsubsection{$\overline{d}\, \overline{l}\, u\, l\, L\, L$}

\begin{table}[H]
\centering
\renewcommand{\arraystretch}{1.4}
\setlength{\tabcolsep}{0.15cm} % Adjust the column separation to make it narrower
\begin{tabular}{|>{\centering\arraybackslash}p{1.5cm}|>{\centering\arraybackslash}p{1.5cm}|>{\centering\arraybackslash}p{1.5cm}|}
\hline
\textbf{$\vec{S_1}$} & \textbf{$\vec{F}$} & \textbf{$\vec{S_2}$} \\
\hhline{|===|}
$\left(\vec{1} , \vec{1} , -1\right)$ & $\left(\vec{3} , \vec{1} , -\frac{1}{3}\right)$ & $\left(\vec{3} , \vec{1} , -\frac{4}{3}\right)$ \\
\hline\end{tabular}
\caption{$1 \rightarrow l$, $2 \rightarrow u$, $3 \rightarrow \overline{l}$, $4 \rightarrow \overline{d}$, $5 \rightarrow L$, $6 \rightarrow L$}
\end{table}

\begin{table}[H]
\centering
\renewcommand{\arraystretch}{1.4}
\setlength{\tabcolsep}{0.15cm} % Adjust the column separation to make it narrower
\begin{tabular}{|>{\centering\arraybackslash}p{1.5cm}|>{\centering\arraybackslash}p{1.5cm}|>{\centering\arraybackslash}p{1.5cm}|}
\hline
\textbf{$\vec{S_1}$} & \textbf{$\vec{F}$} & \textbf{$\vec{S_2}$} \\
\hhline{|===|}
$\left(\vec{\overline{3}} , \vec{2} , -\frac{1}{6}\right)$ & $\left(\vec{1} , \vec{2} , \frac{1}{2}\right)$ & $\left(\vec{1} , \vec{2} , -\frac{1}{2}\right)$ \\
\hline\end{tabular}
\caption{$1 \rightarrow l$, $2 \rightarrow u$, $3 \rightarrow \overline{l}$, $4 \rightarrow L$, $5 \rightarrow \overline{d}$, $6 \rightarrow L$}
\end{table}

\begin{table}[H]
\centering
\renewcommand{\arraystretch}{1.4}
\setlength{\tabcolsep}{0.15cm} % Adjust the column separation to make it narrower
\begin{tabular}{|>{\centering\arraybackslash}p{1.5cm}|>{\centering\arraybackslash}p{1.5cm}|>{\centering\arraybackslash}p{1.5cm}|}
\hline
\textbf{$\vec{S_1}$} & \textbf{$\vec{F}$} & \textbf{$\vec{S_2}$} \\
\hhline{|===|}
$\left(\vec{1} , \vec{2} , \frac{1}{2}\right)$ & $\left(\vec{3} , \vec{2} , \frac{7}{6}\right)$ & $\left(\vec{3} , \vec{2} , \frac{1}{6}\right)$ \\
\hline\end{tabular}
\caption{$1 \rightarrow l$, $2 \rightarrow u$, $3 \rightarrow \overline{d}$, $4 \rightarrow L$, $5 \rightarrow \overline{l}$, $6 \rightarrow L$}
\end{table}

\begin{table}[H]
\centering
\renewcommand{\arraystretch}{1.4}
\setlength{\tabcolsep}{0.15cm} % Adjust the column separation to make it narrower
\begin{tabular}{|>{\centering\arraybackslash}p{1.5cm}|>{\centering\arraybackslash}p{1.5cm}|>{\centering\arraybackslash}p{1.5cm}|}
\hline
\textbf{$\vec{S_1}$} & \textbf{$\vec{F}$} & \textbf{$\vec{S_2}$} \\
\hhline{|===|}
$\left(\vec{\overline{3}} , \vec{1} , \frac{4}{3}\right)$ & $\left(\vec{1} , \vec{1} , 2\right)$ & $\left(\vec{1} , \vec{1} , 1\right)$ \\
\hline\end{tabular}
\caption{$1 \rightarrow l$, $2 \rightarrow u$, $3 \rightarrow L$, $4 \rightarrow L$, $5 \rightarrow \overline{l}$, $6 \rightarrow \overline{d}$}
\end{table}

\begin{table}[H]
\centering
\renewcommand{\arraystretch}{1.4}
\setlength{\tabcolsep}{0.15cm} % Adjust the column separation to make it narrower
\begin{tabular}{|>{\centering\arraybackslash}p{1.5cm}|>{\centering\arraybackslash}p{1.5cm}|>{\centering\arraybackslash}p{1.5cm}|}
\hline
\textbf{$\vec{S_1}$} & \textbf{$\vec{F}$} & \textbf{$\vec{S_2}$} \\
\hhline{|===|}
$\left(\vec{1} , \vec{1} , -1\right)$ & $\left(\vec{\overline{3}} , \vec{1} , -\frac{2}{3}\right)$ & $\left(\vec{\overline{3}} , \vec{1} , \frac{1}{3}\right)$ \\
\hline\end{tabular}
\caption{$1 \rightarrow \overline{l}$, $2 \rightarrow \overline{d}$, $3 \rightarrow l$, $4 \rightarrow u$, $5 \rightarrow L$, $6 \rightarrow L$}
\end{table}

\begin{table}[H]
\centering
\renewcommand{\arraystretch}{1.4}
\setlength{\tabcolsep}{0.15cm} % Adjust the column separation to make it narrower
\begin{tabular}{|>{\centering\arraybackslash}p{1.5cm}|>{\centering\arraybackslash}p{1.5cm}|>{\centering\arraybackslash}p{1.5cm}|}
\hline
\textbf{$\vec{S_1}$} & \textbf{$\vec{F}$} & \textbf{$\vec{S_2}$} \\
\hhline{|===|}
$\left(\vec{\overline{3}} , \vec{2} , -\frac{1}{6}\right)$ & $\left(\vec{\overline{3}} , \vec{1} , -\frac{2}{3}\right)$ & $\left(\vec{\overline{3}} , \vec{1} , \frac{1}{3}\right)$ \\
\hline\end{tabular}
\caption{$1 \rightarrow \overline{l}$, $2 \rightarrow L$, $3 \rightarrow l$, $4 \rightarrow u$, $5 \rightarrow \overline{d}$, $6 \rightarrow L$}
\end{table}

\begin{table}[H]
\centering
\renewcommand{\arraystretch}{1.4}
\setlength{\tabcolsep}{0.15cm} % Adjust the column separation to make it narrower
\begin{tabular}{|>{\centering\arraybackslash}p{1.5cm}|>{\centering\arraybackslash}p{1.5cm}|>{\centering\arraybackslash}p{1.5cm}|}
\hline
\textbf{$\vec{S_1}$} & \textbf{$\vec{F}$} & \textbf{$\vec{S_2}$} \\
\hhline{|===|}
$\left(\vec{1} , \vec{1} , -1\right)$ & $\left(\vec{1} , \vec{1} , 0\right)$ & $\left(\vec{\overline{3}} , \vec{1} , \frac{1}{3}\right)$ \\
\hline\end{tabular}
\caption{$1 \rightarrow \overline{d}$, $2 \rightarrow \overline{l}$, $3 \rightarrow l$, $4 \rightarrow u$, $5 \rightarrow L$, $6 \rightarrow L$}
\end{table}

\begin{table}[H]
\centering
\renewcommand{\arraystretch}{1.4}
\setlength{\tabcolsep}{0.15cm} % Adjust the column separation to make it narrower
\begin{tabular}{|>{\centering\arraybackslash}p{1.5cm}|>{\centering\arraybackslash}p{1.5cm}|>{\centering\arraybackslash}p{1.5cm}|}
\hline
\textbf{$\vec{S_1}$} & \textbf{$\vec{F}$} & \textbf{$\vec{S_2}$} \\
\hhline{|===|}
$\left(\vec{1} , \vec{2} , \frac{1}{2}\right)$ & $\left(\vec{1} , \vec{1} , 0\right)$ & $\left(\vec{\overline{3}} , \vec{1} , \frac{1}{3}\right)$ \\
\hline\end{tabular}
\caption{$1 \rightarrow \overline{d}$, $2 \rightarrow L$, $3 \rightarrow l$, $4 \rightarrow u$, $5 \rightarrow \overline{l}$, $6 \rightarrow L$}
\end{table}

\begin{table}[H]
\centering
\renewcommand{\arraystretch}{1.4}
\setlength{\tabcolsep}{0.15cm} % Adjust the column separation to make it narrower
\begin{tabular}{|>{\centering\arraybackslash}p{1.5cm}|>{\centering\arraybackslash}p{1.5cm}|>{\centering\arraybackslash}p{1.5cm}|}
\hline
\textbf{$\vec{S_1}$} & \textbf{$\vec{F}$} & \textbf{$\vec{S_2}$} \\
\hhline{|===|}
$\left(\vec{\overline{3}} , \vec{2} , -\frac{1}{6}\right)$ & $\left(\vec{\overline{3}} , \vec{2} , \frac{5}{6}\right)$ & $\left(\vec{\overline{3}} , \vec{1} , \frac{1}{3}\right)$ \\
\hline\end{tabular}
\caption{$1 \rightarrow L$, $2 \rightarrow \overline{l}$, $3 \rightarrow l$, $4 \rightarrow u$, $5 \rightarrow \overline{d}$, $6 \rightarrow L$}
\end{table}

\begin{table}[H]
\centering
\renewcommand{\arraystretch}{1.4}
\setlength{\tabcolsep}{0.15cm} % Adjust the column separation to make it narrower
\begin{tabular}{|>{\centering\arraybackslash}p{1.5cm}|>{\centering\arraybackslash}p{1.5cm}|>{\centering\arraybackslash}p{1.5cm}|}
\hline
\textbf{$\vec{S_1}$} & \textbf{$\vec{F}$} & \textbf{$\vec{S_2}$} \\
\hhline{|===|}
$\left(\vec{1} , \vec{2} , \frac{1}{2}\right)$ & $\left(\vec{\overline{3}} , \vec{2} , \frac{5}{6}\right)$ & $\left(\vec{\overline{3}} , \vec{1} , \frac{1}{3}\right)$ \\
\hline\end{tabular}
\caption{$1 \rightarrow L$, $2 \rightarrow \overline{d}$, $3 \rightarrow l$, $4 \rightarrow u$, $5 \rightarrow \overline{l}$, $6 \rightarrow L$}
\end{table}

\begin{table}[H]
\centering
\renewcommand{\arraystretch}{1.4}
\setlength{\tabcolsep}{0.15cm} % Adjust the column separation to make it narrower
\begin{tabular}{|>{\centering\arraybackslash}p{1.5cm}|>{\centering\arraybackslash}p{1.5cm}|>{\centering\arraybackslash}p{1.5cm}|}
\hline
\textbf{$\vec{S_1}$} & \textbf{$\vec{F}$} & \textbf{$\vec{S_2}$} \\
\hhline{|===|}
$\left(\vec{\overline{3}} , \vec{1} , \frac{4}{3}\right)$ & $\left(\vec{\overline{3}} , \vec{2} , \frac{5}{6}\right)$ & $\left(\vec{\overline{3}} , \vec{1} , \frac{1}{3}\right)$ \\
\hline\end{tabular}
\caption{$1 \rightarrow L$, $2 \rightarrow L$, $3 \rightarrow l$, $4 \rightarrow u$, $5 \rightarrow \overline{l}$, $6 \rightarrow \overline{d}$}
\label{tab:d9_second_last}
\end{table}

\section{UV completions for \texorpdfstring{$d=10$}{d=10}  operators}
\label{sec:UV_dim10}

Here, we list tree-level UV completions for the $d = 10$ operators of interest in Sec.~\ref{sec:dim10}. Notation and organization follow App.~\ref{sec:UV_dim7}.

\subsection{First Topology}
\begin{figure}[H]
    \centering
    \includegraphics[width=0.35\textwidth]{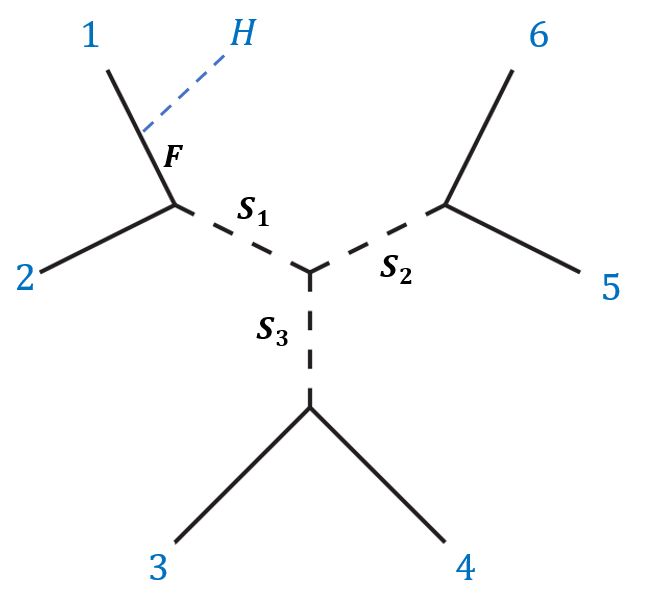}
    \caption{Feynman diagram for the cases where $H$ is attached to an external leg. The corresponding tables are Tab.~\ref{tab:first_first} - Tab.~\ref{tab:first_last}.}
\end{figure}

\subsubsection{$\bar{L} \, \bar{L} \, \bar{L} \, l \, L \, L \, H$}

\begin{table}[H]
\centering
\renewcommand{\arraystretch}{1.4}
\setlength{\tabcolsep}{0.15cm} % Adjust the column separation to make it narrower
\begin{tabular}{|>{\centering\arraybackslash}p{1.5cm}|>{\centering\arraybackslash}p{1.5cm}|>{\centering\arraybackslash}p{1.5cm}|>{\centering\arraybackslash}p{1.5cm}|}
\hline
\textbf{$\vec{S_1}$} & \textbf{$\vec{S_2}$} & \textbf{$\vec{S_3}$} & \textbf{$\vec{F}$} \\
\hhline{|====|}
$\left(\vec{1} , \vec{4} , \frac{3}{2}\right)$ & $\left(\vec{1} , \vec{3} , -1\right)$ & $\left(\vec{1} , \vec{2} , -\frac{1}{2}\right)$ & $\left(\vec{1} , \vec{3} , 1\right)$ \\
\hline
$\left(\vec{1} , \vec{2} , \frac{3}{2}\right)$ & $\left(\vec{1} , \vec{3} , -1\right)$ & $\left(\vec{1} , \vec{2} , -\frac{1}{2}\right)$ & $\left(\vec{1} , \vec{3} , 1\right)$ \\
\hline
$\left(\vec{1} , \vec{2} , \frac{3}{2}\right)$ & $\left(\vec{1} , \vec{1} , -1\right)$ & $\left(\vec{1} , \vec{2} , -\frac{1}{2}\right)$ & $\left(\vec{1} , \vec{3} , 1\right)$ \\
\hline
$\left(\vec{1} , \vec{2} , \frac{3}{2}\right)$ & $\left(\vec{1} , \vec{3} , -1\right)$ & $\left(\vec{1} , \vec{2} , -\frac{1}{2}\right)$ & $\left(\vec{1} , \vec{1} , 1\right)$ \\
\hline
$\left(\vec{1} , \vec{2} , \frac{3}{2}\right)$ & $\left(\vec{1} , \vec{1} , -1\right)$ & $\left(\vec{1} , \vec{2} , -\frac{1}{2}\right)$ & $\left(\vec{1} , \vec{1} , 1\right)$ \\
\hline\end{tabular}
\caption{$1 \rightarrow \overline{L}$, $2 \rightarrow \overline{L}$, $3 \rightarrow l$, $4 \rightarrow \overline{L}$, $5 \rightarrow L$, $6 \rightarrow L$}
\label{tab:first_first}
\end{table}

\begin{table}[H]
\centering
\renewcommand{\arraystretch}{1.4}
\setlength{\tabcolsep}{0.15cm} % Adjust the column separation to make it narrower
\begin{tabular}{|>{\centering\arraybackslash}p{1.5cm}|>{\centering\arraybackslash}p{1.5cm}|>{\centering\arraybackslash}p{1.5cm}|>{\centering\arraybackslash}p{1.5cm}|}
\hline
\textbf{$\vec{S_1}$} & \textbf{$\vec{S_2}$} & \textbf{$\vec{S_3}$} & \textbf{$\vec{F}$} \\
\hhline{|====|}
$\left(\vec{1} , \vec{3} , 0\right)$ & $\left(\vec{1} , \vec{3} , -1\right)$ & $\left(\vec{1} , \vec{3} , 1\right)$ & $\left(\vec{1} , \vec{3} , 1\right)$ \\
\hline
$\left(\vec{1} , \vec{3} , 0\right)$ & $\left(\vec{1} , \vec{3} , -1\right)$ & $\left(\vec{1} , \vec{1} , 1\right)$ & $\left(\vec{1} , \vec{3} , 1\right)$ \\
\hline
$\left(\vec{1} , \vec{3} , 0\right)$ & $\left(\vec{1} , \vec{1} , -1\right)$ & $\left(\vec{1} , \vec{3} , 1\right)$ & $\left(\vec{1} , \vec{3} , 1\right)$ \\
\hline
$\left(\vec{1} , \vec{1} , 0\right)$ & $\left(\vec{1} , \vec{3} , -1\right)$ & $\left(\vec{1} , \vec{3} , 1\right)$ & $\left(\vec{1} , \vec{1} , 1\right)$ \\
\hline
$\left(\vec{1} , \vec{1} , 0\right)$ & $\left(\vec{1} , \vec{1} , -1\right)$ & $\left(\vec{1} , \vec{1} , 1\right)$ & $\left(\vec{1} , \vec{1} , 1\right)$ \\
\hline\end{tabular}
\caption{$1 \rightarrow \overline{L}$, $2 \rightarrow l$, $3 \rightarrow \overline{L}$, $4 \rightarrow \overline{L}$, $5 \rightarrow L$, $6 \rightarrow L$}
\end{table}

\begin{table}[H]
\centering
\renewcommand{\arraystretch}{1.4}
\setlength{\tabcolsep}{0.15cm} % Adjust the column separation to make it narrower
\begin{tabular}{|>{\centering\arraybackslash}p{1.5cm}|>{\centering\arraybackslash}p{1.5cm}|>{\centering\arraybackslash}p{1.5cm}|>{\centering\arraybackslash}p{1.5cm}|}
\hline
\textbf{$\vec{S_1}$} & \textbf{$\vec{S_2}$} & \textbf{$\vec{S_3}$} & \textbf{$\vec{F}$} \\
\hhline{|====|}
$\left(\vec{1} , \vec{4} , -\frac{1}{2}\right)$ & $\left(\vec{1} , \vec{3} , 1\right)$ & $\left(\vec{1} , \vec{2} , -\frac{1}{2}\right)$ & $\left(\vec{1} , \vec{3} , 0\right)$ \\
\hline
$\left(\vec{1} , \vec{2} , -\frac{1}{2}\right)$ & $\left(\vec{1} , \vec{3} , 1\right)$ & $\left(\vec{1} , \vec{2} , -\frac{1}{2}\right)$ & $\left(\vec{1} , \vec{3} , 0\right)$ \\
\hline
$\left(\vec{1} , \vec{2} , -\frac{1}{2}\right)$ & $\left(\vec{1} , \vec{1} , 1\right)$ & $\left(\vec{1} , \vec{2} , -\frac{1}{2}\right)$ & $\left(\vec{1} , \vec{3} , 0\right)$ \\
\hline
$\left(\vec{1} , \vec{2} , -\frac{1}{2}\right)$ & $\left(\vec{1} , \vec{3} , 1\right)$ & $\left(\vec{1} , \vec{2} , -\frac{1}{2}\right)$ & $\left(\vec{1} , \vec{1} , 0\right)$ \\
\hline
$\left(\vec{1} , \vec{2} , -\frac{1}{2}\right)$ & $\left(\vec{1} , \vec{1} , 1\right)$ & $\left(\vec{1} , \vec{2} , -\frac{1}{2}\right)$ & $\left(\vec{1} , \vec{1} , 0\right)$ \\
\hline\end{tabular}
\caption{$1 \rightarrow L$, $2 \rightarrow L$, $3 \rightarrow l$, $4 \rightarrow \overline{L}$, $5 \rightarrow \overline{L}$, $6 \rightarrow \overline{L}$}
\end{table}

\begin{table}[H]
\centering
\renewcommand{\arraystretch}{1.4}
\setlength{\tabcolsep}{0.15cm} % Adjust the column separation to make it narrower
\begin{tabular}{|>{\centering\arraybackslash}p{1.5cm}|>{\centering\arraybackslash}p{1.5cm}|>{\centering\arraybackslash}p{1.5cm}|>{\centering\arraybackslash}p{1.5cm}|}
\hline
\textbf{$\vec{S_1}$} & \textbf{$\vec{S_2}$} & \textbf{$\vec{S_3}$} & \textbf{$\vec{F}$} \\
\hhline{|====|}
$\left(\vec{1} , \vec{3} , 0\right)$ & $\left(\vec{1} , \vec{3} , -1\right)$ & $\left(\vec{1} , \vec{3} , 1\right)$ & $\left(\vec{1} , \vec{2} , -\frac{1}{2}\right)$ \\
\hline
$\left(\vec{1} , \vec{3} , 0\right)$ & $\left(\vec{1} , \vec{3} , -1\right)$ & $\left(\vec{1} , \vec{1} , 1\right)$ & $\left(\vec{1} , \vec{2} , -\frac{1}{2}\right)$ \\
\hline
$\left(\vec{1} , \vec{3} , 0\right)$ & $\left(\vec{1} , \vec{1} , -1\right)$ & $\left(\vec{1} , \vec{3} , 1\right)$ & $\left(\vec{1} , \vec{2} , -\frac{1}{2}\right)$ \\
\hline
$\left(\vec{1} , \vec{1} , 0\right)$ & $\left(\vec{1} , \vec{3} , -1\right)$ & $\left(\vec{1} , \vec{3} , 1\right)$ & $\left(\vec{1} , \vec{2} , -\frac{1}{2}\right)$ \\
\hline
$\left(\vec{1} , \vec{1} , 0\right)$ & $\left(\vec{1} , \vec{1} , -1\right)$ & $\left(\vec{1} , \vec{1} , 1\right)$ & $\left(\vec{1} , \vec{2} , -\frac{1}{2}\right)$ \\
\hline\end{tabular}
\caption{$1 \rightarrow l$, $2 \rightarrow \overline{L}$, $3 \rightarrow \overline{L}$, $4 \rightarrow \overline{L}$, $5 \rightarrow L$, $6 \rightarrow L$}
\end{table}

\subsubsection{$\bar{L} \, \bar{L} \, \bar{l} \, l \, l \, L \, H$}

\begin{table}[H]
\centering
\renewcommand{\arraystretch}{1.4}
\setlength{\tabcolsep}{0.15cm} % Adjust the column separation to make it narrower
\begin{tabular}{|>{\centering\arraybackslash}p{1.5cm}|>{\centering\arraybackslash}p{1.5cm}|>{\centering\arraybackslash}p{1.5cm}|>{\centering\arraybackslash}p{1.5cm}|}
\hline
\textbf{$\vec{S_1}$} & \textbf{$\vec{S_2}$} & \textbf{$\vec{S_3}$} & \textbf{$\vec{F}$} \\
\hhline{|====|}
$\left(\vec{1} , \vec{2} , \frac{3}{2}\right)$ & $\left(\vec{1} , \vec{2} , \frac{1}{2}\right)$ & $\left(\vec{1} , \vec{1} , -2\right)$ & $\left(\vec{1} , \vec{3} , 1\right)$ \\
\hline
$\left(\vec{1} , \vec{2} , \frac{3}{2}\right)$ & $\left(\vec{1} , \vec{2} , \frac{1}{2}\right)$ & $\left(\vec{1} , \vec{1} , -2\right)$ & $\left(\vec{1} , \vec{1} , 1\right)$ \\
\hline\end{tabular}
\caption{$1 \rightarrow \overline{L}$, $2 \rightarrow \overline{L}$, $3 \rightarrow l$, $4 \rightarrow l$, $5 \rightarrow \overline{l}$, $6 \rightarrow L$}
\end{table}

\begin{table}[H]
\centering
\renewcommand{\arraystretch}{1.4}
\setlength{\tabcolsep}{0.15cm} % Adjust the column separation to make it narrower
\begin{tabular}{|>{\centering\arraybackslash}p{1.5cm}|>{\centering\arraybackslash}p{1.5cm}|>{\centering\arraybackslash}p{1.5cm}|>{\centering\arraybackslash}p{1.5cm}|}
\hline
\textbf{$\vec{S_1}$} & \textbf{$\vec{S_2}$} & \textbf{$\vec{S_3}$} & \textbf{$\vec{F}$} \\
\hhline{|====|}
$\left(\vec{1} , \vec{3} , 0\right)$ & $\left(\vec{1} , \vec{2} , \frac{1}{2}\right)$ & $\left(\vec{1} , \vec{2} , -\frac{1}{2}\right)$ & $\left(\vec{1} , \vec{3} , 1\right)$ \\
\hline
$\left(\vec{1} , \vec{1} , 0\right)$ & $\left(\vec{1} , \vec{2} , \frac{1}{2}\right)$ & $\left(\vec{1} , \vec{2} , -\frac{1}{2}\right)$ & $\left(\vec{1} , \vec{1} , 1\right)$ \\
\hline\end{tabular}
\caption{$1 \rightarrow \overline{L}$, $2 \rightarrow l$, $3 \rightarrow l$, $4 \rightarrow \overline{L}$, $5 \rightarrow \overline{l}$, $6 \rightarrow L$}
\end{table}

\begin{table}[H]
\centering
\renewcommand{\arraystretch}{1.4}
\setlength{\tabcolsep}{0.15cm} % Adjust the column separation to make it narrower
\begin{tabular}{|>{\centering\arraybackslash}p{1.5cm}|>{\centering\arraybackslash}p{1.5cm}|>{\centering\arraybackslash}p{1.5cm}|>{\centering\arraybackslash}p{1.5cm}|}
\hline
\textbf{$\vec{S_1}$} & \textbf{$\vec{S_2}$} & \textbf{$\vec{S_3}$} & \textbf{$\vec{F}$} \\
\hhline{|====|}
$\left(\vec{1} , \vec{3} , 1\right)$ & $\left(\vec{1} , \vec{3} , 1\right)$ & $\left(\vec{1} , \vec{1} , -2\right)$ & $\left(\vec{1} , \vec{2} , \frac{3}{2}\right)$ \\
\hline
$\left(\vec{1} , \vec{1} , 1\right)$ & $\left(\vec{1} , \vec{1} , 1\right)$ & $\left(\vec{1} , \vec{1} , -2\right)$ & $\left(\vec{1} , \vec{2} , \frac{3}{2}\right)$ \\
\hline\end{tabular}
\caption{$1 \rightarrow \overline{l}$, $2 \rightarrow L$, $3 \rightarrow l$, $4 \rightarrow l$, $5 \rightarrow \overline{L}$, $6 \rightarrow \overline{L}$}
\end{table}

\begin{table}[H]
\centering
\renewcommand{\arraystretch}{1.4}
\setlength{\tabcolsep}{0.15cm} % Adjust the column separation to make it narrower
\begin{tabular}{|>{\centering\arraybackslash}p{1.5cm}|>{\centering\arraybackslash}p{1.5cm}|>{\centering\arraybackslash}p{1.5cm}|>{\centering\arraybackslash}p{1.5cm}|}
\hline
\textbf{$\vec{S_1}$} & \textbf{$\vec{S_2}$} & \textbf{$\vec{S_3}$} & \textbf{$\vec{F}$} \\
\hhline{|====|}
$\left(\vec{1} , \vec{3} , 1\right)$ & $\left(\vec{1} , \vec{2} , -\frac{1}{2}\right)$ & $\left(\vec{1} , \vec{2} , -\frac{1}{2}\right)$ & $\left(\vec{1} , \vec{2} , \frac{3}{2}\right)$ \\
\hline
$\left(\vec{1} , \vec{1} , 1\right)$ & $\left(\vec{1} , \vec{2} , -\frac{1}{2}\right)$ & $\left(\vec{1} , \vec{2} , -\frac{1}{2}\right)$ & $\left(\vec{1} , \vec{2} , \frac{3}{2}\right)$ \\
\hline\end{tabular}
\caption{$1 \rightarrow \overline{l}$, $2 \rightarrow L$, $3 \rightarrow l$, $4 \rightarrow \overline{L}$, $5 \rightarrow l$, $6 \rightarrow \overline{L}$}
\end{table}

\begin{table}[H]
\centering
\renewcommand{\arraystretch}{1.4}
\setlength{\tabcolsep}{0.15cm} % Adjust the column separation to make it narrower
\begin{tabular}{|>{\centering\arraybackslash}p{1.5cm}|>{\centering\arraybackslash}p{1.5cm}|>{\centering\arraybackslash}p{1.5cm}|>{\centering\arraybackslash}p{1.5cm}|}
\hline
\textbf{$\vec{S_1}$} & \textbf{$\vec{S_2}$} & \textbf{$\vec{S_3}$} & \textbf{$\vec{F}$} \\
\hhline{|====|}
$\left(\vec{1} , \vec{3} , 0\right)$ & $\left(\vec{1} , \vec{2} , \frac{1}{2}\right)$ & $\left(\vec{1} , \vec{2} , -\frac{1}{2}\right)$ & $\left(\vec{1} , \vec{2} , -\frac{1}{2}\right)$ \\
\hline
$\left(\vec{1} , \vec{1} , 0\right)$ & $\left(\vec{1} , \vec{2} , \frac{1}{2}\right)$ & $\left(\vec{1} , \vec{2} , -\frac{1}{2}\right)$ & $\left(\vec{1} , \vec{2} , -\frac{1}{2}\right)$ \\
\hline\end{tabular}
\caption{$1 \rightarrow l$, $2 \rightarrow \overline{L}$, $3 \rightarrow l$, $4 \rightarrow \overline{L}$, $5 \rightarrow \overline{l}$, $6 \rightarrow L$}
\end{table}

\begin{table}[H]
\centering
\renewcommand{\arraystretch}{1.4}
\setlength{\tabcolsep}{0.15cm} % Adjust the column separation to make it narrower
\begin{tabular}{|>{\centering\arraybackslash}p{1.5cm}|>{\centering\arraybackslash}p{1.5cm}|>{\centering\arraybackslash}p{1.5cm}|>{\centering\arraybackslash}p{1.5cm}|}
\hline
\textbf{$\vec{S_1}$} & \textbf{$\vec{S_2}$} & \textbf{$\vec{S_3}$} & \textbf{$\vec{F}$} \\
\hhline{|====|}
$\left(\vec{1} , \vec{2} , -\frac{3}{2}\right)$ & $\left(\vec{1} , \vec{2} , \frac{1}{2}\right)$ & $\left(\vec{1} , \vec{3} , 1\right)$ & $\left(\vec{1} , \vec{2} , -\frac{1}{2}\right)$ \\
\hline
$\left(\vec{1} , \vec{2} , -\frac{3}{2}\right)$ & $\left(\vec{1} , \vec{2} , \frac{1}{2}\right)$ & $\left(\vec{1} , \vec{1} , 1\right)$ & $\left(\vec{1} , \vec{2} , -\frac{1}{2}\right)$ \\
\hline\end{tabular}
\caption{$1 \rightarrow l$, $2 \rightarrow l$, $3 \rightarrow \overline{L}$, $4 \rightarrow \overline{L}$, $5 \rightarrow \overline{l}$, $6 \rightarrow L$}
\end{table}

\begin{table}[H]
\centering
\renewcommand{\arraystretch}{1.4}
\setlength{\tabcolsep}{0.15cm} % Adjust the column separation to make it narrower
\begin{tabular}{|>{\centering\arraybackslash}p{1.5cm}|>{\centering\arraybackslash}p{1.5cm}|>{\centering\arraybackslash}p{1.5cm}|>{\centering\arraybackslash}p{1.5cm}|}
\hline
\textbf{$\vec{S_1}$} & \textbf{$\vec{S_2}$} & \textbf{$\vec{S_3}$} & \textbf{$\vec{F}$} \\
\hhline{|====|}
$\left(\vec{1} , \vec{3} , 1\right)$ & $\left(\vec{1} , \vec{3} , 1\right)$ & $\left(\vec{1} , \vec{1} , -2\right)$ & $\left(\vec{1} , \vec{3} , 0\right)$ \\
\hline
$\left(\vec{1} , \vec{1} , 1\right)$ & $\left(\vec{1} , \vec{1} , 1\right)$ & $\left(\vec{1} , \vec{1} , -2\right)$ & $\left(\vec{1} , \vec{1} , 0\right)$ \\
\hline\end{tabular}
\caption{$1 \rightarrow L$, $2 \rightarrow \overline{l}$, $3 \rightarrow l$, $4 \rightarrow l$, $5 \rightarrow \overline{L}$, $6 \rightarrow \overline{L}$}
\end{table}

\begin{table}[H]
\centering
\renewcommand{\arraystretch}{1.4}
\setlength{\tabcolsep}{0.15cm} % Adjust the column separation to make it narrower
\begin{tabular}{|>{\centering\arraybackslash}p{1.5cm}|>{\centering\arraybackslash}p{1.5cm}|>{\centering\arraybackslash}p{1.5cm}|>{\centering\arraybackslash}p{1.5cm}|}
\hline
\textbf{$\vec{S_1}$} & \textbf{$\vec{S_2}$} & \textbf{$\vec{S_3}$} & \textbf{$\vec{F}$} \\
\hhline{|====|}
$\left(\vec{1} , \vec{3} , 1\right)$ & $\left(\vec{1} , \vec{2} , -\frac{1}{2}\right)$ & $\left(\vec{1} , \vec{2} , -\frac{1}{2}\right)$ & $\left(\vec{1} , \vec{3} , 0\right)$ \\
\hline
$\left(\vec{1} , \vec{1} , 1\right)$ & $\left(\vec{1} , \vec{2} , -\frac{1}{2}\right)$ & $\left(\vec{1} , \vec{2} , -\frac{1}{2}\right)$ & $\left(\vec{1} , \vec{1} , 0\right)$ \\
\hline\end{tabular}
\caption{$1 \rightarrow L$, $2 \rightarrow \overline{l}$, $3 \rightarrow l$, $4 \rightarrow \overline{L}$, $5 \rightarrow l$, $6 \rightarrow \overline{L}$}
\end{table}

\subsubsection{$\bar{L} \, \bar{l} \, \bar{l} \, l \, l \, l \, H$}

\begin{table}[H]
\centering
\renewcommand{\arraystretch}{1.4}
\setlength{\tabcolsep}{0.15cm} % Adjust the column separation to make it narrower
\begin{tabular}{|>{\centering\arraybackslash}p{1.5cm}|>{\centering\arraybackslash}p{1.5cm}|>{\centering\arraybackslash}p{1.5cm}|>{\centering\arraybackslash}p{1.5cm}|}
\hline
\textbf{$\vec{S_1}$} & \textbf{$\vec{S_2}$} & \textbf{$\vec{S_3}$} & \textbf{$\vec{F}$} \\
\hhline{|====|}
$\left(\vec{1} , \vec{1} , 0\right)$ & $\left(\vec{1} , \vec{1} , 2\right)$ & $\left(\vec{1} , \vec{1} , -2\right)$ & $\left(\vec{1} , \vec{1} , 1\right)$ \\
\hline\end{tabular}
\caption{$1 \rightarrow \overline{L}$, $2 \rightarrow l$, $3 \rightarrow l$, $4 \rightarrow l$, $5 \rightarrow \overline{l}$, $6 \rightarrow \overline{l}$}
\end{table}

\begin{table}[H]
\centering
\renewcommand{\arraystretch}{1.4}
\setlength{\tabcolsep}{0.15cm} % Adjust the column separation to make it narrower
\begin{tabular}{|>{\centering\arraybackslash}p{1.5cm}|>{\centering\arraybackslash}p{1.5cm}|>{\centering\arraybackslash}p{1.5cm}|>{\centering\arraybackslash}p{1.5cm}|}
\hline
\textbf{$\vec{S_1}$} & \textbf{$\vec{S_2}$} & \textbf{$\vec{S_3}$} & \textbf{$\vec{F}$} \\
\hhline{|====|}
$\left(\vec{1} , \vec{2} , \frac{5}{2}\right)$ & $\left(\vec{1} , \vec{2} , -\frac{1}{2}\right)$ & $\left(\vec{1} , \vec{1} , -2\right)$ & $\left(\vec{1} , \vec{2} , \frac{3}{2}\right)$ \\
\hline\end{tabular}
\caption{$1 \rightarrow \overline{l}$, $2 \rightarrow \overline{l}$, $3 \rightarrow l$, $4 \rightarrow l$, $5 \rightarrow l$, $6 \rightarrow \overline{L}$}
\end{table}

\begin{table}[H]
\centering
\renewcommand{\arraystretch}{1.4}
\setlength{\tabcolsep}{0.15cm} % Adjust the column separation to make it narrower
\begin{tabular}{|>{\centering\arraybackslash}p{1.5cm}|>{\centering\arraybackslash}p{1.5cm}|>{\centering\arraybackslash}p{1.5cm}|>{\centering\arraybackslash}p{1.5cm}|}
\hline
\textbf{$\vec{S_1}$} & \textbf{$\vec{S_2}$} & \textbf{$\vec{S_3}$} & \textbf{$\vec{F}$} \\
\hhline{|====|}
$\left(\vec{1} , \vec{1} , 0\right)$ & $\left(\vec{1} , \vec{1} , 2\right)$ & $\left(\vec{1} , \vec{1} , -2\right)$ & $\left(\vec{1} , \vec{2} , -\frac{1}{2}\right)$ \\
\hline\end{tabular}
\caption{$1 \rightarrow l$, $2 \rightarrow \overline{L}$, $3 \rightarrow l$, $4 \rightarrow l$, $5 \rightarrow \overline{l}$, $6 \rightarrow \overline{l}$}
\end{table}

\begin{table}[H]
\centering
\renewcommand{\arraystretch}{1.4}
\setlength{\tabcolsep}{0.15cm} % Adjust the column separation to make it narrower
\begin{tabular}{|>{\centering\arraybackslash}p{1.5cm}|>{\centering\arraybackslash}p{1.5cm}|>{\centering\arraybackslash}p{1.5cm}|>{\centering\arraybackslash}p{1.5cm}|}
\hline
\textbf{$\vec{S_1}$} & \textbf{$\vec{S_2}$} & \textbf{$\vec{S_3}$} & \textbf{$\vec{F}$} \\
\hhline{|====|}
$\left(\vec{1} , \vec{2} , -\frac{3}{2}\right)$ & $\left(\vec{1} , \vec{1} , 2\right)$ & $\left(\vec{1} , \vec{2} , -\frac{1}{2}\right)$ & $\left(\vec{1} , \vec{2} , -\frac{1}{2}\right)$ \\
\hline\end{tabular}
\caption{$1 \rightarrow l$, $2 \rightarrow l$, $3 \rightarrow l$, $4 \rightarrow \overline{L}$, $5 \rightarrow \overline{l}$, $6 \rightarrow \overline{l}$}
\label{tab:first_last}
\end{table}

\subsection{Second Topology}
\begin{figure}[H]
    \centering
    \includegraphics[width=0.35\textwidth]{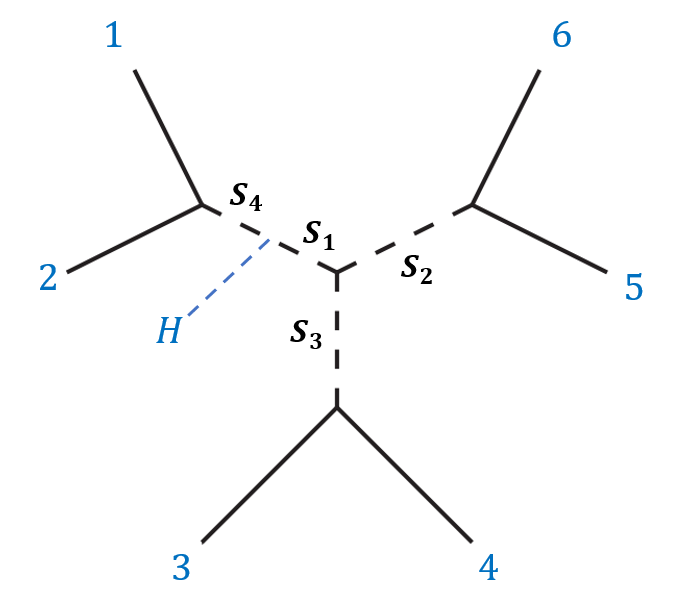}
    \caption{Feynman diagram for the cases where $H$ is attached to an internal mediator. The corresponding tables are Tab.~\ref{tab:second_first} - Tab.~\ref{tab:second_last}.}
\end{figure}

\subsubsection{$\bar{L} \, \bar{L} \, \bar{L} \, l \, L \, L \, H$}

\begin{table}[H]
\centering
\renewcommand{\arraystretch}{1.4}
\setlength{\tabcolsep}{0.15cm} % Adjust the column separation to make it narrower
\begin{tabular}{|>{\centering\arraybackslash}p{1.5cm}|>{\centering\arraybackslash}p{1.5cm}|>{\centering\arraybackslash}p{1.5cm}|>{\centering\arraybackslash}p{1.5cm}|}
\hline
\textbf{$\vec{S_1}$} & \textbf{$\vec{S_2}$} & \textbf{$\vec{S_3}$} & \textbf{$\vec{S_4}$} \\
\hhline{|====|}
$\left(\vec{1} , \vec{4} , \frac{3}{2}\right)$ & $\left(\vec{1} , \vec{3} , -1\right)$ & $\left(\vec{1} , \vec{2} , -\frac{1}{2}\right)$ & $\left(\vec{1} , \vec{3} , 1\right)$ \\
\hline
$\left(\vec{1} , \vec{2} , \frac{3}{2}\right)$ & $\left(\vec{1} , \vec{3} , -1\right)$ & $\left(\vec{1} , \vec{2} , -\frac{1}{2}\right)$ & $\left(\vec{1} , \vec{3} , 1\right)$ \\
\hline
$\left(\vec{1} , \vec{2} , \frac{3}{2}\right)$ & $\left(\vec{1} , \vec{1} , -1\right)$ & $\left(\vec{1} , \vec{2} , -\frac{1}{2}\right)$ & $\left(\vec{1} , \vec{3} , 1\right)$ \\
\hline
$\left(\vec{1} , \vec{2} , \frac{3}{2}\right)$ & $\left(\vec{1} , \vec{3} , -1\right)$ & $\left(\vec{1} , \vec{2} , -\frac{1}{2}\right)$ & $\left(\vec{1} , \vec{1} , 1\right)$ \\
\hline
$\left(\vec{1} , \vec{2} , \frac{3}{2}\right)$ & $\left(\vec{1} , \vec{1} , -1\right)$ & $\left(\vec{1} , \vec{2} , -\frac{1}{2}\right)$ & $\left(\vec{1} , \vec{1} , 1\right)$ \\
\hline\end{tabular}
\caption{$1 \rightarrow \overline{L}$, $2 \rightarrow \overline{L}$, $3 \rightarrow l$, $4 \rightarrow \overline{L}$, $5 \rightarrow L$, $6 \rightarrow L$}
\label{tab:second_first}
\end{table}

\begin{table}[H]
\centering
\renewcommand{\arraystretch}{1.4}
\setlength{\tabcolsep}{0.15cm} % Adjust the column separation to make it narrower
\begin{tabular}{|>{\centering\arraybackslash}p{1.5cm}|>{\centering\arraybackslash}p{1.5cm}|>{\centering\arraybackslash}p{1.5cm}|>{\centering\arraybackslash}p{1.5cm}|}
\hline
\textbf{$\vec{S_1}$} & \textbf{$\vec{S_2}$} & \textbf{$\vec{S_3}$} & \textbf{$\vec{S_4}$} \\
\hhline{|====|}
$\left(\vec{1} , \vec{3} , 0\right)$ & $\left(\vec{1} , \vec{3} , -1\right)$ & $\left(\vec{1} , \vec{3} , 1\right)$ & $\left(\vec{1} , \vec{2} , -\frac{1}{2}\right)$ \\
\hline
$\left(\vec{1} , \vec{3} , 0\right)$ & $\left(\vec{1} , \vec{3} , -1\right)$ & $\left(\vec{1} , \vec{1} , 1\right)$ & $\left(\vec{1} , \vec{2} , -\frac{1}{2}\right)$ \\
\hline
$\left(\vec{1} , \vec{3} , 0\right)$ & $\left(\vec{1} , \vec{1} , -1\right)$ & $\left(\vec{1} , \vec{3} , 1\right)$ & $\left(\vec{1} , \vec{2} , -\frac{1}{2}\right)$ \\
\hline
$\left(\vec{1} , \vec{1} , 0\right)$ & $\left(\vec{1} , \vec{3} , -1\right)$ & $\left(\vec{1} , \vec{3} , 1\right)$ & $\left(\vec{1} , \vec{2} , -\frac{1}{2}\right)$ \\
\hline
$\left(\vec{1} , \vec{1} , 0\right)$ & $\left(\vec{1} , \vec{1} , -1\right)$ & $\left(\vec{1} , \vec{1} , 1\right)$ & $\left(\vec{1} , \vec{2} , -\frac{1}{2}\right)$ \\
\hline\end{tabular}
\caption{$1 \rightarrow l$, $2 \rightarrow \overline{L}$, $3 \rightarrow \overline{L}$, $4 \rightarrow \overline{L}$, $5 \rightarrow L$, $6 \rightarrow L$}
\end{table}

\begin{table}[H]
\centering
\renewcommand{\arraystretch}{1.4}
\setlength{\tabcolsep}{0.15cm} % Adjust the column separation to make it narrower
\begin{tabular}{|>{\centering\arraybackslash}p{1.5cm}|>{\centering\arraybackslash}p{1.5cm}|>{\centering\arraybackslash}p{1.5cm}|>{\centering\arraybackslash}p{1.5cm}|}
\hline
\textbf{$\vec{S_1}$} & \textbf{$\vec{S_2}$} & \textbf{$\vec{S_3}$} & \textbf{$\vec{S_4}$} \\
\hhline{|====|}
$\left(\vec{1} , \vec{4} , -\frac{1}{2}\right)$ & $\left(\vec{1} , \vec{3} , 1\right)$ & $\left(\vec{1} , \vec{2} , -\frac{1}{2}\right)$ & $\left(\vec{1} , \vec{3} , -1\right)$ \\
\hline
$\left(\vec{1} , \vec{2} , -\frac{1}{2}\right)$ & $\left(\vec{1} , \vec{3} , 1\right)$ & $\left(\vec{1} , \vec{2} , -\frac{1}{2}\right)$ & $\left(\vec{1} , \vec{3} , -1\right)$ \\
\hline
$\left(\vec{1} , \vec{2} , -\frac{1}{2}\right)$ & $\left(\vec{1} , \vec{1} , 1\right)$ & $\left(\vec{1} , \vec{2} , -\frac{1}{2}\right)$ & $\left(\vec{1} , \vec{3} , -1\right)$ \\
\hline
$\left(\vec{1} , \vec{2} , -\frac{1}{2}\right)$ & $\left(\vec{1} , \vec{3} , 1\right)$ & $\left(\vec{1} , \vec{2} , -\frac{1}{2}\right)$ & $\left(\vec{1} , \vec{1} , -1\right)$ \\
\hline
$\left(\vec{1} , \vec{2} , -\frac{1}{2}\right)$ & $\left(\vec{1} , \vec{1} , 1\right)$ & $\left(\vec{1} , \vec{2} , -\frac{1}{2}\right)$ & $\left(\vec{1} , \vec{1} , -1\right)$ \\
\hline\end{tabular}
\caption{$1 \rightarrow L$, $2 \rightarrow L$, $3 \rightarrow l$, $4 \rightarrow \overline{L}$, $5 \rightarrow \overline{L}$, $6 \rightarrow \overline{L}$}
\end{table}

\subsubsection{$\bar{L} \, \bar{L} \, \bar{l} \, l \, l \, L \, H$}

\begin{table}[H]
\centering
\renewcommand{\arraystretch}{1.4}
\setlength{\tabcolsep}{0.15cm} % Adjust the column separation to make it narrower
\begin{tabular}{|>{\centering\arraybackslash}p{1.5cm}|>{\centering\arraybackslash}p{1.5cm}|>{\centering\arraybackslash}p{1.5cm}|>{\centering\arraybackslash}p{1.5cm}|}
\hline
\textbf{$\vec{S_1}$} & \textbf{$\vec{S_2}$} & \textbf{$\vec{S_3}$} & \textbf{$\vec{S_4}$} \\
\hhline{|====|}
$\left(\vec{1} , \vec{2} , \frac{3}{2}\right)$ & $\left(\vec{1} , \vec{2} , \frac{1}{2}\right)$ & $\left(\vec{1} , \vec{1} , -2\right)$ & $\left(\vec{1} , \vec{3} , 1\right)$ \\
\hline
$\left(\vec{1} , \vec{2} , \frac{3}{2}\right)$ & $\left(\vec{1} , \vec{2} , \frac{1}{2}\right)$ & $\left(\vec{1} , \vec{1} , -2\right)$ & $\left(\vec{1} , \vec{1} , 1\right)$ \\
\hline\end{tabular}
\caption{$1 \rightarrow \overline{L}$, $2 \rightarrow \overline{L}$, $3 \rightarrow l$, $4 \rightarrow l$, $5 \rightarrow \overline{l}$, $6 \rightarrow L$}
\end{table}

\begin{table}[H]
\centering
\renewcommand{\arraystretch}{1.4}
\setlength{\tabcolsep}{0.15cm} % Adjust the column separation to make it narrower
\begin{tabular}{|>{\centering\arraybackslash}p{1.5cm}|>{\centering\arraybackslash}p{1.5cm}|>{\centering\arraybackslash}p{1.5cm}|>{\centering\arraybackslash}p{1.5cm}|}
\hline
\textbf{$\vec{S_1}$} & \textbf{$\vec{S_2}$} & \textbf{$\vec{S_3}$} & \textbf{$\vec{S_4}$} \\
\hhline{|====|}
$\left(\vec{1} , \vec{3} , 0\right)$ & $\left(\vec{1} , \vec{2} , \frac{1}{2}\right)$ & $\left(\vec{1} , \vec{2} , -\frac{1}{2}\right)$ & $\left(\vec{1} , \vec{2} , -\frac{1}{2}\right)$ \\
\hline
$\left(\vec{1} , \vec{1} , 0\right)$ & $\left(\vec{1} , \vec{2} , \frac{1}{2}\right)$ & $\left(\vec{1} , \vec{2} , -\frac{1}{2}\right)$ & $\left(\vec{1} , \vec{2} , -\frac{1}{2}\right)$ \\
\hline\end{tabular}
\caption{$1 \rightarrow l$, $2 \rightarrow \overline{L}$, $3 \rightarrow l$, $4 \rightarrow \overline{L}$, $5 \rightarrow \overline{l}$, $6 \rightarrow L$}
\end{table}

\begin{table}[H]
\centering
\renewcommand{\arraystretch}{1.4}
\setlength{\tabcolsep}{0.15cm} % Adjust the column separation to make it narrower
\begin{tabular}{|>{\centering\arraybackslash}p{1.5cm}|>{\centering\arraybackslash}p{1.5cm}|>{\centering\arraybackslash}p{1.5cm}|>{\centering\arraybackslash}p{1.5cm}|}
\hline
\textbf{$\vec{S_1}$} & \textbf{$\vec{S_2}$} & \textbf{$\vec{S_3}$} & \textbf{$\vec{S_4}$} \\
\hhline{|====|}
$\left(\vec{1} , \vec{3} , 1\right)$ & $\left(\vec{1} , \vec{3} , 1\right)$ & $\left(\vec{1} , \vec{1} , -2\right)$ & $\left(\vec{1} , \vec{2} , \frac{1}{2}\right)$ \\
\hline
$\left(\vec{1} , \vec{1} , 1\right)$ & $\left(\vec{1} , \vec{1} , 1\right)$ & $\left(\vec{1} , \vec{1} , -2\right)$ & $\left(\vec{1} , \vec{2} , \frac{1}{2}\right)$ \\
\hline\end{tabular}
\caption{$1 \rightarrow \overline{l}$, $2 \rightarrow L$, $3 \rightarrow l$, $4 \rightarrow l$, $5 \rightarrow \overline{L}$, $6 \rightarrow \overline{L}$}
\end{table}

\begin{table}[H]
\centering
\renewcommand{\arraystretch}{1.4}
\setlength{\tabcolsep}{0.15cm} % Adjust the column separation to make it narrower
\begin{tabular}{|>{\centering\arraybackslash}p{1.5cm}|>{\centering\arraybackslash}p{1.5cm}|>{\centering\arraybackslash}p{1.5cm}|>{\centering\arraybackslash}p{1.5cm}|}
\hline
\textbf{$\vec{S_1}$} & \textbf{$\vec{S_2}$} & \textbf{$\vec{S_3}$} & \textbf{$\vec{S_4}$} \\
\hhline{|====|}
$\left(\vec{1} , \vec{3} , 1\right)$ & $\left(\vec{1} , \vec{2} , -\frac{1}{2}\right)$ & $\left(\vec{1} , \vec{2} , -\frac{1}{2}\right)$ & $\left(\vec{1} , \vec{2} , \frac{1}{2}\right)$ \\
\hline
$\left(\vec{1} , \vec{1} , 1\right)$ & $\left(\vec{1} , \vec{2} , -\frac{1}{2}\right)$ & $\left(\vec{1} , \vec{2} , -\frac{1}{2}\right)$ & $\left(\vec{1} , \vec{2} , \frac{1}{2}\right)$ \\
\hline\end{tabular}
\caption{$1 \rightarrow \overline{l}$, $2 \rightarrow L$, $3 \rightarrow l$, $4 \rightarrow \overline{L}$, $5 \rightarrow l$, $6 \rightarrow \overline{L}$}
\end{table}

\begin{table}[H]
\centering
\renewcommand{\arraystretch}{1.4}
\setlength{\tabcolsep}{0.15cm} % Adjust the column separation to make it narrower
\begin{tabular}{|>{\centering\arraybackslash}p{1.5cm}|>{\centering\arraybackslash}p{1.5cm}|>{\centering\arraybackslash}p{1.5cm}|>{\centering\arraybackslash}p{1.5cm}|}
\hline
\textbf{$\vec{S_1}$} & \textbf{$\vec{S_2}$} & \textbf{$\vec{S_3}$} & \textbf{$\vec{S_4}$} \\
\hhline{|====|}
$\left(\vec{1} , \vec{2} , -\frac{3}{2}\right)$ & $\left(\vec{1} , \vec{2} , \frac{1}{2}\right)$ & $\left(\vec{1} , \vec{3} , 1\right)$ & $\left(\vec{1} , \vec{1} , -2\right)$ \\
\hline
$\left(\vec{1} , \vec{2} , -\frac{3}{2}\right)$ & $\left(\vec{1} , \vec{2} , \frac{1}{2}\right)$ & $\left(\vec{1} , \vec{1} , 1\right)$ & $\left(\vec{1} , \vec{1} , -2\right)$ \\
\hline\end{tabular}
\caption{$1 \rightarrow l$, $2 \rightarrow l$, $3 \rightarrow \overline{L}$, $4 \rightarrow \overline{L}$, $5 \rightarrow \overline{l}$, $6 \rightarrow L$}
\end{table}

\subsubsection{$\bar{L} \, \bar{l} \, \bar{l} \, l \, l \, l \, H$}

\begin{table}[H]
\centering
\renewcommand{\arraystretch}{1.4}
\setlength{\tabcolsep}{0.15cm} % Adjust the column separation to make it narrower
\begin{tabular}{|>{\centering\arraybackslash}p{1.5cm}|>{\centering\arraybackslash}p{1.5cm}|>{\centering\arraybackslash}p{1.5cm}|>{\centering\arraybackslash}p{1.5cm}|}
\hline
\textbf{$\vec{S_1}$} & \textbf{$\vec{S_2}$} & \textbf{$\vec{S_3}$} & \textbf{$\vec{S_4}$} \\
\hhline{|====|}
$\left(\vec{1} , \vec{1} , 0\right)$ & $\left(\vec{1} , \vec{1} , 2\right)$ & $\left(\vec{1} , \vec{1} , -2\right)$ & $\left(\vec{1} , \vec{2} , -\frac{1}{2}\right)$ \\
\hline\end{tabular}
\caption{$1 \rightarrow l$, $2 \rightarrow \overline{L}$, $3 \rightarrow l$, $4 \rightarrow l$, $5 \rightarrow \overline{l}$, $6 \rightarrow \overline{l}$}
\end{table}

\begin{table}[H]
\centering
\renewcommand{\arraystretch}{1.4}
\setlength{\tabcolsep}{0.15cm} % Adjust the column separation to make it narrower
\begin{tabular}{|>{\centering\arraybackslash}p{1.5cm}|>{\centering\arraybackslash}p{1.5cm}|>{\centering\arraybackslash}p{1.5cm}|>{\centering\arraybackslash}p{1.5cm}|}
\hline
\textbf{$\vec{S_1}$} & \textbf{$\vec{S_2}$} & \textbf{$\vec{S_3}$} & \textbf{$\vec{S_4}$} \\
\hhline{|====|}
$\left(\vec{1} , \vec{2} , \frac{5}{2}\right)$ & $\left(\vec{1} , \vec{2} , -\frac{1}{2}\right)$ & $\left(\vec{1} , \vec{1} , -2\right)$ & $\left(\vec{1} , \vec{1} , 2\right)$ \\
\hline\end{tabular}
\caption{$1 \rightarrow \overline{l}$, $2 \rightarrow \overline{l}$, $3 \rightarrow l$, $4 \rightarrow l$, $5 \rightarrow l$, $6 \rightarrow \overline{L}$}
\end{table}

\begin{table}[H]
\centering
\renewcommand{\arraystretch}{1.4}
\setlength{\tabcolsep}{0.15cm} % Adjust the column separation to make it narrower
\begin{tabular}{|>{\centering\arraybackslash}p{1.5cm}|>{\centering\arraybackslash}p{1.5cm}|>{\centering\arraybackslash}p{1.5cm}|>{\centering\arraybackslash}p{1.5cm}|}
\hline
\textbf{$\vec{S_1}$} & \textbf{$\vec{S_2}$} & \textbf{$\vec{S_3}$} & \textbf{$\vec{S_4}$} \\
\hhline{|====|}
$\left(\vec{1} , \vec{2} , -\frac{3}{2}\right)$ & $\left(\vec{1} , \vec{1} , 2\right)$ & $\left(\vec{1} , \vec{2} , -\frac{1}{2}\right)$ & $\left(\vec{1} , \vec{1} , -2\right)$ \\
\hline\end{tabular}
\caption{$1 \rightarrow l$, $2 \rightarrow l$, $3 \rightarrow l$, $4 \rightarrow \overline{L}$, $5 \rightarrow \overline{l}$, $6 \rightarrow \overline{l}$}
\label{tab:second_last}
\end{table}

\subsection{Third Topology}
\begin{figure}[H]
    \centering
    \includegraphics[width=0.35\textwidth]{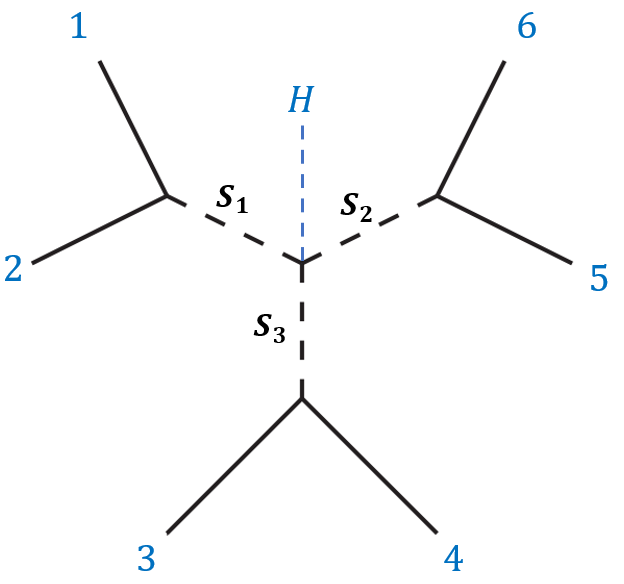}
    \caption{Feynman diagram for the cases where $H$ is attached to a central vertex of mediators. The corresponding tables are Tab.~\ref{tab:third_first} - Tab.~\ref{tab:third_last}.}
\end{figure}

\subsubsection{$\bar{L} \, \bar{L} \, \bar{L} \, l \, L \, L \, H$}

\begin{table}[H]
\centering
\renewcommand{\arraystretch}{1.4}
\setlength{\tabcolsep}{0.15cm} % Adjust the column separation to make it narrower
\begin{tabular}{|>{\centering\arraybackslash}p{1.5cm}|>{\centering\arraybackslash}p{1.5cm}|>{\centering\arraybackslash}p{1.5cm}|}
\hline
\textbf{$\vec{S_1}$} & \textbf{$\vec{S_2}$} & \textbf{$\vec{S_3}$} \\
\hhline{|===|}
$\left(\vec{1} , \vec{2} , -\frac{1}{2}\right)$ & $\left(\vec{1} , \vec{3} , -1\right)$ & $\left(\vec{1} , \vec{3} , 1\right)$ \\
\hline
$\left(\vec{1} , \vec{2} , -\frac{1}{2}\right)$ & $\left(\vec{1} , \vec{3} , -1\right)$ & $\left(\vec{1} , \vec{1} , 1\right)$ \\
\hline
$\left(\vec{1} , \vec{2} , -\frac{1}{2}\right)$ & $\left(\vec{1} , \vec{1} , -1\right)$ & $\left(\vec{1} , \vec{3} , 1\right)$ \\
\hline
$\left(\vec{1} , \vec{2} , -\frac{1}{2}\right)$ & $\left(\vec{1} , \vec{1} , -1\right)$ & $\left(\vec{1} , \vec{1} , 1\right)$ \\
\hline\end{tabular}
\caption{$1 \rightarrow l$, $2 \rightarrow \overline{L}$, $3 \rightarrow \overline{L}$, $4 \rightarrow \overline{L}$, $5 \rightarrow L$, $6 \rightarrow L$}
\label{tab:third_first}
\end{table}

\subsubsection{$\bar{L} \, \bar{L} \, \bar{l} \, l \, l \, L \, H$}

\begin{table}[H]
\centering
\renewcommand{\arraystretch}{1.4}
\setlength{\tabcolsep}{0.15cm} % Adjust the column separation to make it narrower
\begin{tabular}{|>{\centering\arraybackslash}p{1.5cm}|>{\centering\arraybackslash}p{1.5cm}|>{\centering\arraybackslash}p{1.5cm}|}
\hline
\textbf{$\vec{S_1}$} & \textbf{$\vec{S_2}$} & \textbf{$\vec{S_3}$} \\
\hhline{|===|}
$\left(\vec{1} , \vec{1} , -2\right)$ & $\left(\vec{1} , \vec{2} , \frac{1}{2}\right)$ & $\left(\vec{1} , \vec{3} , 1\right)$ \\
\hline
$\left(\vec{1} , \vec{1} , -2\right)$ & $\left(\vec{1} , \vec{2} , \frac{1}{2}\right)$ & $\left(\vec{1} , \vec{1} , 1\right)$ \\
\hline\end{tabular}
\caption{$1 \rightarrow l$, $2 \rightarrow l$, $3 \rightarrow \overline{L}$, $4 \rightarrow \overline{L}$, $5 \rightarrow \overline{l}$, $6 \rightarrow L$}
\end{table}

\begin{table}[H]
\centering
\renewcommand{\arraystretch}{1.4}
\setlength{\tabcolsep}{0.15cm} % Adjust the column separation to make it narrower
\begin{tabular}{|>{\centering\arraybackslash}p{1.5cm}|>{\centering\arraybackslash}p{1.5cm}|>{\centering\arraybackslash}p{1.5cm}|}
\hline
\textbf{$\vec{S_1}$} & \textbf{$\vec{S_2}$} & \textbf{$\vec{S_3}$} \\
\hhline{|===|}
$\left(\vec{1} , \vec{2} , -\frac{1}{2}\right)$ & $\left(\vec{1} , \vec{2} , \frac{1}{2}\right)$ & $\left(\vec{1} , \vec{2} , -\frac{1}{2}\right)$ \\
\hline\end{tabular}
\caption{$1 \rightarrow l$, $2 \rightarrow \overline{L}$, $3 \rightarrow l$, $4 \rightarrow \overline{L}$, $5 \rightarrow \overline{l}$, $6 \rightarrow L$}
\end{table}

\subsubsection{$\bar{L} \, \bar{l} \, \bar{l} \, l \, l \, l \, H$}

\begin{table}[H]
\centering
\renewcommand{\arraystretch}{1.4}
\setlength{\tabcolsep}{0.15cm} % Adjust the column separation to make it narrower
\begin{tabular}{|>{\centering\arraybackslash}p{1.5cm}|>{\centering\arraybackslash}p{1.5cm}|>{\centering\arraybackslash}p{1.5cm}|}
\hline
\textbf{$\vec{S_1}$} & \textbf{$\vec{S_2}$} & \textbf{$\vec{S_3}$} \\
\hhline{|===|}
$\left(\vec{1} , \vec{1} , -2\right)$ & $\left(\vec{1} , \vec{1} , 2\right)$ & $\left(\vec{1} , \vec{2} , -\frac{1}{2}\right)$ \\
\hline\end{tabular}
\caption{$1 \rightarrow l$, $2 \rightarrow l$, $3 \rightarrow l$, $4 \rightarrow \overline{L}$, $5 \rightarrow \overline{l}$, $6 \rightarrow \overline{l}$}
\label{tab:third_last}
\end{table}

\subsection{Fourth Topology}
\begin{figure}[H]
    \centering
    \includegraphics[width=0.35\textwidth]{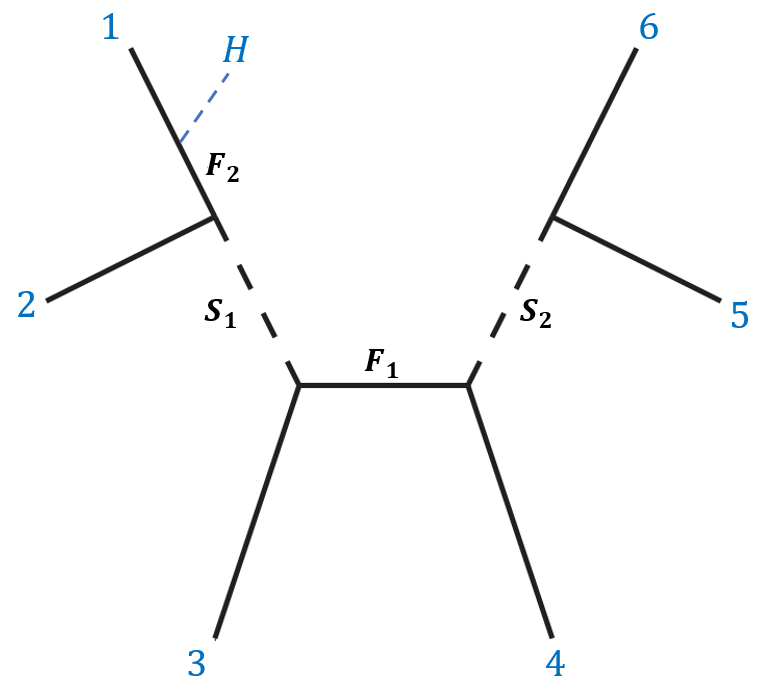}
    \caption{Feynman diagram for the cases where $H$ is attached to an external leg. The corresponding tables are Tab.~\ref{tab:fourth_first} - Tab.~\ref{tab:fourth_last}.}
\end{figure}

\subsubsection{$\bar{L} \, \bar{L} \, \bar{L} \, l \, L \, L \, H$}

\begin{table}[H]
\centering
\renewcommand{\arraystretch}{1.4}
\setlength{\tabcolsep}{0.15cm} % Adjust the column separation to make it narrower
\begin{tabular}{|>{\centering\arraybackslash}p{1.5cm}|>{\centering\arraybackslash}p{1.5cm}|>{\centering\arraybackslash}p{1.5cm}|>{\centering\arraybackslash}p{1.5cm}|}
\hline
\textbf{$\vec{S_1}$} & \textbf{$\vec{S_2}$} & \textbf{$\vec{F_1}$} & \textbf{$\vec{F_2}$} \\
\hhline{|====|}
$\left(\vec{1} , \vec{4} , \frac{3}{2}\right)$ & $\left(\vec{1} , \vec{3} , -1\right)$ & $\left(\vec{1} , \vec{3} , -2\right)$ & $\left(\vec{1} , \vec{3} , 1\right)$ \\
\hline
$\left(\vec{1} , \vec{2} , \frac{3}{2}\right)$ & $\left(\vec{1} , \vec{3} , -1\right)$ & $\left(\vec{1} , \vec{3} , -2\right)$ & $\left(\vec{1} , \vec{3} , 1\right)$ \\
\hline
$\left(\vec{1} , \vec{2} , \frac{3}{2}\right)$ & $\left(\vec{1} , \vec{1} , -1\right)$ & $\left(\vec{1} , \vec{1} , -2\right)$ & $\left(\vec{1} , \vec{3} , 1\right)$ \\
\hline
$\left(\vec{1} , \vec{2} , \frac{3}{2}\right)$ & $\left(\vec{1} , \vec{3} , -1\right)$ & $\left(\vec{1} , \vec{3} , -2\right)$ & $\left(\vec{1} , \vec{1} , 1\right)$ \\
\hline
$\left(\vec{1} , \vec{2} , \frac{3}{2}\right)$ & $\left(\vec{1} , \vec{1} , -1\right)$ & $\left(\vec{1} , \vec{1} , -2\right)$ & $\left(\vec{1} , \vec{1} , 1\right)$ \\
\hline\end{tabular}
\caption{$1 \rightarrow \overline{L}$, $2 \rightarrow \overline{L}$, $3 \rightarrow \overline{L}$, $4 \rightarrow l$, $5 \rightarrow L$, $6 \rightarrow L$}
\label{tab:fourth_first}
\end{table}

\begin{table}[H]
\centering
\renewcommand{\arraystretch}{1.4}
\setlength{\tabcolsep}{0.15cm} % Adjust the column separation to make it narrower
\begin{tabular}{|>{\centering\arraybackslash}p{1.5cm}|>{\centering\arraybackslash}p{1.5cm}|>{\centering\arraybackslash}p{1.5cm}|>{\centering\arraybackslash}p{1.5cm}|}
\hline
\textbf{$\vec{S_1}$} & \textbf{$\vec{S_2}$} & \textbf{$\vec{F_1}$} & \textbf{$\vec{F_2}$} \\
\hhline{|====|}
$\left(\vec{1} , \vec{4} , \frac{3}{2}\right)$ & $\left(\vec{1} , \vec{2} , -\frac{1}{2}\right)$ & $\left(\vec{1} , \vec{3} , -1\right)$ & $\left(\vec{1} , \vec{3} , 1\right)$ \\
\hline
$\left(\vec{1} , \vec{2} , \frac{3}{2}\right)$ & $\left(\vec{1} , \vec{2} , -\frac{1}{2}\right)$ & $\left(\vec{1} , \vec{3} , -1\right)$ & $\left(\vec{1} , \vec{3} , 1\right)$ \\
\hline
$\left(\vec{1} , \vec{2} , \frac{3}{2}\right)$ & $\left(\vec{1} , \vec{2} , -\frac{1}{2}\right)$ & $\left(\vec{1} , \vec{1} , -1\right)$ & $\left(\vec{1} , \vec{3} , 1\right)$ \\
\hline
$\left(\vec{1} , \vec{2} , \frac{3}{2}\right)$ & $\left(\vec{1} , \vec{2} , -\frac{1}{2}\right)$ & $\left(\vec{1} , \vec{3} , -1\right)$ & $\left(\vec{1} , \vec{1} , 1\right)$ \\
\hline
$\left(\vec{1} , \vec{2} , \frac{3}{2}\right)$ & $\left(\vec{1} , \vec{2} , -\frac{1}{2}\right)$ & $\left(\vec{1} , \vec{1} , -1\right)$ & $\left(\vec{1} , \vec{1} , 1\right)$ \\
\hline\end{tabular}
\caption{$1 \rightarrow \overline{L}$, $2 \rightarrow \overline{L}$, $3 \rightarrow L$, $4 \rightarrow L$, $5 \rightarrow l$, $6 \rightarrow \overline{L}$}
\end{table}

\begin{table}[H]
\centering
\renewcommand{\arraystretch}{1.4}
\setlength{\tabcolsep}{0.15cm} % Adjust the column separation to make it narrower
\begin{tabular}{|>{\centering\arraybackslash}p{1.5cm}|>{\centering\arraybackslash}p{1.5cm}|>{\centering\arraybackslash}p{1.5cm}|>{\centering\arraybackslash}p{1.5cm}|}
\hline
\textbf{$\vec{S_1}$} & \textbf{$\vec{S_2}$} & \textbf{$\vec{F_1}$} & \textbf{$\vec{F_2}$} \\
\hhline{|====|}
$\left(\vec{1} , \vec{4} , \frac{3}{2}\right)$ & $\left(\vec{1} , \vec{3} , -1\right)$ & $\left(\vec{1} , \vec{4} , -\frac{1}{2}\right)$ & $\left(\vec{1} , \vec{3} , 1\right)$ \\
\hline
$\left(\vec{1} , \vec{2} , \frac{3}{2}\right)$ & $\left(\vec{1} , \vec{3} , -1\right)$ & $\left(\vec{1} , \vec{2} , -\frac{1}{2}\right)$ & $\left(\vec{1} , \vec{3} , 1\right)$ \\
\hline
$\left(\vec{1} , \vec{2} , \frac{3}{2}\right)$ & $\left(\vec{1} , \vec{1} , -1\right)$ & $\left(\vec{1} , \vec{2} , -\frac{1}{2}\right)$ & $\left(\vec{1} , \vec{3} , 1\right)$ \\
\hline
$\left(\vec{1} , \vec{2} , \frac{3}{2}\right)$ & $\left(\vec{1} , \vec{3} , -1\right)$ & $\left(\vec{1} , \vec{2} , -\frac{1}{2}\right)$ & $\left(\vec{1} , \vec{1} , 1\right)$ \\
\hline
$\left(\vec{1} , \vec{2} , \frac{3}{2}\right)$ & $\left(\vec{1} , \vec{1} , -1\right)$ & $\left(\vec{1} , \vec{2} , -\frac{1}{2}\right)$ & $\left(\vec{1} , \vec{1} , 1\right)$ \\
\hline\end{tabular}
\caption{$1 \rightarrow \overline{L}$, $2 \rightarrow \overline{L}$, $3 \rightarrow l$, $4 \rightarrow \overline{L}$, $5 \rightarrow L$, $6 \rightarrow L$}
\end{table}

\begin{table}[H]
\centering
\renewcommand{\arraystretch}{1.4}
\setlength{\tabcolsep}{0.15cm} % Adjust the column separation to make it narrower
\begin{tabular}{|>{\centering\arraybackslash}p{1.5cm}|>{\centering\arraybackslash}p{1.5cm}|>{\centering\arraybackslash}p{1.5cm}|>{\centering\arraybackslash}p{1.5cm}|}
\hline
\textbf{$\vec{S_1}$} & \textbf{$\vec{S_2}$} & \textbf{$\vec{F_1}$} & \textbf{$\vec{F_2}$} \\
\hhline{|====|}
$\left(\vec{1} , \vec{3} , 0\right)$ & $\left(\vec{1} , \vec{3} , -1\right)$ & $\left(\vec{1} , \vec{4} , -\frac{1}{2}\right)$ & $\left(\vec{1} , \vec{3} , 1\right)$ \\
\hline
$\left(\vec{1} , \vec{3} , 0\right)$ & $\left(\vec{1} , \vec{3} , -1\right)$ & $\left(\vec{1} , \vec{2} , -\frac{1}{2}\right)$ & $\left(\vec{1} , \vec{3} , 1\right)$ \\
\hline
$\left(\vec{1} , \vec{3} , 0\right)$ & $\left(\vec{1} , \vec{1} , -1\right)$ & $\left(\vec{1} , \vec{2} , -\frac{1}{2}\right)$ & $\left(\vec{1} , \vec{3} , 1\right)$ \\
\hline
$\left(\vec{1} , \vec{1} , 0\right)$ & $\left(\vec{1} , \vec{3} , -1\right)$ & $\left(\vec{1} , \vec{2} , -\frac{1}{2}\right)$ & $\left(\vec{1} , \vec{1} , 1\right)$ \\
\hline
$\left(\vec{1} , \vec{1} , 0\right)$ & $\left(\vec{1} , \vec{1} , -1\right)$ & $\left(\vec{1} , \vec{2} , -\frac{1}{2}\right)$ & $\left(\vec{1} , \vec{1} , 1\right)$ \\
\hline\end{tabular}
\caption{$1 \rightarrow \overline{L}$, $2 \rightarrow l$, $3 \rightarrow \overline{L}$, $4 \rightarrow \overline{L}$, $5 \rightarrow L$, $6 \rightarrow L$}
\end{table}

\begin{table}[H]
\centering
\renewcommand{\arraystretch}{1.4}
\setlength{\tabcolsep}{0.15cm} % Adjust the column separation to make it narrower
\begin{tabular}{|>{\centering\arraybackslash}p{1.5cm}|>{\centering\arraybackslash}p{1.5cm}|>{\centering\arraybackslash}p{1.5cm}|>{\centering\arraybackslash}p{1.5cm}|}
\hline
\textbf{$\vec{S_1}$} & \textbf{$\vec{S_2}$} & \textbf{$\vec{F_1}$} & \textbf{$\vec{F_2}$} \\
\hhline{|====|}
$\left(\vec{1} , \vec{3} , 0\right)$ & $\left(\vec{1} , \vec{3} , 1\right)$ & $\left(\vec{1} , \vec{4} , \frac{1}{2}\right)$ & $\left(\vec{1} , \vec{3} , 1\right)$ \\
\hline
$\left(\vec{1} , \vec{3} , 0\right)$ & $\left(\vec{1} , \vec{3} , 1\right)$ & $\left(\vec{1} , \vec{2} , \frac{1}{2}\right)$ & $\left(\vec{1} , \vec{3} , 1\right)$ \\
\hline
$\left(\vec{1} , \vec{3} , 0\right)$ & $\left(\vec{1} , \vec{1} , 1\right)$ & $\left(\vec{1} , \vec{2} , \frac{1}{2}\right)$ & $\left(\vec{1} , \vec{3} , 1\right)$ \\
\hline
$\left(\vec{1} , \vec{1} , 0\right)$ & $\left(\vec{1} , \vec{3} , 1\right)$ & $\left(\vec{1} , \vec{2} , \frac{1}{2}\right)$ & $\left(\vec{1} , \vec{1} , 1\right)$ \\
\hline
$\left(\vec{1} , \vec{1} , 0\right)$ & $\left(\vec{1} , \vec{1} , 1\right)$ & $\left(\vec{1} , \vec{2} , \frac{1}{2}\right)$ & $\left(\vec{1} , \vec{1} , 1\right)$ \\
\hline\end{tabular}
\caption{$1 \rightarrow \overline{L}$, $2 \rightarrow l$, $3 \rightarrow L$, $4 \rightarrow L$, $5 \rightarrow \overline{L}$, $6 \rightarrow \overline{L}$}
\end{table}

\begin{table}[H]
\centering
\renewcommand{\arraystretch}{1.4}
\setlength{\tabcolsep}{0.15cm} % Adjust the column separation to make it narrower
\begin{tabular}{|>{\centering\arraybackslash}p{1.5cm}|>{\centering\arraybackslash}p{1.5cm}|>{\centering\arraybackslash}p{1.5cm}|>{\centering\arraybackslash}p{1.5cm}|}
\hline
\textbf{$\vec{S_1}$} & \textbf{$\vec{S_2}$} & \textbf{$\vec{F_1}$} & \textbf{$\vec{F_2}$} \\
\hhline{|====|}
$\left(\vec{1} , \vec{4} , -\frac{1}{2}\right)$ & $\left(\vec{1} , \vec{2} , -\frac{1}{2}\right)$ & $\left(\vec{1} , \vec{3} , 0\right)$ & $\left(\vec{1} , \vec{3} , 0\right)$ \\
\hline
$\left(\vec{1} , \vec{2} , -\frac{1}{2}\right)$ & $\left(\vec{1} , \vec{2} , -\frac{1}{2}\right)$ & $\left(\vec{1} , \vec{3} , 0\right)$ & $\left(\vec{1} , \vec{3} , 0\right)$ \\
\hline
$\left(\vec{1} , \vec{2} , -\frac{1}{2}\right)$ & $\left(\vec{1} , \vec{2} , -\frac{1}{2}\right)$ & $\left(\vec{1} , \vec{1} , 0\right)$ & $\left(\vec{1} , \vec{3} , 0\right)$ \\
\hline
$\left(\vec{1} , \vec{2} , -\frac{1}{2}\right)$ & $\left(\vec{1} , \vec{2} , -\frac{1}{2}\right)$ & $\left(\vec{1} , \vec{3} , 0\right)$ & $\left(\vec{1} , \vec{1} , 0\right)$ \\
\hline
$\left(\vec{1} , \vec{2} , -\frac{1}{2}\right)$ & $\left(\vec{1} , \vec{2} , -\frac{1}{2}\right)$ & $\left(\vec{1} , \vec{1} , 0\right)$ & $\left(\vec{1} , \vec{1} , 0\right)$ \\
\hline\end{tabular}
\caption{$1 \rightarrow L$, $2 \rightarrow L$, $3 \rightarrow \overline{L}$, $4 \rightarrow \overline{L}$, $5 \rightarrow l$, $6 \rightarrow \overline{L}$}
\end{table}

\begin{table}[H]
\centering
\renewcommand{\arraystretch}{1.4}
\setlength{\tabcolsep}{0.15cm} % Adjust the column separation to make it narrower
\begin{tabular}{|>{\centering\arraybackslash}p{1.5cm}|>{\centering\arraybackslash}p{1.5cm}|>{\centering\arraybackslash}p{1.5cm}|>{\centering\arraybackslash}p{1.5cm}|}
\hline
\textbf{$\vec{S_1}$} & \textbf{$\vec{S_2}$} & \textbf{$\vec{F_1}$} & \textbf{$\vec{F_2}$} \\
\hhline{|====|}
$\left(\vec{1} , \vec{4} , -\frac{1}{2}\right)$ & $\left(\vec{1} , \vec{3} , 1\right)$ & $\left(\vec{1} , \vec{3} , 0\right)$ & $\left(\vec{1} , \vec{3} , 0\right)$ \\
\hline
$\left(\vec{1} , \vec{2} , -\frac{1}{2}\right)$ & $\left(\vec{1} , \vec{3} , 1\right)$ & $\left(\vec{1} , \vec{3} , 0\right)$ & $\left(\vec{1} , \vec{3} , 0\right)$ \\
\hline
$\left(\vec{1} , \vec{2} , -\frac{1}{2}\right)$ & $\left(\vec{1} , \vec{1} , 1\right)$ & $\left(\vec{1} , \vec{1} , 0\right)$ & $\left(\vec{1} , \vec{3} , 0\right)$ \\
\hline
$\left(\vec{1} , \vec{2} , -\frac{1}{2}\right)$ & $\left(\vec{1} , \vec{3} , 1\right)$ & $\left(\vec{1} , \vec{3} , 0\right)$ & $\left(\vec{1} , \vec{1} , 0\right)$ \\
\hline
$\left(\vec{1} , \vec{2} , -\frac{1}{2}\right)$ & $\left(\vec{1} , \vec{1} , 1\right)$ & $\left(\vec{1} , \vec{1} , 0\right)$ & $\left(\vec{1} , \vec{1} , 0\right)$ \\
\hline\end{tabular}
\caption{$1 \rightarrow L$, $2 \rightarrow L$, $3 \rightarrow \overline{L}$, $4 \rightarrow l$, $5 \rightarrow \overline{L}$, $6 \rightarrow \overline{L}$}
\end{table}

\begin{table}[H]
\centering
\renewcommand{\arraystretch}{1.4}
\setlength{\tabcolsep}{0.15cm} % Adjust the column separation to make it narrower
\begin{tabular}{|>{\centering\arraybackslash}p{1.5cm}|>{\centering\arraybackslash}p{1.5cm}|>{\centering\arraybackslash}p{1.5cm}|>{\centering\arraybackslash}p{1.5cm}|}
\hline
\textbf{$\vec{S_1}$} & \textbf{$\vec{S_2}$} & \textbf{$\vec{F_1}$} & \textbf{$\vec{F_2}$} \\
\hhline{|====|}
$\left(\vec{1} , \vec{4} , -\frac{1}{2}\right)$ & $\left(\vec{1} , \vec{3} , 1\right)$ & $\left(\vec{1} , \vec{4} , \frac{3}{2}\right)$ & $\left(\vec{1} , \vec{3} , 0\right)$ \\
\hline
$\left(\vec{1} , \vec{2} , -\frac{1}{2}\right)$ & $\left(\vec{1} , \vec{3} , 1\right)$ & $\left(\vec{1} , \vec{2} , \frac{3}{2}\right)$ & $\left(\vec{1} , \vec{3} , 0\right)$ \\
\hline
$\left(\vec{1} , \vec{2} , -\frac{1}{2}\right)$ & $\left(\vec{1} , \vec{1} , 1\right)$ & $\left(\vec{1} , \vec{2} , \frac{3}{2}\right)$ & $\left(\vec{1} , \vec{3} , 0\right)$ \\
\hline
$\left(\vec{1} , \vec{2} , -\frac{1}{2}\right)$ & $\left(\vec{1} , \vec{3} , 1\right)$ & $\left(\vec{1} , \vec{2} , \frac{3}{2}\right)$ & $\left(\vec{1} , \vec{1} , 0\right)$ \\
\hline
$\left(\vec{1} , \vec{2} , -\frac{1}{2}\right)$ & $\left(\vec{1} , \vec{1} , 1\right)$ & $\left(\vec{1} , \vec{2} , \frac{3}{2}\right)$ & $\left(\vec{1} , \vec{1} , 0\right)$ \\
\hline\end{tabular}
\caption{$1 \rightarrow L$, $2 \rightarrow L$, $3 \rightarrow l$, $4 \rightarrow \overline{L}$, $5 \rightarrow \overline{L}$, $6 \rightarrow \overline{L}$}
\end{table}

\begin{table}[H]
\centering
\renewcommand{\arraystretch}{1.4}
\setlength{\tabcolsep}{0.15cm} % Adjust the column separation to make it narrower
\begin{tabular}{|>{\centering\arraybackslash}p{1.5cm}|>{\centering\arraybackslash}p{1.5cm}|>{\centering\arraybackslash}p{1.5cm}|>{\centering\arraybackslash}p{1.5cm}|}
\hline
\textbf{$\vec{S_1}$} & \textbf{$\vec{S_2}$} & \textbf{$\vec{F_1}$} & \textbf{$\vec{F_2}$} \\
\hhline{|====|}
$\left(\vec{1} , \vec{3} , 0\right)$ & $\left(\vec{1} , \vec{3} , -1\right)$ & $\left(\vec{1} , \vec{4} , -\frac{1}{2}\right)$ & $\left(\vec{1} , \vec{2} , -\frac{1}{2}\right)$ \\
\hline
$\left(\vec{1} , \vec{3} , 0\right)$ & $\left(\vec{1} , \vec{3} , -1\right)$ & $\left(\vec{1} , \vec{2} , -\frac{1}{2}\right)$ & $\left(\vec{1} , \vec{2} , -\frac{1}{2}\right)$ \\
\hline
$\left(\vec{1} , \vec{1} , 0\right)$ & $\left(\vec{1} , \vec{3} , -1\right)$ & $\left(\vec{1} , \vec{2} , -\frac{1}{2}\right)$ & $\left(\vec{1} , \vec{2} , -\frac{1}{2}\right)$ \\
\hline
$\left(\vec{1} , \vec{3} , 0\right)$ & $\left(\vec{1} , \vec{1} , -1\right)$ & $\left(\vec{1} , \vec{2} , -\frac{1}{2}\right)$ & $\left(\vec{1} , \vec{2} , -\frac{1}{2}\right)$ \\
\hline
$\left(\vec{1} , \vec{1} , 0\right)$ & $\left(\vec{1} , \vec{1} , -1\right)$ & $\left(\vec{1} , \vec{2} , -\frac{1}{2}\right)$ & $\left(\vec{1} , \vec{2} , -\frac{1}{2}\right)$ \\
\hline\end{tabular}
\caption{$1 \rightarrow l$, $2 \rightarrow \overline{L}$, $3 \rightarrow \overline{L}$, $4 \rightarrow \overline{L}$, $5 \rightarrow L$, $6 \rightarrow L$}
\end{table}

\begin{table}[H]
\centering
\renewcommand{\arraystretch}{1.4}
\setlength{\tabcolsep}{0.15cm} % Adjust the column separation to make it narrower
\begin{tabular}{|>{\centering\arraybackslash}p{1.5cm}|>{\centering\arraybackslash}p{1.5cm}|>{\centering\arraybackslash}p{1.5cm}|>{\centering\arraybackslash}p{1.5cm}|}
\hline
\textbf{$\vec{S_1}$} & \textbf{$\vec{S_2}$} & \textbf{$\vec{F_1}$} & \textbf{$\vec{F_2}$} \\
\hhline{|====|}
$\left(\vec{1} , \vec{3} , 0\right)$ & $\left(\vec{1} , \vec{3} , 1\right)$ & $\left(\vec{1} , \vec{4} , \frac{1}{2}\right)$ & $\left(\vec{1} , \vec{2} , -\frac{1}{2}\right)$ \\
\hline
$\left(\vec{1} , \vec{3} , 0\right)$ & $\left(\vec{1} , \vec{3} , 1\right)$ & $\left(\vec{1} , \vec{2} , \frac{1}{2}\right)$ & $\left(\vec{1} , \vec{2} , -\frac{1}{2}\right)$ \\
\hline
$\left(\vec{1} , \vec{1} , 0\right)$ & $\left(\vec{1} , \vec{3} , 1\right)$ & $\left(\vec{1} , \vec{2} , \frac{1}{2}\right)$ & $\left(\vec{1} , \vec{2} , -\frac{1}{2}\right)$ \\
\hline
$\left(\vec{1} , \vec{3} , 0\right)$ & $\left(\vec{1} , \vec{1} , 1\right)$ & $\left(\vec{1} , \vec{2} , \frac{1}{2}\right)$ & $\left(\vec{1} , \vec{2} , -\frac{1}{2}\right)$ \\
\hline
$\left(\vec{1} , \vec{1} , 0\right)$ & $\left(\vec{1} , \vec{1} , 1\right)$ & $\left(\vec{1} , \vec{2} , \frac{1}{2}\right)$ & $\left(\vec{1} , \vec{2} , -\frac{1}{2}\right)$ \\
\hline\end{tabular}
\caption{$1 \rightarrow l$, $2 \rightarrow \overline{L}$, $3 \rightarrow L$, $4 \rightarrow L$, $5 \rightarrow \overline{L}$, $6 \rightarrow \overline{L}$}
\end{table}

\subsubsection{$\bar{L} \, \bar{L} \, \bar{l} \, l \, l \, L \, H$}

\begin{table}[H]
\centering
\renewcommand{\arraystretch}{1.4}
\setlength{\tabcolsep}{0.15cm} % Adjust the column separation to make it narrower
\begin{tabular}{|>{\centering\arraybackslash}p{1.5cm}|>{\centering\arraybackslash}p{1.5cm}|>{\centering\arraybackslash}p{1.5cm}|>{\centering\arraybackslash}p{1.5cm}|}
\hline
\textbf{$\vec{S_1}$} & \textbf{$\vec{S_2}$} & \textbf{$\vec{F_1}$} & \textbf{$\vec{F_2}$} \\
\hhline{|====|}
$\left(\vec{1} , \vec{2} , \frac{3}{2}\right)$ & $\left(\vec{1} , \vec{1} , -2\right)$ & $\left(\vec{1} , \vec{2} , -\frac{5}{2}\right)$ & $\left(\vec{1} , \vec{3} , 1\right)$ \\
\hline
$\left(\vec{1} , \vec{2} , \frac{3}{2}\right)$ & $\left(\vec{1} , \vec{1} , -2\right)$ & $\left(\vec{1} , \vec{2} , -\frac{5}{2}\right)$ & $\left(\vec{1} , \vec{1} , 1\right)$ \\
\hline\end{tabular}
\caption{$1 \rightarrow \overline{L}$, $2 \rightarrow \overline{L}$, $3 \rightarrow \overline{l}$, $4 \rightarrow L$, $5 \rightarrow l$, $6 \rightarrow l$}
\end{table}

\begin{table}[H]
\centering
\renewcommand{\arraystretch}{1.4}
\setlength{\tabcolsep}{0.15cm} % Adjust the column separation to make it narrower
\begin{tabular}{|>{\centering\arraybackslash}p{1.5cm}|>{\centering\arraybackslash}p{1.5cm}|>{\centering\arraybackslash}p{1.5cm}|>{\centering\arraybackslash}p{1.5cm}|}
\hline
\textbf{$\vec{S_1}$} & \textbf{$\vec{S_2}$} & \textbf{$\vec{F_1}$} & \textbf{$\vec{F_2}$} \\
\hhline{|====|}
$\left(\vec{1} , \vec{2} , \frac{3}{2}\right)$ & $\left(\vec{1} , \vec{2} , \frac{1}{2}\right)$ & $\left(\vec{1} , \vec{2} , -\frac{1}{2}\right)$ & $\left(\vec{1} , \vec{3} , 1\right)$ \\
\hline
$\left(\vec{1} , \vec{2} , \frac{3}{2}\right)$ & $\left(\vec{1} , \vec{2} , \frac{1}{2}\right)$ & $\left(\vec{1} , \vec{2} , -\frac{1}{2}\right)$ & $\left(\vec{1} , \vec{1} , 1\right)$ \\
\hline\end{tabular}
\caption{$1 \rightarrow \overline{L}$, $2 \rightarrow \overline{L}$, $3 \rightarrow l$, $4 \rightarrow l$, $5 \rightarrow \overline{l}$, $6 \rightarrow L$}
\end{table}

\begin{table}[H]
\centering
\renewcommand{\arraystretch}{1.4}
\setlength{\tabcolsep}{0.15cm} % Adjust the column separation to make it narrower
\begin{tabular}{|>{\centering\arraybackslash}p{1.5cm}|>{\centering\arraybackslash}p{1.5cm}|>{\centering\arraybackslash}p{1.5cm}|>{\centering\arraybackslash}p{1.5cm}|}
\hline
\textbf{$\vec{S_1}$} & \textbf{$\vec{S_2}$} & \textbf{$\vec{F_1}$} & \textbf{$\vec{F_2}$} \\
\hhline{|====|}
$\left(\vec{1} , \vec{2} , \frac{3}{2}\right)$ & $\left(\vec{1} , \vec{1} , -2\right)$ & $\left(\vec{1} , \vec{1} , -1\right)$ & $\left(\vec{1} , \vec{3} , 1\right)$ \\
\hline
$\left(\vec{1} , \vec{2} , \frac{3}{2}\right)$ & $\left(\vec{1} , \vec{1} , -2\right)$ & $\left(\vec{1} , \vec{1} , -1\right)$ & $\left(\vec{1} , \vec{1} , 1\right)$ \\
\hline\end{tabular}
\caption{$1 \rightarrow \overline{L}$, $2 \rightarrow \overline{L}$, $3 \rightarrow L$, $4 \rightarrow \overline{l}$, $5 \rightarrow l$, $6 \rightarrow l$}
\end{table}

\begin{table}[H]
\centering
\renewcommand{\arraystretch}{1.4}
\setlength{\tabcolsep}{0.15cm} % Adjust the column separation to make it narrower
\begin{tabular}{|>{\centering\arraybackslash}p{1.5cm}|>{\centering\arraybackslash}p{1.5cm}|>{\centering\arraybackslash}p{1.5cm}|>{\centering\arraybackslash}p{1.5cm}|}
\hline
\textbf{$\vec{S_1}$} & \textbf{$\vec{S_2}$} & \textbf{$\vec{F_1}$} & \textbf{$\vec{F_2}$} \\
\hhline{|====|}
$\left(\vec{1} , \vec{3} , 0\right)$ & $\left(\vec{1} , \vec{2} , \frac{1}{2}\right)$ & $\left(\vec{1} , \vec{2} , -\frac{1}{2}\right)$ & $\left(\vec{1} , \vec{3} , 1\right)$ \\
\hline
$\left(\vec{1} , \vec{1} , 0\right)$ & $\left(\vec{1} , \vec{2} , \frac{1}{2}\right)$ & $\left(\vec{1} , \vec{2} , -\frac{1}{2}\right)$ & $\left(\vec{1} , \vec{1} , 1\right)$ \\
\hline\end{tabular}
\caption{$1 \rightarrow \overline{L}$, $2 \rightarrow l$, $3 \rightarrow \overline{L}$, $4 \rightarrow l$, $5 \rightarrow \overline{l}$, $6 \rightarrow L$}
\end{table}

\begin{table}[H]
\centering
\renewcommand{\arraystretch}{1.4}
\setlength{\tabcolsep}{0.15cm} % Adjust the column separation to make it narrower
\begin{tabular}{|>{\centering\arraybackslash}p{1.5cm}|>{\centering\arraybackslash}p{1.5cm}|>{\centering\arraybackslash}p{1.5cm}|>{\centering\arraybackslash}p{1.5cm}|}
\hline
\textbf{$\vec{S_1}$} & \textbf{$\vec{S_2}$} & \textbf{$\vec{F_1}$} & \textbf{$\vec{F_2}$} \\
\hhline{|====|}
$\left(\vec{1} , \vec{3} , 0\right)$ & $\left(\vec{1} , \vec{2} , -\frac{1}{2}\right)$ & $\left(\vec{1} , \vec{3} , -1\right)$ & $\left(\vec{1} , \vec{3} , 1\right)$ \\
\hline
$\left(\vec{1} , \vec{1} , 0\right)$ & $\left(\vec{1} , \vec{2} , -\frac{1}{2}\right)$ & $\left(\vec{1} , \vec{1} , -1\right)$ & $\left(\vec{1} , \vec{1} , 1\right)$ \\
\hline\end{tabular}
\caption{$1 \rightarrow \overline{L}$, $2 \rightarrow l$, $3 \rightarrow \overline{l}$, $4 \rightarrow L$, $5 \rightarrow l$, $6 \rightarrow \overline{L}$}
\end{table}

\begin{table}[H]
\centering
\renewcommand{\arraystretch}{1.4}
\setlength{\tabcolsep}{0.15cm} % Adjust the column separation to make it narrower
\begin{tabular}{|>{\centering\arraybackslash}p{1.5cm}|>{\centering\arraybackslash}p{1.5cm}|>{\centering\arraybackslash}p{1.5cm}|>{\centering\arraybackslash}p{1.5cm}|}
\hline
\textbf{$\vec{S_1}$} & \textbf{$\vec{S_2}$} & \textbf{$\vec{F_1}$} & \textbf{$\vec{F_2}$} \\
\hhline{|====|}
$\left(\vec{1} , \vec{3} , 0\right)$ & $\left(\vec{1} , \vec{2} , \frac{1}{2}\right)$ & $\left(\vec{1} , \vec{3} , 1\right)$ & $\left(\vec{1} , \vec{3} , 1\right)$ \\
\hline
$\left(\vec{1} , \vec{1} , 0\right)$ & $\left(\vec{1} , \vec{2} , \frac{1}{2}\right)$ & $\left(\vec{1} , \vec{1} , 1\right)$ & $\left(\vec{1} , \vec{1} , 1\right)$ \\
\hline\end{tabular}
\caption{$1 \rightarrow \overline{L}$, $2 \rightarrow l$, $3 \rightarrow l$, $4 \rightarrow \overline{L}$, $5 \rightarrow \overline{l}$, $6 \rightarrow L$}
\end{table}

\begin{table}[H]
\centering
\renewcommand{\arraystretch}{1.4}
\setlength{\tabcolsep}{0.15cm} % Adjust the column separation to make it narrower
\begin{tabular}{|>{\centering\arraybackslash}p{1.5cm}|>{\centering\arraybackslash}p{1.5cm}|>{\centering\arraybackslash}p{1.5cm}|>{\centering\arraybackslash}p{1.5cm}|}
\hline
\textbf{$\vec{S_1}$} & \textbf{$\vec{S_2}$} & \textbf{$\vec{F_1}$} & \textbf{$\vec{F_2}$} \\
\hhline{|====|}
$\left(\vec{1} , \vec{3} , 0\right)$ & $\left(\vec{1} , \vec{2} , -\frac{1}{2}\right)$ & $\left(\vec{1} , \vec{2} , \frac{1}{2}\right)$ & $\left(\vec{1} , \vec{3} , 1\right)$ \\
\hline
$\left(\vec{1} , \vec{1} , 0\right)$ & $\left(\vec{1} , \vec{2} , -\frac{1}{2}\right)$ & $\left(\vec{1} , \vec{2} , \frac{1}{2}\right)$ & $\left(\vec{1} , \vec{1} , 1\right)$ \\
\hline\end{tabular}
\caption{$1 \rightarrow \overline{L}$, $2 \rightarrow l$, $3 \rightarrow L$, $4 \rightarrow \overline{l}$, $5 \rightarrow l$, $6 \rightarrow \overline{L}$}
\end{table}

\begin{table}[H]
\centering
\renewcommand{\arraystretch}{1.4}
\setlength{\tabcolsep}{0.15cm} % Adjust the column separation to make it narrower
\begin{tabular}{|>{\centering\arraybackslash}p{1.5cm}|>{\centering\arraybackslash}p{1.5cm}|>{\centering\arraybackslash}p{1.5cm}|>{\centering\arraybackslash}p{1.5cm}|}
\hline
\textbf{$\vec{S_1}$} & \textbf{$\vec{S_2}$} & \textbf{$\vec{F_1}$} & \textbf{$\vec{F_2}$} \\
\hhline{|====|}
$\left(\vec{1} , \vec{3} , 1\right)$ & $\left(\vec{1} , \vec{1} , -2\right)$ & $\left(\vec{1} , \vec{2} , -\frac{3}{2}\right)$ & $\left(\vec{1} , \vec{2} , \frac{3}{2}\right)$ \\
\hline
$\left(\vec{1} , \vec{1} , 1\right)$ & $\left(\vec{1} , \vec{1} , -2\right)$ & $\left(\vec{1} , \vec{2} , -\frac{3}{2}\right)$ & $\left(\vec{1} , \vec{2} , \frac{3}{2}\right)$ \\
\hline\end{tabular}
\caption{$1 \rightarrow \overline{l}$, $2 \rightarrow L$, $3 \rightarrow \overline{L}$, $4 \rightarrow \overline{L}$, $5 \rightarrow l$, $6 \rightarrow l$}
\end{table}

\begin{table}[H]
\centering
\renewcommand{\arraystretch}{1.4}
\setlength{\tabcolsep}{0.15cm} % Adjust the column separation to make it narrower
\begin{tabular}{|>{\centering\arraybackslash}p{1.5cm}|>{\centering\arraybackslash}p{1.5cm}|>{\centering\arraybackslash}p{1.5cm}|>{\centering\arraybackslash}p{1.5cm}|}
\hline
\textbf{$\vec{S_1}$} & \textbf{$\vec{S_2}$} & \textbf{$\vec{F_1}$} & \textbf{$\vec{F_2}$} \\
\hhline{|====|}
$\left(\vec{1} , \vec{3} , 1\right)$ & $\left(\vec{1} , \vec{2} , -\frac{1}{2}\right)$ & $\left(\vec{1} , \vec{2} , -\frac{3}{2}\right)$ & $\left(\vec{1} , \vec{2} , \frac{3}{2}\right)$ \\
\hline
$\left(\vec{1} , \vec{1} , 1\right)$ & $\left(\vec{1} , \vec{2} , -\frac{1}{2}\right)$ & $\left(\vec{1} , \vec{2} , -\frac{3}{2}\right)$ & $\left(\vec{1} , \vec{2} , \frac{3}{2}\right)$ \\
\hline\end{tabular}
\caption{$1 \rightarrow \overline{l}$, $2 \rightarrow L$, $3 \rightarrow \overline{L}$, $4 \rightarrow l$, $5 \rightarrow l$, $6 \rightarrow \overline{L}$}
\end{table}

\begin{table}[H]
\centering
\renewcommand{\arraystretch}{1.4}
\setlength{\tabcolsep}{0.15cm} % Adjust the column separation to make it narrower
\begin{tabular}{|>{\centering\arraybackslash}p{1.5cm}|>{\centering\arraybackslash}p{1.5cm}|>{\centering\arraybackslash}p{1.5cm}|>{\centering\arraybackslash}p{1.5cm}|}
\hline
\textbf{$\vec{S_1}$} & \textbf{$\vec{S_2}$} & \textbf{$\vec{F_1}$} & \textbf{$\vec{F_2}$} \\
\hhline{|====|}
$\left(\vec{1} , \vec{3} , 1\right)$ & $\left(\vec{1} , \vec{2} , -\frac{1}{2}\right)$ & $\left(\vec{1} , \vec{3} , 0\right)$ & $\left(\vec{1} , \vec{2} , \frac{3}{2}\right)$ \\
\hline
$\left(\vec{1} , \vec{1} , 1\right)$ & $\left(\vec{1} , \vec{2} , -\frac{1}{2}\right)$ & $\left(\vec{1} , \vec{1} , 0\right)$ & $\left(\vec{1} , \vec{2} , \frac{3}{2}\right)$ \\
\hline\end{tabular}
\caption{$1 \rightarrow \overline{l}$, $2 \rightarrow L$, $3 \rightarrow l$, $4 \rightarrow \overline{L}$, $5 \rightarrow l$, $6 \rightarrow \overline{L}$}
\end{table}

\begin{table}[H]
\centering
\renewcommand{\arraystretch}{1.4}
\setlength{\tabcolsep}{0.15cm} % Adjust the column separation to make it narrower
\begin{tabular}{|>{\centering\arraybackslash}p{1.5cm}|>{\centering\arraybackslash}p{1.5cm}|>{\centering\arraybackslash}p{1.5cm}|>{\centering\arraybackslash}p{1.5cm}|}
\hline
\textbf{$\vec{S_1}$} & \textbf{$\vec{S_2}$} & \textbf{$\vec{F_1}$} & \textbf{$\vec{F_2}$} \\
\hhline{|====|}
$\left(\vec{1} , \vec{3} , 1\right)$ & $\left(\vec{1} , \vec{3} , 1\right)$ & $\left(\vec{1} , \vec{3} , 0\right)$ & $\left(\vec{1} , \vec{2} , \frac{3}{2}\right)$ \\
\hline
$\left(\vec{1} , \vec{1} , 1\right)$ & $\left(\vec{1} , \vec{1} , 1\right)$ & $\left(\vec{1} , \vec{1} , 0\right)$ & $\left(\vec{1} , \vec{2} , \frac{3}{2}\right)$ \\
\hline\end{tabular}
\caption{$1 \rightarrow \overline{l}$, $2 \rightarrow L$, $3 \rightarrow l$, $4 \rightarrow l$, $5 \rightarrow \overline{L}$, $6 \rightarrow \overline{L}$}
\end{table}

\begin{table}[H]
\centering
\renewcommand{\arraystretch}{1.4}
\setlength{\tabcolsep}{0.15cm} % Adjust the column separation to make it narrower
\begin{tabular}{|>{\centering\arraybackslash}p{1.5cm}|>{\centering\arraybackslash}p{1.5cm}|>{\centering\arraybackslash}p{1.5cm}|>{\centering\arraybackslash}p{1.5cm}|}
\hline
\textbf{$\vec{S_1}$} & \textbf{$\vec{S_2}$} & \textbf{$\vec{F_1}$} & \textbf{$\vec{F_2}$} \\
\hhline{|====|}
$\left(\vec{1} , \vec{3} , 0\right)$ & $\left(\vec{1} , \vec{2} , \frac{1}{2}\right)$ & $\left(\vec{1} , \vec{2} , -\frac{1}{2}\right)$ & $\left(\vec{1} , \vec{2} , -\frac{1}{2}\right)$ \\
\hline
$\left(\vec{1} , \vec{1} , 0\right)$ & $\left(\vec{1} , \vec{2} , \frac{1}{2}\right)$ & $\left(\vec{1} , \vec{2} , -\frac{1}{2}\right)$ & $\left(\vec{1} , \vec{2} , -\frac{1}{2}\right)$ \\
\hline\end{tabular}
\caption{$1 \rightarrow l$, $2 \rightarrow \overline{L}$, $3 \rightarrow \overline{L}$, $4 \rightarrow l$, $5 \rightarrow \overline{l}$, $6 \rightarrow L$}
\end{table}

\begin{table}[H]
\centering
\renewcommand{\arraystretch}{1.4}
\setlength{\tabcolsep}{0.15cm} % Adjust the column separation to make it narrower
\begin{tabular}{|>{\centering\arraybackslash}p{1.5cm}|>{\centering\arraybackslash}p{1.5cm}|>{\centering\arraybackslash}p{1.5cm}|>{\centering\arraybackslash}p{1.5cm}|}
\hline
\textbf{$\vec{S_1}$} & \textbf{$\vec{S_2}$} & \textbf{$\vec{F_1}$} & \textbf{$\vec{F_2}$} \\
\hhline{|====|}
$\left(\vec{1} , \vec{3} , 0\right)$ & $\left(\vec{1} , \vec{2} , -\frac{1}{2}\right)$ & $\left(\vec{1} , \vec{3} , -1\right)$ & $\left(\vec{1} , \vec{2} , -\frac{1}{2}\right)$ \\
\hline
$\left(\vec{1} , \vec{1} , 0\right)$ & $\left(\vec{1} , \vec{2} , -\frac{1}{2}\right)$ & $\left(\vec{1} , \vec{1} , -1\right)$ & $\left(\vec{1} , \vec{2} , -\frac{1}{2}\right)$ \\
\hline\end{tabular}
\caption{$1 \rightarrow l$, $2 \rightarrow \overline{L}$, $3 \rightarrow \overline{l}$, $4 \rightarrow L$, $5 \rightarrow l$, $6 \rightarrow \overline{L}$}
\end{table}

\begin{table}[H]
\centering
\renewcommand{\arraystretch}{1.4}
\setlength{\tabcolsep}{0.15cm} % Adjust the column separation to make it narrower
\begin{tabular}{|>{\centering\arraybackslash}p{1.5cm}|>{\centering\arraybackslash}p{1.5cm}|>{\centering\arraybackslash}p{1.5cm}|>{\centering\arraybackslash}p{1.5cm}|}
\hline
\textbf{$\vec{S_1}$} & \textbf{$\vec{S_2}$} & \textbf{$\vec{F_1}$} & \textbf{$\vec{F_2}$} \\
\hhline{|====|}
$\left(\vec{1} , \vec{3} , 0\right)$ & $\left(\vec{1} , \vec{2} , \frac{1}{2}\right)$ & $\left(\vec{1} , \vec{3} , 1\right)$ & $\left(\vec{1} , \vec{2} , -\frac{1}{2}\right)$ \\
\hline
$\left(\vec{1} , \vec{1} , 0\right)$ & $\left(\vec{1} , \vec{2} , \frac{1}{2}\right)$ & $\left(\vec{1} , \vec{1} , 1\right)$ & $\left(\vec{1} , \vec{2} , -\frac{1}{2}\right)$ \\
\hline\end{tabular}
\caption{$1 \rightarrow l$, $2 \rightarrow \overline{L}$, $3 \rightarrow l$, $4 \rightarrow \overline{L}$, $5 \rightarrow \overline{l}$, $6 \rightarrow L$}
\end{table}

\begin{table}[H]
\centering
\renewcommand{\arraystretch}{1.4}
\setlength{\tabcolsep}{0.15cm} % Adjust the column separation to make it narrower
\begin{tabular}{|>{\centering\arraybackslash}p{1.5cm}|>{\centering\arraybackslash}p{1.5cm}|>{\centering\arraybackslash}p{1.5cm}|>{\centering\arraybackslash}p{1.5cm}|}
\hline
\textbf{$\vec{S_1}$} & \textbf{$\vec{S_2}$} & \textbf{$\vec{F_1}$} & \textbf{$\vec{F_2}$} \\
\hhline{|====|}
$\left(\vec{1} , \vec{3} , 0\right)$ & $\left(\vec{1} , \vec{2} , -\frac{1}{2}\right)$ & $\left(\vec{1} , \vec{2} , \frac{1}{2}\right)$ & $\left(\vec{1} , \vec{2} , -\frac{1}{2}\right)$ \\
\hline
$\left(\vec{1} , \vec{1} , 0\right)$ & $\left(\vec{1} , \vec{2} , -\frac{1}{2}\right)$ & $\left(\vec{1} , \vec{2} , \frac{1}{2}\right)$ & $\left(\vec{1} , \vec{2} , -\frac{1}{2}\right)$ \\
\hline\end{tabular}
\caption{$1 \rightarrow l$, $2 \rightarrow \overline{L}$, $3 \rightarrow L$, $4 \rightarrow \overline{l}$, $5 \rightarrow l$, $6 \rightarrow \overline{L}$}
\end{table}

\begin{table}[H]
\centering
\renewcommand{\arraystretch}{1.4}
\setlength{\tabcolsep}{0.15cm} % Adjust the column separation to make it narrower
\begin{tabular}{|>{\centering\arraybackslash}p{1.5cm}|>{\centering\arraybackslash}p{1.5cm}|>{\centering\arraybackslash}p{1.5cm}|>{\centering\arraybackslash}p{1.5cm}|}
\hline
\textbf{$\vec{S_1}$} & \textbf{$\vec{S_2}$} & \textbf{$\vec{F_1}$} & \textbf{$\vec{F_2}$} \\
\hhline{|====|}
$\left(\vec{1} , \vec{2} , -\frac{3}{2}\right)$ & $\left(\vec{1} , \vec{2} , \frac{1}{2}\right)$ & $\left(\vec{1} , \vec{3} , 1\right)$ & $\left(\vec{1} , \vec{2} , -\frac{1}{2}\right)$ \\
\hline
$\left(\vec{1} , \vec{2} , -\frac{3}{2}\right)$ & $\left(\vec{1} , \vec{2} , \frac{1}{2}\right)$ & $\left(\vec{1} , \vec{1} , 1\right)$ & $\left(\vec{1} , \vec{2} , -\frac{1}{2}\right)$ \\
\hline\end{tabular}
\caption{$1 \rightarrow l$, $2 \rightarrow l$, $3 \rightarrow \overline{L}$, $4 \rightarrow \overline{L}$, $5 \rightarrow \overline{l}$, $6 \rightarrow L$}
\end{table}

\begin{table}[H]
\centering
\renewcommand{\arraystretch}{1.4}
\setlength{\tabcolsep}{0.15cm} % Adjust the column separation to make it narrower
\begin{tabular}{|>{\centering\arraybackslash}p{1.5cm}|>{\centering\arraybackslash}p{1.5cm}|>{\centering\arraybackslash}p{1.5cm}|>{\centering\arraybackslash}p{1.5cm}|}
\hline
\textbf{$\vec{S_1}$} & \textbf{$\vec{S_2}$} & \textbf{$\vec{F_1}$} & \textbf{$\vec{F_2}$} \\
\hhline{|====|}
$\left(\vec{1} , \vec{2} , -\frac{3}{2}\right)$ & $\left(\vec{1} , \vec{3} , 1\right)$ & $\left(\vec{1} , \vec{2} , \frac{1}{2}\right)$ & $\left(\vec{1} , \vec{2} , -\frac{1}{2}\right)$ \\
\hline
$\left(\vec{1} , \vec{2} , -\frac{3}{2}\right)$ & $\left(\vec{1} , \vec{1} , 1\right)$ & $\left(\vec{1} , \vec{2} , \frac{1}{2}\right)$ & $\left(\vec{1} , \vec{2} , -\frac{1}{2}\right)$ \\
\hline\end{tabular}
\caption{$1 \rightarrow l$, $2 \rightarrow l$, $3 \rightarrow \overline{l}$, $4 \rightarrow L$, $5 \rightarrow \overline{L}$, $6 \rightarrow \overline{L}$}
\end{table}

\begin{table}[H]
\centering
\renewcommand{\arraystretch}{1.4}
\setlength{\tabcolsep}{0.15cm} % Adjust the column separation to make it narrower
\begin{tabular}{|>{\centering\arraybackslash}p{1.5cm}|>{\centering\arraybackslash}p{1.5cm}|>{\centering\arraybackslash}p{1.5cm}|>{\centering\arraybackslash}p{1.5cm}|}
\hline
\textbf{$\vec{S_1}$} & \textbf{$\vec{S_2}$} & \textbf{$\vec{F_1}$} & \textbf{$\vec{F_2}$} \\
\hhline{|====|}
$\left(\vec{1} , \vec{2} , -\frac{3}{2}\right)$ & $\left(\vec{1} , \vec{3} , 1\right)$ & $\left(\vec{1} , \vec{3} , 2\right)$ & $\left(\vec{1} , \vec{2} , -\frac{1}{2}\right)$ \\
\hline
$\left(\vec{1} , \vec{2} , -\frac{3}{2}\right)$ & $\left(\vec{1} , \vec{1} , 1\right)$ & $\left(\vec{1} , \vec{1} , 2\right)$ & $\left(\vec{1} , \vec{2} , -\frac{1}{2}\right)$ \\
\hline\end{tabular}
\caption{$1 \rightarrow l$, $2 \rightarrow l$, $3 \rightarrow L$, $4 \rightarrow \overline{l}$, $5 \rightarrow \overline{L}$, $6 \rightarrow \overline{L}$}
\end{table}

\begin{table}[H]
\centering
\renewcommand{\arraystretch}{1.4}
\setlength{\tabcolsep}{0.15cm} % Adjust the column separation to make it narrower
\begin{tabular}{|>{\centering\arraybackslash}p{1.5cm}|>{\centering\arraybackslash}p{1.5cm}|>{\centering\arraybackslash}p{1.5cm}|>{\centering\arraybackslash}p{1.5cm}|}
\hline
\textbf{$\vec{S_1}$} & \textbf{$\vec{S_2}$} & \textbf{$\vec{F_1}$} & \textbf{$\vec{F_2}$} \\
\hhline{|====|}
$\left(\vec{1} , \vec{3} , 1\right)$ & $\left(\vec{1} , \vec{1} , -2\right)$ & $\left(\vec{1} , \vec{2} , -\frac{3}{2}\right)$ & $\left(\vec{1} , \vec{3} , 0\right)$ \\
\hline
$\left(\vec{1} , \vec{1} , 1\right)$ & $\left(\vec{1} , \vec{1} , -2\right)$ & $\left(\vec{1} , \vec{2} , -\frac{3}{2}\right)$ & $\left(\vec{1} , \vec{1} , 0\right)$ \\
\hline\end{tabular}
\caption{$1 \rightarrow L$, $2 \rightarrow \overline{l}$, $3 \rightarrow \overline{L}$, $4 \rightarrow \overline{L}$, $5 \rightarrow l$, $6 \rightarrow l$}
\end{table}

\begin{table}[H]
\centering
\renewcommand{\arraystretch}{1.4}
\setlength{\tabcolsep}{0.15cm} % Adjust the column separation to make it narrower
\begin{tabular}{|>{\centering\arraybackslash}p{1.5cm}|>{\centering\arraybackslash}p{1.5cm}|>{\centering\arraybackslash}p{1.5cm}|>{\centering\arraybackslash}p{1.5cm}|}
\hline
\textbf{$\vec{S_1}$} & \textbf{$\vec{S_2}$} & \textbf{$\vec{F_1}$} & \textbf{$\vec{F_2}$} \\
\hhline{|====|}
$\left(\vec{1} , \vec{3} , 1\right)$ & $\left(\vec{1} , \vec{2} , -\frac{1}{2}\right)$ & $\left(\vec{1} , \vec{2} , -\frac{3}{2}\right)$ & $\left(\vec{1} , \vec{3} , 0\right)$ \\
\hline
$\left(\vec{1} , \vec{1} , 1\right)$ & $\left(\vec{1} , \vec{2} , -\frac{1}{2}\right)$ & $\left(\vec{1} , \vec{2} , -\frac{3}{2}\right)$ & $\left(\vec{1} , \vec{1} , 0\right)$ \\
\hline\end{tabular}
\caption{$1 \rightarrow L$, $2 \rightarrow \overline{l}$, $3 \rightarrow \overline{L}$, $4 \rightarrow l$, $5 \rightarrow l$, $6 \rightarrow \overline{L}$}
\end{table}

\begin{table}[H]
\centering
\renewcommand{\arraystretch}{1.4}
\setlength{\tabcolsep}{0.15cm} % Adjust the column separation to make it narrower
\begin{tabular}{|>{\centering\arraybackslash}p{1.5cm}|>{\centering\arraybackslash}p{1.5cm}|>{\centering\arraybackslash}p{1.5cm}|>{\centering\arraybackslash}p{1.5cm}|}
\hline
\textbf{$\vec{S_1}$} & \textbf{$\vec{S_2}$} & \textbf{$\vec{F_1}$} & \textbf{$\vec{F_2}$} \\
\hhline{|====|}
$\left(\vec{1} , \vec{3} , 1\right)$ & $\left(\vec{1} , \vec{2} , -\frac{1}{2}\right)$ & $\left(\vec{1} , \vec{3} , 0\right)$ & $\left(\vec{1} , \vec{3} , 0\right)$ \\
\hline
$\left(\vec{1} , \vec{1} , 1\right)$ & $\left(\vec{1} , \vec{2} , -\frac{1}{2}\right)$ & $\left(\vec{1} , \vec{1} , 0\right)$ & $\left(\vec{1} , \vec{1} , 0\right)$ \\
\hline\end{tabular}
\caption{$1 \rightarrow L$, $2 \rightarrow \overline{l}$, $3 \rightarrow l$, $4 \rightarrow \overline{L}$, $5 \rightarrow l$, $6 \rightarrow \overline{L}$}
\end{table}

\begin{table}[H]
\centering
\renewcommand{\arraystretch}{1.4}
\setlength{\tabcolsep}{0.15cm} % Adjust the column separation to make it narrower
\begin{tabular}{|>{\centering\arraybackslash}p{1.5cm}|>{\centering\arraybackslash}p{1.5cm}|>{\centering\arraybackslash}p{1.5cm}|>{\centering\arraybackslash}p{1.5cm}|}
\hline
\textbf{$\vec{S_1}$} & \textbf{$\vec{S_2}$} & \textbf{$\vec{F_1}$} & \textbf{$\vec{F_2}$} \\
\hhline{|====|}
$\left(\vec{1} , \vec{3} , 1\right)$ & $\left(\vec{1} , \vec{3} , 1\right)$ & $\left(\vec{1} , \vec{3} , 0\right)$ & $\left(\vec{1} , \vec{3} , 0\right)$ \\
\hline
$\left(\vec{1} , \vec{1} , 1\right)$ & $\left(\vec{1} , \vec{1} , 1\right)$ & $\left(\vec{1} , \vec{1} , 0\right)$ & $\left(\vec{1} , \vec{1} , 0\right)$ \\
\hline\end{tabular}
\caption{$1 \rightarrow L$, $2 \rightarrow \overline{l}$, $3 \rightarrow l$, $4 \rightarrow l$, $5 \rightarrow \overline{L}$, $6 \rightarrow \overline{L}$}
\end{table}

\subsubsection{$\bar{L} \, \bar{l} \, \bar{l} \, l \, l \, l \, H$}

\begin{table}[H]
\centering
\renewcommand{\arraystretch}{1.4}
\setlength{\tabcolsep}{0.15cm} % Adjust the column separation to make it narrower
\begin{tabular}{|>{\centering\arraybackslash}p{1.5cm}|>{\centering\arraybackslash}p{1.5cm}|>{\centering\arraybackslash}p{1.5cm}|>{\centering\arraybackslash}p{1.5cm}|}
\hline
\textbf{$\vec{S_1}$} & \textbf{$\vec{S_2}$} & \textbf{$\vec{F_1}$} & \textbf{$\vec{F_2}$} \\
\hhline{|====|}
$\left(\vec{1} , \vec{1} , 0\right)$ & $\left(\vec{1} , \vec{1} , -2\right)$ & $\left(\vec{1} , \vec{1} , -1\right)$ & $\left(\vec{1} , \vec{1} , 1\right)$ \\
\hline\end{tabular}
\caption{$1 \rightarrow \overline{L}$, $2 \rightarrow l$, $3 \rightarrow \overline{l}$, $4 \rightarrow \overline{l}$, $5 \rightarrow l$, $6 \rightarrow l$}
\end{table}

\begin{table}[H]
\centering
\renewcommand{\arraystretch}{1.4}
\setlength{\tabcolsep}{0.15cm} % Adjust the column separation to make it narrower
\begin{tabular}{|>{\centering\arraybackslash}p{1.5cm}|>{\centering\arraybackslash}p{1.5cm}|>{\centering\arraybackslash}p{1.5cm}|>{\centering\arraybackslash}p{1.5cm}|}
\hline
\textbf{$\vec{S_1}$} & \textbf{$\vec{S_2}$} & \textbf{$\vec{F_1}$} & \textbf{$\vec{F_2}$} \\
\hhline{|====|}
$\left(\vec{1} , \vec{1} , 0\right)$ & $\left(\vec{1} , \vec{1} , 2\right)$ & $\left(\vec{1} , \vec{1} , 1\right)$ & $\left(\vec{1} , \vec{1} , 1\right)$ \\
\hline\end{tabular}
\caption{$1 \rightarrow \overline{L}$, $2 \rightarrow l$, $3 \rightarrow l$, $4 \rightarrow l$, $5 \rightarrow \overline{l}$, $6 \rightarrow \overline{l}$}
\end{table}

\begin{table}[H]
\centering
\renewcommand{\arraystretch}{1.4}
\setlength{\tabcolsep}{0.15cm} % Adjust the column separation to make it narrower
\begin{tabular}{|>{\centering\arraybackslash}p{1.5cm}|>{\centering\arraybackslash}p{1.5cm}|>{\centering\arraybackslash}p{1.5cm}|>{\centering\arraybackslash}p{1.5cm}|}
\hline
\textbf{$\vec{S_1}$} & \textbf{$\vec{S_2}$} & \textbf{$\vec{F_1}$} & \textbf{$\vec{F_2}$} \\
\hhline{|====|}
$\left(\vec{1} , \vec{2} , \frac{5}{2}\right)$ & $\left(\vec{1} , \vec{1} , -2\right)$ & $\left(\vec{1} , \vec{1} , -3\right)$ & $\left(\vec{1} , \vec{2} , \frac{3}{2}\right)$ \\
\hline\end{tabular}
\caption{$1 \rightarrow \overline{l}$, $2 \rightarrow \overline{l}$, $3 \rightarrow \overline{L}$, $4 \rightarrow l$, $5 \rightarrow l$, $6 \rightarrow l$}
\end{table}

\begin{table}[H]
\centering
\renewcommand{\arraystretch}{1.4}
\setlength{\tabcolsep}{0.15cm} % Adjust the column separation to make it narrower
\begin{tabular}{|>{\centering\arraybackslash}p{1.5cm}|>{\centering\arraybackslash}p{1.5cm}|>{\centering\arraybackslash}p{1.5cm}|>{\centering\arraybackslash}p{1.5cm}|}
\hline
\textbf{$\vec{S_1}$} & \textbf{$\vec{S_2}$} & \textbf{$\vec{F_1}$} & \textbf{$\vec{F_2}$} \\
\hhline{|====|}
$\left(\vec{1} , \vec{2} , \frac{5}{2}\right)$ & $\left(\vec{1} , \vec{1} , -2\right)$ & $\left(\vec{1} , \vec{2} , -\frac{3}{2}\right)$ & $\left(\vec{1} , \vec{2} , \frac{3}{2}\right)$ \\
\hline\end{tabular}
\caption{$1 \rightarrow \overline{l}$, $2 \rightarrow \overline{l}$, $3 \rightarrow l$, $4 \rightarrow \overline{L}$, $5 \rightarrow l$, $6 \rightarrow l$}
\end{table}

\begin{table}[H]
\centering
\renewcommand{\arraystretch}{1.4}
\setlength{\tabcolsep}{0.15cm} % Adjust the column separation to make it narrower
\begin{tabular}{|>{\centering\arraybackslash}p{1.5cm}|>{\centering\arraybackslash}p{1.5cm}|>{\centering\arraybackslash}p{1.5cm}|>{\centering\arraybackslash}p{1.5cm}|}
\hline
\textbf{$\vec{S_1}$} & \textbf{$\vec{S_2}$} & \textbf{$\vec{F_1}$} & \textbf{$\vec{F_2}$} \\
\hhline{|====|}
$\left(\vec{1} , \vec{2} , \frac{5}{2}\right)$ & $\left(\vec{1} , \vec{2} , -\frac{1}{2}\right)$ & $\left(\vec{1} , \vec{2} , -\frac{3}{2}\right)$ & $\left(\vec{1} , \vec{2} , \frac{3}{2}\right)$ \\
\hline\end{tabular}
\caption{$1 \rightarrow \overline{l}$, $2 \rightarrow \overline{l}$, $3 \rightarrow l$, $4 \rightarrow l$, $5 \rightarrow l$, $6 \rightarrow \overline{L}$}
\end{table}

\begin{table}[H]
\centering
\renewcommand{\arraystretch}{1.4}
\setlength{\tabcolsep}{0.15cm} % Adjust the column separation to make it narrower
\begin{tabular}{|>{\centering\arraybackslash}p{1.5cm}|>{\centering\arraybackslash}p{1.5cm}|>{\centering\arraybackslash}p{1.5cm}|>{\centering\arraybackslash}p{1.5cm}|}
\hline
\textbf{$\vec{S_1}$} & \textbf{$\vec{S_2}$} & \textbf{$\vec{F_1}$} & \textbf{$\vec{F_2}$} \\
\hhline{|====|}
$\left(\vec{1} , \vec{1} , 0\right)$ & $\left(\vec{1} , \vec{1} , -2\right)$ & $\left(\vec{1} , \vec{1} , -1\right)$ & $\left(\vec{1} , \vec{2} , -\frac{1}{2}\right)$ \\
\hline\end{tabular}
\caption{$1 \rightarrow l$, $2 \rightarrow \overline{L}$, $3 \rightarrow \overline{l}$, $4 \rightarrow \overline{l}$, $5 \rightarrow l$, $6 \rightarrow l$}
\end{table}

\begin{table}[H]
\centering
\renewcommand{\arraystretch}{1.4}
\setlength{\tabcolsep}{0.15cm} % Adjust the column separation to make it narrower
\begin{tabular}{|>{\centering\arraybackslash}p{1.5cm}|>{\centering\arraybackslash}p{1.5cm}|>{\centering\arraybackslash}p{1.5cm}|>{\centering\arraybackslash}p{1.5cm}|}
\hline
\textbf{$\vec{S_1}$} & \textbf{$\vec{S_2}$} & \textbf{$\vec{F_1}$} & \textbf{$\vec{F_2}$} \\
\hhline{|====|}
$\left(\vec{1} , \vec{1} , 0\right)$ & $\left(\vec{1} , \vec{1} , 2\right)$ & $\left(\vec{1} , \vec{1} , 1\right)$ & $\left(\vec{1} , \vec{2} , -\frac{1}{2}\right)$ \\
\hline\end{tabular}
\caption{$1 \rightarrow l$, $2 \rightarrow \overline{L}$, $3 \rightarrow l$, $4 \rightarrow l$, $5 \rightarrow \overline{l}$, $6 \rightarrow \overline{l}$}
\end{table}

\begin{table}[H]
\centering
\renewcommand{\arraystretch}{1.4}
\setlength{\tabcolsep}{0.15cm} % Adjust the column separation to make it narrower
\begin{tabular}{|>{\centering\arraybackslash}p{1.5cm}|>{\centering\arraybackslash}p{1.5cm}|>{\centering\arraybackslash}p{1.5cm}|>{\centering\arraybackslash}p{1.5cm}|}
\hline
\textbf{$\vec{S_1}$} & \textbf{$\vec{S_2}$} & \textbf{$\vec{F_1}$} & \textbf{$\vec{F_2}$} \\
\hhline{|====|}
$\left(\vec{1} , \vec{2} , -\frac{3}{2}\right)$ & $\left(\vec{1} , \vec{1} , 2\right)$ & $\left(\vec{1} , \vec{1} , 1\right)$ & $\left(\vec{1} , \vec{2} , -\frac{1}{2}\right)$ \\
\hline\end{tabular}
\caption{$1 \rightarrow l$, $2 \rightarrow l$, $3 \rightarrow \overline{L}$, $4 \rightarrow l$, $5 \rightarrow \overline{l}$, $6 \rightarrow \overline{l}$}
\end{table}

\begin{table}[H]
\centering
\renewcommand{\arraystretch}{1.4}
\setlength{\tabcolsep}{0.15cm} % Adjust the column separation to make it narrower
\begin{tabular}{|>{\centering\arraybackslash}p{1.5cm}|>{\centering\arraybackslash}p{1.5cm}|>{\centering\arraybackslash}p{1.5cm}|>{\centering\arraybackslash}p{1.5cm}|}
\hline
\textbf{$\vec{S_1}$} & \textbf{$\vec{S_2}$} & \textbf{$\vec{F_1}$} & \textbf{$\vec{F_2}$} \\
\hhline{|====|}
$\left(\vec{1} , \vec{2} , -\frac{3}{2}\right)$ & $\left(\vec{1} , \vec{2} , -\frac{1}{2}\right)$ & $\left(\vec{1} , \vec{2} , \frac{1}{2}\right)$ & $\left(\vec{1} , \vec{2} , -\frac{1}{2}\right)$ \\
\hline\end{tabular}
\caption{$1 \rightarrow l$, $2 \rightarrow l$, $3 \rightarrow \overline{l}$, $4 \rightarrow \overline{l}$, $5 \rightarrow l$, $6 \rightarrow \overline{L}$}
\end{table}

\begin{table}[H]
\centering
\renewcommand{\arraystretch}{1.4}
\setlength{\tabcolsep}{0.15cm} % Adjust the column separation to make it narrower
\begin{tabular}{|>{\centering\arraybackslash}p{1.5cm}|>{\centering\arraybackslash}p{1.5cm}|>{\centering\arraybackslash}p{1.5cm}|>{\centering\arraybackslash}p{1.5cm}|}
\hline
\textbf{$\vec{S_1}$} & \textbf{$\vec{S_2}$} & \textbf{$\vec{F_1}$} & \textbf{$\vec{F_2}$} \\
\hhline{|====|}
$\left(\vec{1} , \vec{2} , -\frac{3}{2}\right)$ & $\left(\vec{1} , \vec{1} , 2\right)$ & $\left(\vec{1} , \vec{2} , \frac{5}{2}\right)$ & $\left(\vec{1} , \vec{2} , -\frac{1}{2}\right)$ \\
\hline\end{tabular}
\caption{$1 \rightarrow l$, $2 \rightarrow l$, $3 \rightarrow l$, $4 \rightarrow \overline{L}$, $5 \rightarrow \overline{l}$, $6 \rightarrow \overline{l}$}
\label{tab:fourth_last}
\end{table}

\subsection{Fifth Topology}
\begin{figure}[H]
    \centering
    \includegraphics[width=0.35\textwidth]{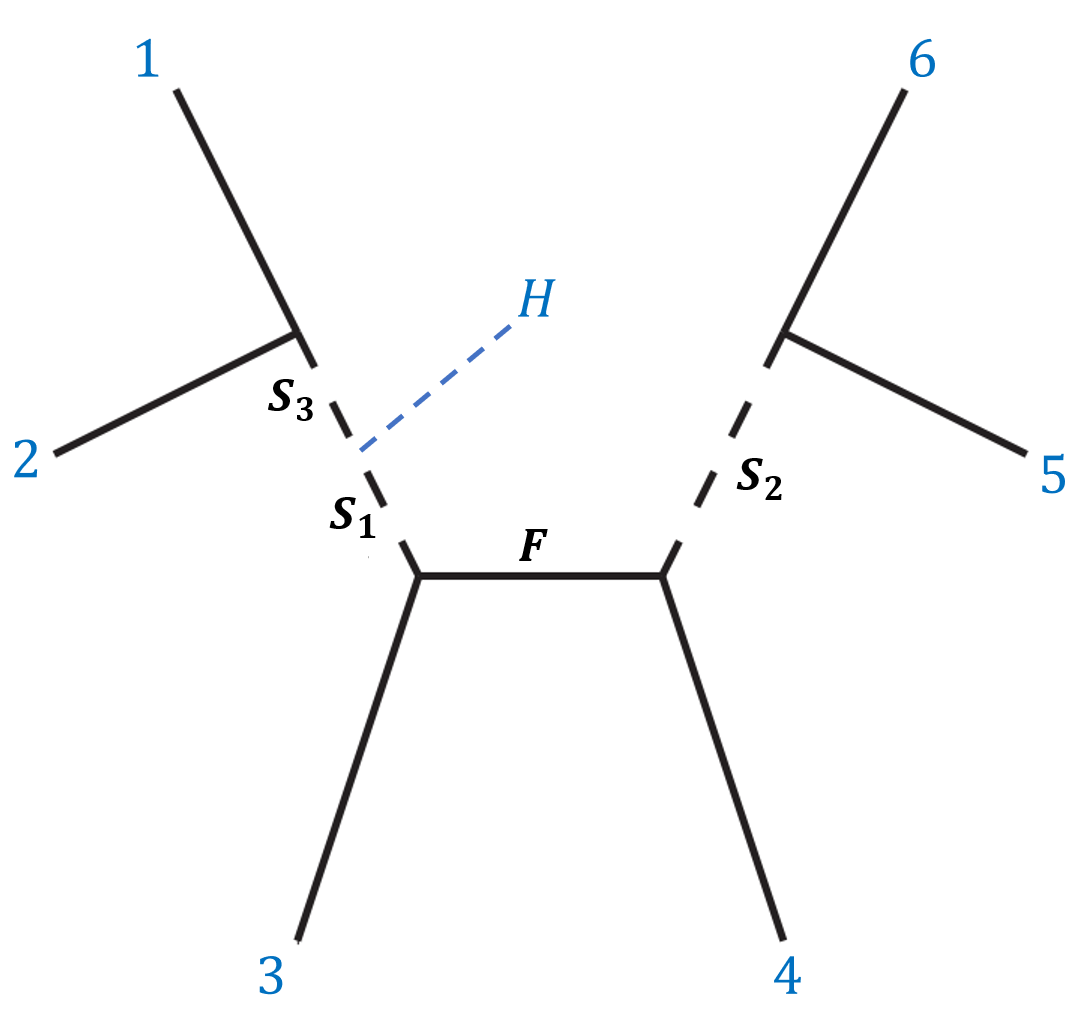}
    \caption{Feynman diagram for the cases where $H$ is attached to an internal scalar. The corresponding tables are Tab.~\ref{tab:fifth_first} - Tab.~\ref{tab:fifth_last}.}
\end{figure}

\subsubsection{$\bar{L} \, \bar{L} \, \bar{L} \, l \, L \, L \, H$}

\begin{table}[H]
\centering
\renewcommand{\arraystretch}{1.4}
\setlength{\tabcolsep}{0.15cm} % Adjust the column separation to make it narrower
\begin{tabular}{|>{\centering\arraybackslash}p{1.5cm}|>{\centering\arraybackslash}p{1.5cm}|>{\centering\arraybackslash}p{1.5cm}|>{\centering\arraybackslash}p{1.5cm}|}
\hline
\textbf{$\vec{S_1}$} & \textbf{$\vec{S_2}$} & \textbf{$\vec{F}$} & \textbf{$\vec{S_3}$} \\
\hhline{|====|}
$\left(\vec{1} , \vec{4} , \frac{3}{2}\right)$ & $\left(\vec{1} , \vec{3} , -1\right)$ & $\left(\vec{1} , \vec{3} , -2\right)$ & $\left(\vec{1} , \vec{3} , 1\right)$ \\
\hline
$\left(\vec{1} , \vec{2} , \frac{3}{2}\right)$ & $\left(\vec{1} , \vec{3} , -1\right)$ & $\left(\vec{1} , \vec{3} , -2\right)$ & $\left(\vec{1} , \vec{3} , 1\right)$ \\
\hline
$\left(\vec{1} , \vec{2} , \frac{3}{2}\right)$ & $\left(\vec{1} , \vec{1} , -1\right)$ & $\left(\vec{1} , \vec{1} , -2\right)$ & $\left(\vec{1} , \vec{3} , 1\right)$ \\
\hline
$\left(\vec{1} , \vec{2} , \frac{3}{2}\right)$ & $\left(\vec{1} , \vec{3} , -1\right)$ & $\left(\vec{1} , \vec{3} , -2\right)$ & $\left(\vec{1} , \vec{1} , 1\right)$ \\
\hline
$\left(\vec{1} , \vec{2} , \frac{3}{2}\right)$ & $\left(\vec{1} , \vec{1} , -1\right)$ & $\left(\vec{1} , \vec{1} , -2\right)$ & $\left(\vec{1} , \vec{1} , 1\right)$ \\
\hline\end{tabular}
\caption{$1 \rightarrow \overline{L}$, $2 \rightarrow \overline{L}$, $3 \rightarrow \overline{L}$, $4 \rightarrow l$, $5 \rightarrow L$, $6 \rightarrow L$}
\label{tab:fifth_first}
\end{table}

\begin{table}[H]
\centering
\renewcommand{\arraystretch}{1.4}
\setlength{\tabcolsep}{0.15cm} % Adjust the column separation to make it narrower
\begin{tabular}{|>{\centering\arraybackslash}p{1.5cm}|>{\centering\arraybackslash}p{1.5cm}|>{\centering\arraybackslash}p{1.5cm}|>{\centering\arraybackslash}p{1.5cm}|}
\hline
\textbf{$\vec{S_1}$} & \textbf{$\vec{S_2}$} & \textbf{$\vec{F}$} & \textbf{$\vec{S_3}$} \\
\hhline{|====|}
$\left(\vec{1} , \vec{4} , \frac{3}{2}\right)$ & $\left(\vec{1} , \vec{2} , -\frac{1}{2}\right)$ & $\left(\vec{1} , \vec{3} , -1\right)$ & $\left(\vec{1} , \vec{3} , 1\right)$ \\
\hline
$\left(\vec{1} , \vec{2} , \frac{3}{2}\right)$ & $\left(\vec{1} , \vec{2} , -\frac{1}{2}\right)$ & $\left(\vec{1} , \vec{3} , -1\right)$ & $\left(\vec{1} , \vec{3} , 1\right)$ \\
\hline
$\left(\vec{1} , \vec{2} , \frac{3}{2}\right)$ & $\left(\vec{1} , \vec{2} , -\frac{1}{2}\right)$ & $\left(\vec{1} , \vec{1} , -1\right)$ & $\left(\vec{1} , \vec{3} , 1\right)$ \\
\hline
$\left(\vec{1} , \vec{2} , \frac{3}{2}\right)$ & $\left(\vec{1} , \vec{2} , -\frac{1}{2}\right)$ & $\left(\vec{1} , \vec{3} , -1\right)$ & $\left(\vec{1} , \vec{1} , 1\right)$ \\
\hline
$\left(\vec{1} , \vec{2} , \frac{3}{2}\right)$ & $\left(\vec{1} , \vec{2} , -\frac{1}{2}\right)$ & $\left(\vec{1} , \vec{1} , -1\right)$ & $\left(\vec{1} , \vec{1} , 1\right)$ \\
\hline\end{tabular}
\caption{$1 \rightarrow \overline{L}$, $2 \rightarrow \overline{L}$, $3 \rightarrow L$, $4 \rightarrow L$, $5 \rightarrow l$, $6 \rightarrow \overline{L}$}
\end{table}

\begin{table}[H]
\centering
\renewcommand{\arraystretch}{1.4}
\setlength{\tabcolsep}{0.15cm} % Adjust the column separation to make it narrower
\begin{tabular}{|>{\centering\arraybackslash}p{1.5cm}|>{\centering\arraybackslash}p{1.5cm}|>{\centering\arraybackslash}p{1.5cm}|>{\centering\arraybackslash}p{1.5cm}|}
\hline
\textbf{$\vec{S_1}$} & \textbf{$\vec{S_2}$} & \textbf{$\vec{F}$} & \textbf{$\vec{S_3}$} \\
\hhline{|====|}
$\left(\vec{1} , \vec{4} , \frac{3}{2}\right)$ & $\left(\vec{1} , \vec{3} , -1\right)$ & $\left(\vec{1} , \vec{4} , -\frac{1}{2}\right)$ & $\left(\vec{1} , \vec{3} , 1\right)$ \\
\hline
$\left(\vec{1} , \vec{2} , \frac{3}{2}\right)$ & $\left(\vec{1} , \vec{3} , -1\right)$ & $\left(\vec{1} , \vec{2} , -\frac{1}{2}\right)$ & $\left(\vec{1} , \vec{3} , 1\right)$ \\
\hline
$\left(\vec{1} , \vec{2} , \frac{3}{2}\right)$ & $\left(\vec{1} , \vec{1} , -1\right)$ & $\left(\vec{1} , \vec{2} , -\frac{1}{2}\right)$ & $\left(\vec{1} , \vec{3} , 1\right)$ \\
\hline
$\left(\vec{1} , \vec{2} , \frac{3}{2}\right)$ & $\left(\vec{1} , \vec{3} , -1\right)$ & $\left(\vec{1} , \vec{2} , -\frac{1}{2}\right)$ & $\left(\vec{1} , \vec{1} , 1\right)$ \\
\hline
$\left(\vec{1} , \vec{2} , \frac{3}{2}\right)$ & $\left(\vec{1} , \vec{1} , -1\right)$ & $\left(\vec{1} , \vec{2} , -\frac{1}{2}\right)$ & $\left(\vec{1} , \vec{1} , 1\right)$ \\
\hline\end{tabular}
\caption{$1 \rightarrow \overline{L}$, $2 \rightarrow \overline{L}$, $3 \rightarrow l$, $4 \rightarrow \overline{L}$, $5 \rightarrow L$, $6 \rightarrow L$}
\end{table}

\begin{table}[H]
\centering
\renewcommand{\arraystretch}{1.4}
\setlength{\tabcolsep}{0.15cm} % Adjust the column separation to make it narrower
\begin{tabular}{|>{\centering\arraybackslash}p{1.5cm}|>{\centering\arraybackslash}p{1.5cm}|>{\centering\arraybackslash}p{1.5cm}|>{\centering\arraybackslash}p{1.5cm}|}
\hline
\textbf{$\vec{S_1}$} & \textbf{$\vec{S_2}$} & \textbf{$\vec{F}$} & \textbf{$\vec{S_3}$} \\
\hhline{|====|}
$\left(\vec{1} , \vec{3} , 0\right)$ & $\left(\vec{1} , \vec{3} , -1\right)$ & $\left(\vec{1} , \vec{4} , -\frac{1}{2}\right)$ & $\left(\vec{1} , \vec{2} , -\frac{1}{2}\right)$ \\
\hline
$\left(\vec{1} , \vec{3} , 0\right)$ & $\left(\vec{1} , \vec{3} , -1\right)$ & $\left(\vec{1} , \vec{2} , -\frac{1}{2}\right)$ & $\left(\vec{1} , \vec{2} , -\frac{1}{2}\right)$ \\
\hline
$\left(\vec{1} , \vec{1} , 0\right)$ & $\left(\vec{1} , \vec{3} , -1\right)$ & $\left(\vec{1} , \vec{2} , -\frac{1}{2}\right)$ & $\left(\vec{1} , \vec{2} , -\frac{1}{2}\right)$ \\
\hline
$\left(\vec{1} , \vec{3} , 0\right)$ & $\left(\vec{1} , \vec{1} , -1\right)$ & $\left(\vec{1} , \vec{2} , -\frac{1}{2}\right)$ & $\left(\vec{1} , \vec{2} , -\frac{1}{2}\right)$ \\
\hline
$\left(\vec{1} , \vec{1} , 0\right)$ & $\left(\vec{1} , \vec{1} , -1\right)$ & $\left(\vec{1} , \vec{2} , -\frac{1}{2}\right)$ & $\left(\vec{1} , \vec{2} , -\frac{1}{2}\right)$ \\
\hline\end{tabular}
\caption{$1 \rightarrow l$, $2 \rightarrow \overline{L}$, $3 \rightarrow \overline{L}$, $4 \rightarrow \overline{L}$, $5 \rightarrow L$, $6 \rightarrow L$}
\end{table}

\begin{table}[H]
\centering
\renewcommand{\arraystretch}{1.4}
\setlength{\tabcolsep}{0.15cm} % Adjust the column separation to make it narrower
\begin{tabular}{|>{\centering\arraybackslash}p{1.5cm}|>{\centering\arraybackslash}p{1.5cm}|>{\centering\arraybackslash}p{1.5cm}|>{\centering\arraybackslash}p{1.5cm}|}
\hline
\textbf{$\vec{S_1}$} & \textbf{$\vec{S_2}$} & \textbf{$\vec{F}$} & \textbf{$\vec{S_3}$} \\
\hhline{|====|}
$\left(\vec{1} , \vec{3} , 0\right)$ & $\left(\vec{1} , \vec{3} , 1\right)$ & $\left(\vec{1} , \vec{4} , \frac{1}{2}\right)$ & $\left(\vec{1} , \vec{2} , -\frac{1}{2}\right)$ \\
\hline
$\left(\vec{1} , \vec{3} , 0\right)$ & $\left(\vec{1} , \vec{3} , 1\right)$ & $\left(\vec{1} , \vec{2} , \frac{1}{2}\right)$ & $\left(\vec{1} , \vec{2} , -\frac{1}{2}\right)$ \\
\hline
$\left(\vec{1} , \vec{1} , 0\right)$ & $\left(\vec{1} , \vec{3} , 1\right)$ & $\left(\vec{1} , \vec{2} , \frac{1}{2}\right)$ & $\left(\vec{1} , \vec{2} , -\frac{1}{2}\right)$ \\
\hline
$\left(\vec{1} , \vec{3} , 0\right)$ & $\left(\vec{1} , \vec{1} , 1\right)$ & $\left(\vec{1} , \vec{2} , \frac{1}{2}\right)$ & $\left(\vec{1} , \vec{2} , -\frac{1}{2}\right)$ \\
\hline
$\left(\vec{1} , \vec{1} , 0\right)$ & $\left(\vec{1} , \vec{1} , 1\right)$ & $\left(\vec{1} , \vec{2} , \frac{1}{2}\right)$ & $\left(\vec{1} , \vec{2} , -\frac{1}{2}\right)$ \\
\hline\end{tabular}
\caption{$1 \rightarrow l$, $2 \rightarrow \overline{L}$, $3 \rightarrow L$, $4 \rightarrow L$, $5 \rightarrow \overline{L}$, $6 \rightarrow \overline{L}$}
\end{table}

\begin{table}[H]
\centering
\renewcommand{\arraystretch}{1.4}
\setlength{\tabcolsep}{0.15cm} % Adjust the column separation to make it narrower
\begin{tabular}{|>{\centering\arraybackslash}p{1.5cm}|>{\centering\arraybackslash}p{1.5cm}|>{\centering\arraybackslash}p{1.5cm}|>{\centering\arraybackslash}p{1.5cm}|}
\hline
\textbf{$\vec{S_1}$} & \textbf{$\vec{S_2}$} & \textbf{$\vec{F}$} & \textbf{$\vec{S_3}$} \\
\hhline{|====|}
$\left(\vec{1} , \vec{4} , -\frac{1}{2}\right)$ & $\left(\vec{1} , \vec{2} , -\frac{1}{2}\right)$ & $\left(\vec{1} , \vec{3} , 0\right)$ & $\left(\vec{1} , \vec{3} , -1\right)$ \\
\hline
$\left(\vec{1} , \vec{2} , -\frac{1}{2}\right)$ & $\left(\vec{1} , \vec{2} , -\frac{1}{2}\right)$ & $\left(\vec{1} , \vec{3} , 0\right)$ & $\left(\vec{1} , \vec{3} , -1\right)$ \\
\hline
$\left(\vec{1} , \vec{2} , -\frac{1}{2}\right)$ & $\left(\vec{1} , \vec{2} , -\frac{1}{2}\right)$ & $\left(\vec{1} , \vec{1} , 0\right)$ & $\left(\vec{1} , \vec{3} , -1\right)$ \\
\hline
$\left(\vec{1} , \vec{2} , -\frac{1}{2}\right)$ & $\left(\vec{1} , \vec{2} , -\frac{1}{2}\right)$ & $\left(\vec{1} , \vec{3} , 0\right)$ & $\left(\vec{1} , \vec{1} , -1\right)$ \\
\hline
$\left(\vec{1} , \vec{2} , -\frac{1}{2}\right)$ & $\left(\vec{1} , \vec{2} , -\frac{1}{2}\right)$ & $\left(\vec{1} , \vec{1} , 0\right)$ & $\left(\vec{1} , \vec{1} , -1\right)$ \\
\hline\end{tabular}
\caption{$1 \rightarrow L$, $2 \rightarrow L$, $3 \rightarrow \overline{L}$, $4 \rightarrow \overline{L}$, $5 \rightarrow l$, $6 \rightarrow \overline{L}$}
\end{table}

\begin{table}[H]
\centering
\renewcommand{\arraystretch}{1.4}
\setlength{\tabcolsep}{0.15cm} % Adjust the column separation to make it narrower
\begin{tabular}{|>{\centering\arraybackslash}p{1.5cm}|>{\centering\arraybackslash}p{1.5cm}|>{\centering\arraybackslash}p{1.5cm}|>{\centering\arraybackslash}p{1.5cm}|}
\hline
\textbf{$\vec{S_1}$} & \textbf{$\vec{S_2}$} & \textbf{$\vec{F}$} & \textbf{$\vec{S_3}$} \\
\hhline{|====|}
$\left(\vec{1} , \vec{4} , -\frac{1}{2}\right)$ & $\left(\vec{1} , \vec{3} , 1\right)$ & $\left(\vec{1} , \vec{3} , 0\right)$ & $\left(\vec{1} , \vec{3} , -1\right)$ \\
\hline
$\left(\vec{1} , \vec{2} , -\frac{1}{2}\right)$ & $\left(\vec{1} , \vec{3} , 1\right)$ & $\left(\vec{1} , \vec{3} , 0\right)$ & $\left(\vec{1} , \vec{3} , -1\right)$ \\
\hline
$\left(\vec{1} , \vec{2} , -\frac{1}{2}\right)$ & $\left(\vec{1} , \vec{1} , 1\right)$ & $\left(\vec{1} , \vec{1} , 0\right)$ & $\left(\vec{1} , \vec{3} , -1\right)$ \\
\hline
$\left(\vec{1} , \vec{2} , -\frac{1}{2}\right)$ & $\left(\vec{1} , \vec{3} , 1\right)$ & $\left(\vec{1} , \vec{3} , 0\right)$ & $\left(\vec{1} , \vec{1} , -1\right)$ \\
\hline
$\left(\vec{1} , \vec{2} , -\frac{1}{2}\right)$ & $\left(\vec{1} , \vec{1} , 1\right)$ & $\left(\vec{1} , \vec{1} , 0\right)$ & $\left(\vec{1} , \vec{1} , -1\right)$ \\
\hline\end{tabular}
\caption{$1 \rightarrow L$, $2 \rightarrow L$, $3 \rightarrow \overline{L}$, $4 \rightarrow l$, $5 \rightarrow \overline{L}$, $6 \rightarrow \overline{L}$}
\end{table}

\begin{table}[H]
\centering
\renewcommand{\arraystretch}{1.4}
\setlength{\tabcolsep}{0.15cm} % Adjust the column separation to make it narrower
\begin{tabular}{|>{\centering\arraybackslash}p{1.5cm}|>{\centering\arraybackslash}p{1.5cm}|>{\centering\arraybackslash}p{1.5cm}|>{\centering\arraybackslash}p{1.5cm}|}
\hline
\textbf{$\vec{S_1}$} & \textbf{$\vec{S_2}$} & \textbf{$\vec{F}$} & \textbf{$\vec{S_3}$} \\
\hhline{|====|}
$\left(\vec{1} , \vec{4} , -\frac{1}{2}\right)$ & $\left(\vec{1} , \vec{3} , 1\right)$ & $\left(\vec{1} , \vec{4} , \frac{3}{2}\right)$ & $\left(\vec{1} , \vec{3} , -1\right)$ \\
\hline
$\left(\vec{1} , \vec{2} , -\frac{1}{2}\right)$ & $\left(\vec{1} , \vec{3} , 1\right)$ & $\left(\vec{1} , \vec{2} , \frac{3}{2}\right)$ & $\left(\vec{1} , \vec{3} , -1\right)$ \\
\hline
$\left(\vec{1} , \vec{2} , -\frac{1}{2}\right)$ & $\left(\vec{1} , \vec{1} , 1\right)$ & $\left(\vec{1} , \vec{2} , \frac{3}{2}\right)$ & $\left(\vec{1} , \vec{3} , -1\right)$ \\
\hline
$\left(\vec{1} , \vec{2} , -\frac{1}{2}\right)$ & $\left(\vec{1} , \vec{3} , 1\right)$ & $\left(\vec{1} , \vec{2} , \frac{3}{2}\right)$ & $\left(\vec{1} , \vec{1} , -1\right)$ \\
\hline
$\left(\vec{1} , \vec{2} , -\frac{1}{2}\right)$ & $\left(\vec{1} , \vec{1} , 1\right)$ & $\left(\vec{1} , \vec{2} , \frac{3}{2}\right)$ & $\left(\vec{1} , \vec{1} , -1\right)$ \\
\hline\end{tabular}
\caption{$1 \rightarrow L$, $2 \rightarrow L$, $3 \rightarrow l$, $4 \rightarrow \overline{L}$, $5 \rightarrow \overline{L}$, $6 \rightarrow \overline{L}$}
\end{table}

\subsubsection{$\bar{L} \, \bar{L} \, \bar{l} \, l \, l \, L \, H$}

\begin{table}[H]
\centering
\renewcommand{\arraystretch}{1.4}
\setlength{\tabcolsep}{0.15cm} % Adjust the column separation to make it narrower
\begin{tabular}{|>{\centering\arraybackslash}p{1.5cm}|>{\centering\arraybackslash}p{1.5cm}|>{\centering\arraybackslash}p{1.5cm}|>{\centering\arraybackslash}p{1.5cm}|}
\hline
\textbf{$\vec{S_1}$} & \textbf{$\vec{S_2}$} & \textbf{$\vec{F}$} & \textbf{$\vec{S_3}$} \\
\hhline{|====|}
$\left(\vec{1} , \vec{2} , \frac{3}{2}\right)$ & $\left(\vec{1} , \vec{1} , -2\right)$ & $\left(\vec{1} , \vec{2} , -\frac{5}{2}\right)$ & $\left(\vec{1} , \vec{3} , 1\right)$ \\
\hline
$\left(\vec{1} , \vec{2} , \frac{3}{2}\right)$ & $\left(\vec{1} , \vec{1} , -2\right)$ & $\left(\vec{1} , \vec{2} , -\frac{5}{2}\right)$ & $\left(\vec{1} , \vec{1} , 1\right)$ \\
\hline\end{tabular}
\caption{$1 \rightarrow \overline{L}$, $2 \rightarrow \overline{L}$, $3 \rightarrow \overline{l}$, $4 \rightarrow L$, $5 \rightarrow l$, $6 \rightarrow l$}
\end{table}

\begin{table}[H]
\centering
\renewcommand{\arraystretch}{1.4}
\setlength{\tabcolsep}{0.15cm} % Adjust the column separation to make it narrower
\begin{tabular}{|>{\centering\arraybackslash}p{1.5cm}|>{\centering\arraybackslash}p{1.5cm}|>{\centering\arraybackslash}p{1.5cm}|>{\centering\arraybackslash}p{1.5cm}|}
\hline
\textbf{$\vec{S_1}$} & \textbf{$\vec{S_2}$} & \textbf{$\vec{F}$} & \textbf{$\vec{S_3}$} \\
\hhline{|====|}
$\left(\vec{1} , \vec{2} , \frac{3}{2}\right)$ & $\left(\vec{1} , \vec{2} , \frac{1}{2}\right)$ & $\left(\vec{1} , \vec{2} , -\frac{1}{2}\right)$ & $\left(\vec{1} , \vec{3} , 1\right)$ \\
\hline
$\left(\vec{1} , \vec{2} , \frac{3}{2}\right)$ & $\left(\vec{1} , \vec{2} , \frac{1}{2}\right)$ & $\left(\vec{1} , \vec{2} , -\frac{1}{2}\right)$ & $\left(\vec{1} , \vec{1} , 1\right)$ \\
\hline\end{tabular}
\caption{$1 \rightarrow \overline{L}$, $2 \rightarrow \overline{L}$, $3 \rightarrow l$, $4 \rightarrow l$, $5 \rightarrow \overline{l}$, $6 \rightarrow L$}
\end{table}

\begin{table}[H]
\centering
\renewcommand{\arraystretch}{1.4}
\setlength{\tabcolsep}{0.15cm} % Adjust the column separation to make it narrower
\begin{tabular}{|>{\centering\arraybackslash}p{1.5cm}|>{\centering\arraybackslash}p{1.5cm}|>{\centering\arraybackslash}p{1.5cm}|>{\centering\arraybackslash}p{1.5cm}|}
\hline
\textbf{$\vec{S_1}$} & \textbf{$\vec{S_2}$} & \textbf{$\vec{F}$} & \textbf{$\vec{S_3}$} \\
\hhline{|====|}
$\left(\vec{1} , \vec{2} , \frac{3}{2}\right)$ & $\left(\vec{1} , \vec{1} , -2\right)$ & $\left(\vec{1} , \vec{1} , -1\right)$ & $\left(\vec{1} , \vec{3} , 1\right)$ \\
\hline
$\left(\vec{1} , \vec{2} , \frac{3}{2}\right)$ & $\left(\vec{1} , \vec{1} , -2\right)$ & $\left(\vec{1} , \vec{1} , -1\right)$ & $\left(\vec{1} , \vec{1} , 1\right)$ \\
\hline\end{tabular}
\caption{$1 \rightarrow \overline{L}$, $2 \rightarrow \overline{L}$, $3 \rightarrow L$, $4 \rightarrow \overline{l}$, $5 \rightarrow l$, $6 \rightarrow l$}
\end{table}

\begin{table}[H]
\centering
\renewcommand{\arraystretch}{1.4}
\setlength{\tabcolsep}{0.15cm} % Adjust the column separation to make it narrower
\begin{tabular}{|>{\centering\arraybackslash}p{1.5cm}|>{\centering\arraybackslash}p{1.5cm}|>{\centering\arraybackslash}p{1.5cm}|>{\centering\arraybackslash}p{1.5cm}|}
\hline
\textbf{$\vec{S_1}$} & \textbf{$\vec{S_2}$} & \textbf{$\vec{F}$} & \textbf{$\vec{S_3}$} \\
\hhline{|====|}
$\left(\vec{1} , \vec{3} , 0\right)$ & $\left(\vec{1} , \vec{2} , \frac{1}{2}\right)$ & $\left(\vec{1} , \vec{2} , -\frac{1}{2}\right)$ & $\left(\vec{1} , \vec{2} , -\frac{1}{2}\right)$ \\
\hline
$\left(\vec{1} , \vec{1} , 0\right)$ & $\left(\vec{1} , \vec{2} , \frac{1}{2}\right)$ & $\left(\vec{1} , \vec{2} , -\frac{1}{2}\right)$ & $\left(\vec{1} , \vec{2} , -\frac{1}{2}\right)$ \\
\hline\end{tabular}
\caption{$1 \rightarrow l$, $2 \rightarrow \overline{L}$, $3 \rightarrow \overline{L}$, $4 \rightarrow l$, $5 \rightarrow \overline{l}$, $6 \rightarrow L$}
\end{table}

\begin{table}[H]
\centering
\renewcommand{\arraystretch}{1.4}
\setlength{\tabcolsep}{0.15cm} % Adjust the column separation to make it narrower
\begin{tabular}{|>{\centering\arraybackslash}p{1.5cm}|>{\centering\arraybackslash}p{1.5cm}|>{\centering\arraybackslash}p{1.5cm}|>{\centering\arraybackslash}p{1.5cm}|}
\hline
\textbf{$\vec{S_1}$} & \textbf{$\vec{S_2}$} & \textbf{$\vec{F}$} & \textbf{$\vec{S_3}$} \\
\hhline{|====|}
$\left(\vec{1} , \vec{3} , 0\right)$ & $\left(\vec{1} , \vec{2} , -\frac{1}{2}\right)$ & $\left(\vec{1} , \vec{3} , -1\right)$ & $\left(\vec{1} , \vec{2} , -\frac{1}{2}\right)$ \\
\hline
$\left(\vec{1} , \vec{1} , 0\right)$ & $\left(\vec{1} , \vec{2} , -\frac{1}{2}\right)$ & $\left(\vec{1} , \vec{1} , -1\right)$ & $\left(\vec{1} , \vec{2} , -\frac{1}{2}\right)$ \\
\hline\end{tabular}
\caption{$1 \rightarrow l$, $2 \rightarrow \overline{L}$, $3 \rightarrow \overline{l}$, $4 \rightarrow L$, $5 \rightarrow l$, $6 \rightarrow \overline{L}$}
\label{tab:ourCase}
\end{table}

\begin{table}[H]
\centering
\renewcommand{\arraystretch}{1.4}
\setlength{\tabcolsep}{0.15cm} % Adjust the column separation to make it narrower
\begin{tabular}{|>{\centering\arraybackslash}p{1.5cm}|>{\centering\arraybackslash}p{1.5cm}|>{\centering\arraybackslash}p{1.5cm}|>{\centering\arraybackslash}p{1.5cm}|}
\hline
\textbf{$\vec{S_1}$} & \textbf{$\vec{S_2}$} & \textbf{$\vec{F}$} & \textbf{$\vec{S_3}$} \\
\hhline{|====|}
$\left(\vec{1} , \vec{3} , 0\right)$ & $\left(\vec{1} , \vec{2} , \frac{1}{2}\right)$ & $\left(\vec{1} , \vec{3} , 1\right)$ & $\left(\vec{1} , \vec{2} , -\frac{1}{2}\right)$ \\
\hline
$\left(\vec{1} , \vec{1} , 0\right)$ & $\left(\vec{1} , \vec{2} , \frac{1}{2}\right)$ & $\left(\vec{1} , \vec{1} , 1\right)$ & $\left(\vec{1} , \vec{2} , -\frac{1}{2}\right)$ \\
\hline\end{tabular}
\caption{$1 \rightarrow l$, $2 \rightarrow \overline{L}$, $3 \rightarrow l$, $4 \rightarrow \overline{L}$, $5 \rightarrow \overline{l}$, $6 \rightarrow L$}
\end{table}

\begin{table}[H]
\centering
\renewcommand{\arraystretch}{1.4}
\setlength{\tabcolsep}{0.15cm} % Adjust the column separation to make it narrower
\begin{tabular}{|>{\centering\arraybackslash}p{1.5cm}|>{\centering\arraybackslash}p{1.5cm}|>{\centering\arraybackslash}p{1.5cm}|>{\centering\arraybackslash}p{1.5cm}|}
\hline
\textbf{$\vec{S_1}$} & \textbf{$\vec{S_2}$} & \textbf{$\vec{F}$} & \textbf{$\vec{S_3}$} \\
\hhline{|====|}
$\left(\vec{1} , \vec{3} , 0\right)$ & $\left(\vec{1} , \vec{2} , -\frac{1}{2}\right)$ & $\left(\vec{1} , \vec{2} , \frac{1}{2}\right)$ & $\left(\vec{1} , \vec{2} , -\frac{1}{2}\right)$ \\
\hline
$\left(\vec{1} , \vec{1} , 0\right)$ & $\left(\vec{1} , \vec{2} , -\frac{1}{2}\right)$ & $\left(\vec{1} , \vec{2} , \frac{1}{2}\right)$ & $\left(\vec{1} , \vec{2} , -\frac{1}{2}\right)$ \\
\hline\end{tabular}
\caption{$1 \rightarrow l$, $2 \rightarrow \overline{L}$, $3 \rightarrow L$, $4 \rightarrow \overline{l}$, $5 \rightarrow l$, $6 \rightarrow \overline{L}$}
\end{table}

\begin{table}[H]
\centering
\renewcommand{\arraystretch}{1.4}
\setlength{\tabcolsep}{0.15cm} % Adjust the column separation to make it narrower
\begin{tabular}{|>{\centering\arraybackslash}p{1.5cm}|>{\centering\arraybackslash}p{1.5cm}|>{\centering\arraybackslash}p{1.5cm}|>{\centering\arraybackslash}p{1.5cm}|}
\hline
\textbf{$\vec{S_1}$} & \textbf{$\vec{S_2}$} & \textbf{$\vec{F}$} & \textbf{$\vec{S_3}$} \\
\hhline{|====|}
$\left(\vec{1} , \vec{3} , 1\right)$ & $\left(\vec{1} , \vec{1} , -2\right)$ & $\left(\vec{1} , \vec{2} , -\frac{3}{2}\right)$ & $\left(\vec{1} , \vec{2} , \frac{1}{2}\right)$ \\
\hline
$\left(\vec{1} , \vec{1} , 1\right)$ & $\left(\vec{1} , \vec{1} , -2\right)$ & $\left(\vec{1} , \vec{2} , -\frac{3}{2}\right)$ & $\left(\vec{1} , \vec{2} , \frac{1}{2}\right)$ \\
\hline\end{tabular}
\caption{$1 \rightarrow \overline{l}$, $2 \rightarrow L$, $3 \rightarrow \overline{L}$, $4 \rightarrow \overline{L}$, $5 \rightarrow l$, $6 \rightarrow l$}
\end{table}

\begin{table}[H]
\centering
\renewcommand{\arraystretch}{1.4}
\setlength{\tabcolsep}{0.15cm} % Adjust the column separation to make it narrower
\begin{tabular}{|>{\centering\arraybackslash}p{1.5cm}|>{\centering\arraybackslash}p{1.5cm}|>{\centering\arraybackslash}p{1.5cm}|>{\centering\arraybackslash}p{1.5cm}|}
\hline
\textbf{$\vec{S_1}$} & \textbf{$\vec{S_2}$} & \textbf{$\vec{F}$} & \textbf{$\vec{S_3}$} \\
\hhline{|====|}
$\left(\vec{1} , \vec{3} , 1\right)$ & $\left(\vec{1} , \vec{2} , -\frac{1}{2}\right)$ & $\left(\vec{1} , \vec{2} , -\frac{3}{2}\right)$ & $\left(\vec{1} , \vec{2} , \frac{1}{2}\right)$ \\
\hline
$\left(\vec{1} , \vec{1} , 1\right)$ & $\left(\vec{1} , \vec{2} , -\frac{1}{2}\right)$ & $\left(\vec{1} , \vec{2} , -\frac{3}{2}\right)$ & $\left(\vec{1} , \vec{2} , \frac{1}{2}\right)$ \\
\hline\end{tabular}
\caption{$1 \rightarrow \overline{l}$, $2 \rightarrow L$, $3 \rightarrow \overline{L}$, $4 \rightarrow l$, $5 \rightarrow l$, $6 \rightarrow \overline{L}$}
\end{table}

\begin{table}[H]
\centering
\renewcommand{\arraystretch}{1.4}
\setlength{\tabcolsep}{0.15cm} % Adjust the column separation to make it narrower
\begin{tabular}{|>{\centering\arraybackslash}p{1.5cm}|>{\centering\arraybackslash}p{1.5cm}|>{\centering\arraybackslash}p{1.5cm}|>{\centering\arraybackslash}p{1.5cm}|}
\hline
\textbf{$\vec{S_1}$} & \textbf{$\vec{S_2}$} & \textbf{$\vec{F}$} & \textbf{$\vec{S_3}$} \\
\hhline{|====|}
$\left(\vec{1} , \vec{3} , 1\right)$ & $\left(\vec{1} , \vec{2} , -\frac{1}{2}\right)$ & $\left(\vec{1} , \vec{3} , 0\right)$ & $\left(\vec{1} , \vec{2} , \frac{1}{2}\right)$ \\
\hline
$\left(\vec{1} , \vec{1} , 1\right)$ & $\left(\vec{1} , \vec{2} , -\frac{1}{2}\right)$ & $\left(\vec{1} , \vec{1} , 0\right)$ & $\left(\vec{1} , \vec{2} , \frac{1}{2}\right)$ \\
\hline\end{tabular}
\caption{$1 \rightarrow \overline{l}$, $2 \rightarrow L$, $3 \rightarrow l$, $4 \rightarrow \overline{L}$, $5 \rightarrow l$, $6 \rightarrow \overline{L}$}
\end{table}

\begin{table}[H]
\centering
\renewcommand{\arraystretch}{1.4}
\setlength{\tabcolsep}{0.15cm} % Adjust the column separation to make it narrower
\begin{tabular}{|>{\centering\arraybackslash}p{1.5cm}|>{\centering\arraybackslash}p{1.5cm}|>{\centering\arraybackslash}p{1.5cm}|>{\centering\arraybackslash}p{1.5cm}|}
\hline
\textbf{$\vec{S_1}$} & \textbf{$\vec{S_2}$} & \textbf{$\vec{F}$} & \textbf{$\vec{S_3}$} \\
\hhline{|====|}
$\left(\vec{1} , \vec{3} , 1\right)$ & $\left(\vec{1} , \vec{3} , 1\right)$ & $\left(\vec{1} , \vec{3} , 0\right)$ & $\left(\vec{1} , \vec{2} , \frac{1}{2}\right)$ \\
\hline
$\left(\vec{1} , \vec{1} , 1\right)$ & $\left(\vec{1} , \vec{1} , 1\right)$ & $\left(\vec{1} , \vec{1} , 0\right)$ & $\left(\vec{1} , \vec{2} , \frac{1}{2}\right)$ \\
\hline\end{tabular}
\caption{$1 \rightarrow \overline{l}$, $2 \rightarrow L$, $3 \rightarrow l$, $4 \rightarrow l$, $5 \rightarrow \overline{L}$, $6 \rightarrow \overline{L}$}
\end{table}

\begin{table}[H]
\centering
\renewcommand{\arraystretch}{1.4}
\setlength{\tabcolsep}{0.15cm} % Adjust the column separation to make it narrower
\begin{tabular}{|>{\centering\arraybackslash}p{1.5cm}|>{\centering\arraybackslash}p{1.5cm}|>{\centering\arraybackslash}p{1.5cm}|>{\centering\arraybackslash}p{1.5cm}|}
\hline
\textbf{$\vec{S_1}$} & \textbf{$\vec{S_2}$} & \textbf{$\vec{F}$} & \textbf{$\vec{S_3}$} \\
\hhline{|====|}
$\left(\vec{1} , \vec{2} , -\frac{3}{2}\right)$ & $\left(\vec{1} , \vec{2} , \frac{1}{2}\right)$ & $\left(\vec{1} , \vec{3} , 1\right)$ & $\left(\vec{1} , \vec{1} , -2\right)$ \\
\hline
$\left(\vec{1} , \vec{2} , -\frac{3}{2}\right)$ & $\left(\vec{1} , \vec{2} , \frac{1}{2}\right)$ & $\left(\vec{1} , \vec{1} , 1\right)$ & $\left(\vec{1} , \vec{1} , -2\right)$ \\
\hline\end{tabular}
\caption{$1 \rightarrow l$, $2 \rightarrow l$, $3 \rightarrow \overline{L}$, $4 \rightarrow \overline{L}$, $5 \rightarrow \overline{l}$, $6 \rightarrow L$}
\end{table}

\begin{table}[H]
\centering
\renewcommand{\arraystretch}{1.4}
\setlength{\tabcolsep}{0.15cm} % Adjust the column separation to make it narrower
\begin{tabular}{|>{\centering\arraybackslash}p{1.5cm}|>{\centering\arraybackslash}p{1.5cm}|>{\centering\arraybackslash}p{1.5cm}|>{\centering\arraybackslash}p{1.5cm}|}
\hline
\textbf{$\vec{S_1}$} & \textbf{$\vec{S_2}$} & \textbf{$\vec{F}$} & \textbf{$\vec{S_3}$} \\
\hhline{|====|}
$\left(\vec{1} , \vec{2} , -\frac{3}{2}\right)$ & $\left(\vec{1} , \vec{3} , 1\right)$ & $\left(\vec{1} , \vec{2} , \frac{1}{2}\right)$ & $\left(\vec{1} , \vec{1} , -2\right)$ \\
\hline
$\left(\vec{1} , \vec{2} , -\frac{3}{2}\right)$ & $\left(\vec{1} , \vec{1} , 1\right)$ & $\left(\vec{1} , \vec{2} , \frac{1}{2}\right)$ & $\left(\vec{1} , \vec{1} , -2\right)$ \\
\hline\end{tabular}
\caption{$1 \rightarrow l$, $2 \rightarrow l$, $3 \rightarrow \overline{l}$, $4 \rightarrow L$, $5 \rightarrow \overline{L}$, $6 \rightarrow \overline{L}$}
\end{table}

\begin{table}[H]
\centering
\renewcommand{\arraystretch}{1.4}
\setlength{\tabcolsep}{0.15cm} % Adjust the column separation to make it narrower
%
\caption{$1 \rightarrow l$, $2 \rightarrow l$, $3 \rightarrow L$, $4 \rightarrow \overline{l}$, $5 \rightarrow \overline{L}$, $6 \rightarrow \overline{L}$}
\end{table}

\subsubsection{$\bar{L} \, \bar{l} \, \bar{l} \, l \, l \, l \, H$}

\begin{table}[H]
\centering
\renewcommand{\arraystretch}{1.4}
\setlength{\tabcolsep}{0.15cm} % Adjust the column separation to make it narrower
%
\caption{$1 \rightarrow l$, $2 \rightarrow \overline{L}$, $3 \rightarrow \overline{l}$, $4 \rightarrow \overline{l}$, $5 \rightarrow l$, $6 \rightarrow l$}
\end{table}

\begin{table}[H]
\centering
\renewcommand{\arraystretch}{1.4}
\setlength{\tabcolsep}{0.15cm} % Adjust the column separation to make it narrower
%
\caption{$1 \rightarrow l$, $2 \rightarrow \overline{L}$, $3 \rightarrow l$, $4 \rightarrow l$, $5 \rightarrow \overline{l}$, $6 \rightarrow \overline{l}$}
\end{table}

\begin{table}[H]
\centering
\renewcommand{\arraystretch}{1.4}
\setlength{\tabcolsep}{0.15cm} % Adjust the column separation to make it narrower
%
\caption{$1 \rightarrow \overline{l}$, $2 \rightarrow \overline{l}$, $3 \rightarrow \overline{L}$, $4 \rightarrow l$, $5 \rightarrow l$, $6 \rightarrow l$}
\end{table}

\begin{table}[H]
\centering
\renewcommand{\arraystretch}{1.4}
\setlength{\tabcolsep}{0.15cm} % Adjust the column separation to make it narrower
%
\caption{$1 \rightarrow \overline{l}$, $2 \rightarrow \overline{l}$, $3 \rightarrow l$, $4 \rightarrow \overline{L}$, $5 \rightarrow l$, $6 \rightarrow l$}
\end{table}

\begin{table}[H]
\centering
\renewcommand{\arraystretch}{1.4}
\setlength{\tabcolsep}{0.15cm} % Adjust the column separation to make it narrower
%
\caption{$1 \rightarrow \overline{l}$, $2 \rightarrow \overline{l}$, $3 \rightarrow l$, $4 \rightarrow l$, $5 \rightarrow l$, $6 \rightarrow \overline{L}$}
\end{table}

\begin{table}[H]
\centering
\renewcommand{\arraystretch}{1.4}
\setlength{\tabcolsep}{0.15cm} % Adjust the column separation to make it narrower
%
\caption{$1 \rightarrow l$, $2 \rightarrow l$, $3 \rightarrow \overline{L}$, $4 \rightarrow l$, $5 \rightarrow \overline{l}$, $6 \rightarrow \overline{l}$}
\end{table}

\begin{table}[H]
\centering
\renewcommand{\arraystretch}{1.4}
\setlength{\tabcolsep}{0.15cm} % Adjust the column separation to make it narrower
%
\caption{$1 \rightarrow l$, $2 \rightarrow l$, $3 \rightarrow \overline{l}$, $4 \rightarrow \overline{l}$, $5 \rightarrow l$, $6 \rightarrow \overline{L}$}
\end{table}

\begin{table}[H]
\centering
\renewcommand{\arraystretch}{1.4}
\setlength{\tabcolsep}{0.15cm} % Adjust the column separation to make it narrower
%
\caption{$1 \rightarrow l$, $2 \rightarrow l$, $3 \rightarrow l$, $4 \rightarrow \overline{L}$, $5 \rightarrow \overline{l}$, $6 \rightarrow \overline{l}$}
\label{tab:fifth_last}
\end{table}

\subsection{Sixth Topology}
\begin{figure}[H]
    \centering
    \includegraphics[width=0.35\textwidth]{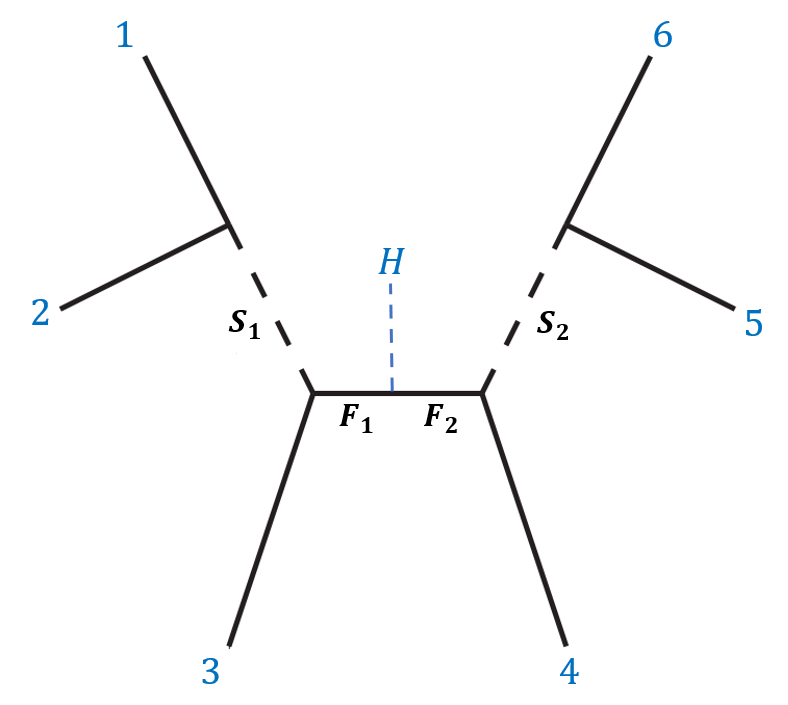}
    \caption{Feynman diagram for the cases where $H$ is attached to a fermion mediator. The corresponding tables are Tab.~\ref{tab:sixth_first} - Tab.~\ref{tab:sixth_last}.}
\end{figure}

\subsubsection{$\bar{L} \, \bar{L} \, \bar{L} \, l \, L \, L \, H$}

\begin{table}[H]
\centering
\renewcommand{\arraystretch}{1.4}
\setlength{\tabcolsep}{0.15cm} % Adjust the column separation to make it narrower

\caption{$1 \rightarrow \overline{L}$, $2 \rightarrow \overline{L}$, $3 \rightarrow \overline{L}$, $4 \rightarrow l$, $5 \rightarrow L$, $6 \rightarrow L$}
\label{tab:sixth_first}
\end{table}

\begin{table}[H]
\centering
\renewcommand{\arraystretch}{1.4}
\setlength{\tabcolsep}{0.15cm} % Adjust the column separation to make it narrower
%
\caption{$1 \rightarrow \overline{L}$, $2 \rightarrow l$, $3 \rightarrow L$, $4 \rightarrow L$, $5 \rightarrow \overline{L}$, $6 \rightarrow \overline{L}$}
\end{table}

\begin{table}[H]
\centering
\renewcommand{\arraystretch}{1.4}
\setlength{\tabcolsep}{0.15cm} % Adjust the column separation to make it narrower
%
\caption{$1 \rightarrow \overline{L}$, $2 \rightarrow \overline{L}$, $3 \rightarrow l$, $4 \rightarrow \overline{L}$, $5 \rightarrow L$, $6 \rightarrow L$}
\end{table}

\begin{table}[H]
\centering
\renewcommand{\arraystretch}{1.4}
\setlength{\tabcolsep}{0.15cm} % Adjust the column separation to make it narrower
%
\caption{$1 \rightarrow l$, $2 \rightarrow \overline{L}$, $3 \rightarrow \overline{L}$, $4 \rightarrow \overline{L}$, $5 \rightarrow L$, $6 \rightarrow L$}
\end{table}

\begin{table}[H]
\centering
\renewcommand{\arraystretch}{1.4}
\setlength{\tabcolsep}{0.15cm} % Adjust the column separation to make it narrower
%
\caption{$1 \rightarrow l$, $2 \rightarrow \overline{L}$, $3 \rightarrow L$, $4 \rightarrow L$, $5 \rightarrow \overline{L}$, $6 \rightarrow \overline{L}$}
\end{table}

\begin{table}[H]
\centering
\renewcommand{\arraystretch}{1.4}
\setlength{\tabcolsep}{0.15cm} % Adjust the column separation to make it narrower
%
\caption{$1 \rightarrow \overline{L}$, $2 \rightarrow l$, $3 \rightarrow \overline{L}$, $4 \rightarrow \overline{L}$, $5 \rightarrow L$, $6 \rightarrow L$}
\end{table}

\subsubsection{$\bar{L} \, \bar{L} \, \bar{l} \, l \, l \, L \, H$}

\begin{table}[H]
\centering
\renewcommand{\arraystretch}{1.4}
\setlength{\tabcolsep}{0.15cm} % Adjust the column separation to make it narrower
%
\caption{$1 \rightarrow l$, $2 \rightarrow l$, $3 \rightarrow L$, $4 \rightarrow \overline{l}$, $5 \rightarrow \overline{L}$, $6 \rightarrow \overline{L}$}
\end{table}

\begin{table}[H]
\centering
\renewcommand{\arraystretch}{1.4}
\setlength{\tabcolsep}{0.15cm} % Adjust the column separation to make it narrower
%
\caption{$1 \rightarrow \overline{L}$, $2 \rightarrow \overline{L}$, $3 \rightarrow l$, $4 \rightarrow l$, $5 \rightarrow \overline{l}$, $6 \rightarrow L$}
\end{table}

\begin{table}[H]
\centering
\renewcommand{\arraystretch}{1.4}
\setlength{\tabcolsep}{0.15cm} % Adjust the column separation to make it narrower
%
\caption{$1 \rightarrow l$, $2 \rightarrow l$, $3 \rightarrow \overline{l}$, $4 \rightarrow L$, $5 \rightarrow \overline{L}$, $6 \rightarrow \overline{L}$}
\end{table}

\begin{table}[H]
\centering
\renewcommand{\arraystretch}{1.4}
\setlength{\tabcolsep}{0.15cm} % Adjust the column separation to make it narrower
%
\caption{$1 \rightarrow l$, $2 \rightarrow \overline{L}$, $3 \rightarrow \overline{L}$, $4 \rightarrow l$, $5 \rightarrow \overline{l}$, $6 \rightarrow L$}
\end{table}

\begin{table}[H]
\centering
\renewcommand{\arraystretch}{1.4}
\setlength{\tabcolsep}{0.15cm} % Adjust the column separation to make it narrower
%
\caption{$1 \rightarrow l$, $2 \rightarrow \overline{L}$, $3 \rightarrow \overline{l}$, $4 \rightarrow L$, $5 \rightarrow l$, $6 \rightarrow \overline{L}$}
\end{table}

\begin{table}[H]
\centering
\renewcommand{\arraystretch}{1.4}
\setlength{\tabcolsep}{0.15cm} % Adjust the column separation to make it narrower
%
\caption{$1 \rightarrow l$, $2 \rightarrow \overline{L}$, $3 \rightarrow l$, $4 \rightarrow \overline{L}$, $5 \rightarrow \overline{l}$, $6 \rightarrow L$}
\end{table}

\begin{table}[H]
\centering
\renewcommand{\arraystretch}{1.4}
\setlength{\tabcolsep}{0.15cm} % Adjust the column separation to make it narrower
%
\caption{$1 \rightarrow l$, $2 \rightarrow \overline{L}$, $3 \rightarrow L$, $4 \rightarrow \overline{l}$, $5 \rightarrow l$, $6 \rightarrow \overline{L}$}
\end{table}

\begin{table}[H]
\centering
\renewcommand{\arraystretch}{1.4}
\setlength{\tabcolsep}{0.15cm} % Adjust the column separation to make it narrower
%
\caption{$1 \rightarrow l$, $2 \rightarrow l$, $3 \rightarrow \overline{L}$, $4 \rightarrow \overline{L}$, $5 \rightarrow L$, $6 \rightarrow \overline{l}$}
\end{table}

\begin{table}[H]
\centering
\renewcommand{\arraystretch}{1.4}
\setlength{\tabcolsep}{0.15cm} % Adjust the column separation to make it narrower
%
\caption{$1 \rightarrow \overline{L}$, $2 \rightarrow l$, $3 \rightarrow l$, $4 \rightarrow \overline{L}$, $5 \rightarrow L$, $6 \rightarrow \overline{l}$}
\end{table}

\begin{table}[H]
\centering
\renewcommand{\arraystretch}{1.4}
\setlength{\tabcolsep}{0.15cm} % Adjust the column separation to make it narrower
%
\caption{$1 \rightarrow \overline{L}$, $2 \rightarrow l$, $3 \rightarrow \overline{L}$, $4 \rightarrow l$, $5 \rightarrow L$, $6 \rightarrow \overline{l}$}
\end{table}

\begin{table}[H]
\centering
\renewcommand{\arraystretch}{1.4}
\setlength{\tabcolsep}{0.15cm} % Adjust the column separation to make it narrower
%
\caption{$1 \rightarrow \overline{L}$, $2 \rightarrow \overline{L}$, $3 \rightarrow l$, $4 \rightarrow l$, $5 \rightarrow L$, $6 \rightarrow \overline{l}$}
\end{table}

\begin{table}[H]
\centering
\renewcommand{\arraystretch}{1.4}
\setlength{\tabcolsep}{0.15cm} % Adjust the column separation to make it narrower
%
\caption{$1 \rightarrow l$, $2 \rightarrow l$, $3 \rightarrow \overline{L}$, $4 \rightarrow \overline{L}$, $5 \rightarrow \overline{l}$, $6 \rightarrow L$}
\end{table}

\subsubsection{$\bar{L} \, \bar{l} \, \bar{l} \, l \, l \, l \, H$}

\begin{table}[H]
\centering
\renewcommand{\arraystretch}{1.4}
\setlength{\tabcolsep}{0.15cm} % Adjust the column separation to make it narrower
%
\caption{$1 \rightarrow l$, $2 \rightarrow l$, $3 \rightarrow \overline{l}$, $4 \rightarrow \overline{l}$, $5 \rightarrow \overline{L}$, $6 \rightarrow l$}
\end{table}

\begin{table}[H]
\centering
\renewcommand{\arraystretch}{1.4}
\setlength{\tabcolsep}{0.15cm} % Adjust the column separation to make it narrower
%
\caption{$1 \rightarrow l$, $2 \rightarrow \overline{L}$, $3 \rightarrow l$, $4 \rightarrow l$, $5 \rightarrow \overline{l}$, $6 \rightarrow \overline{l}$}
\end{table}

\begin{table}[H]
\centering
\renewcommand{\arraystretch}{1.4}
\setlength{\tabcolsep}{0.15cm} % Adjust the column separation to make it narrower
%
\caption{$1 \rightarrow l$, $2 \rightarrow l$, $3 \rightarrow l$, $4 \rightarrow \overline{L}$, $5 \rightarrow \overline{l}$, $6 \rightarrow \overline{l}$}
\end{table}

\begin{table}[H]
\centering
\renewcommand{\arraystretch}{1.4}
\setlength{\tabcolsep}{0.15cm} % Adjust the column separation to make it narrower
%
\caption{$1 \rightarrow l$, $2 \rightarrow l$, $3 \rightarrow \overline{L}$, $4 \rightarrow l$, $5 \rightarrow \overline{l}$, $6 \rightarrow \overline{l}$}
\end{table}

\begin{table}[H]
\centering
\renewcommand{\arraystretch}{1.4}
\setlength{\tabcolsep}{0.15cm} % Adjust the column separation to make it narrower
%
\caption{$1 \rightarrow \overline{L}$, $2 \rightarrow l$, $3 \rightarrow l$, $4 \rightarrow l$, $5 \rightarrow \overline{l}$, $6 \rightarrow \overline{l}$}
\end{table}

\begin{table}[H]
\centering
\renewcommand{\arraystretch}{1.4}
\setlength{\tabcolsep}{0.15cm} % Adjust the column separation to make it narrower
%
\caption{$1 \rightarrow l$, $2 \rightarrow l$, $3 \rightarrow \overline{l}$, $4 \rightarrow \overline{l}$, $5 \rightarrow l$, $6 \rightarrow \overline{L}$}
\label{tab:sixth_last}
\end{table}

\subsection{Seventh Topology}
\begin{figure}[H]
    \centering
    \includegraphics[width=0.35\textwidth]{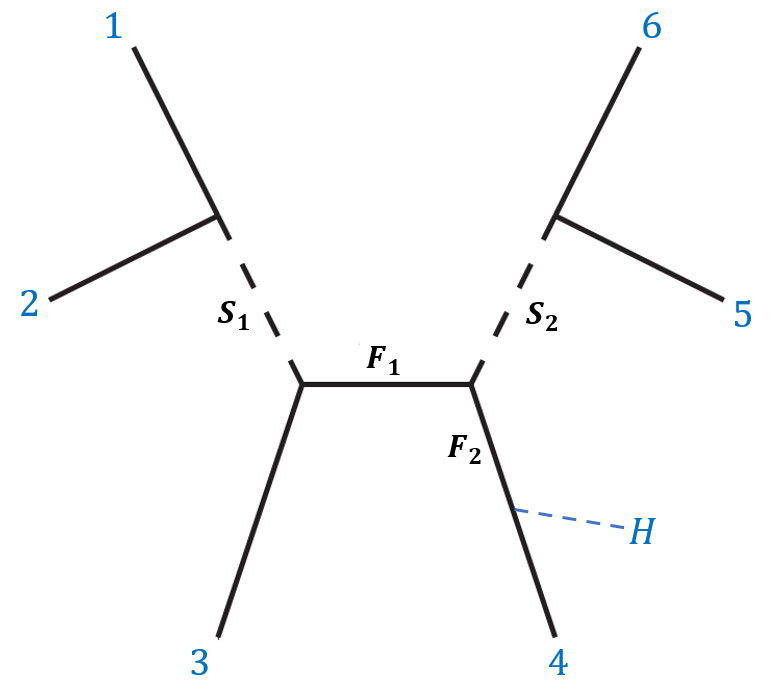}
    \caption{Feynman diagram for the cases where $H$ is attached to an external leg that connects to a fermion mediaator. The corresponding tables are Tab.~\ref{tab:seventh_first} - Tab.~\ref{tab:seventh_last}.}
\end{figure}

\subsubsection{$\bar{L} \, \bar{L} \, \bar{L} \, l \, L \, L \, H$}

\begin{table}[H]
\centering
\renewcommand{\arraystretch}{1.4}
\setlength{\tabcolsep}{0.15cm} % Adjust the column separation to make it narrower

\caption{$1 \rightarrow \overline{L}$, $2 \rightarrow \overline{L}$, $3 \rightarrow \overline{L}$, $4 \rightarrow l$, $5 \rightarrow L$, $6 \rightarrow L$}
\label{tab:seventh_first}
\end{table}

\begin{table}[H]
\centering
\renewcommand{\arraystretch}{1.4}
\setlength{\tabcolsep}{0.15cm} % Adjust the column separation to make it narrower
%
\caption{$1 \rightarrow \overline{L}$, $2 \rightarrow l$, $3 \rightarrow L$, $4 \rightarrow L$, $5 \rightarrow \overline{L}$, $6 \rightarrow \overline{L}$}
\end{table}

\begin{table}[H]
\centering
\renewcommand{\arraystretch}{1.4}
\setlength{\tabcolsep}{0.15cm} % Adjust the column separation to make it narrower
%
\caption{$1 \rightarrow \overline{L}$, $2 \rightarrow \overline{L}$, $3 \rightarrow l$, $4 \rightarrow \overline{L}$, $5 \rightarrow L$, $6 \rightarrow L$}
\end{table}

\begin{table}[H]
\centering
\renewcommand{\arraystretch}{1.4}
\setlength{\tabcolsep}{0.15cm} % Adjust the column separation to make it narrower
%
\caption{$1 \rightarrow l$, $2 \rightarrow \overline{L}$, $3 \rightarrow \overline{L}$, $4 \rightarrow \overline{L}$, $5 \rightarrow L$, $6 \rightarrow L$}
\end{table}

\begin{table}[H]
\centering
\renewcommand{\arraystretch}{1.4}
\setlength{\tabcolsep}{0.15cm} % Adjust the column separation to make it narrower
%
\caption{$1 \rightarrow l$, $2 \rightarrow \overline{L}$, $3 \rightarrow L$, $4 \rightarrow L$, $5 \rightarrow \overline{L}$, $6 \rightarrow \overline{L}$}
\end{table}

\begin{table}[H]
\centering
\renewcommand{\arraystretch}{1.4}
\setlength{\tabcolsep}{0.15cm} % Adjust the column separation to make it narrower
%
\caption{$1 \rightarrow \overline{L}$, $2 \rightarrow l$, $3 \rightarrow \overline{L}$, $4 \rightarrow \overline{L}$, $5 \rightarrow L$, $6 \rightarrow L$}
\end{table}

\subsubsection{$\bar{L} \, \bar{L} \, \bar{l} \, l \, l \, L \, H$}

\begin{table}[H]
\centering
\renewcommand{\arraystretch}{1.4}
\setlength{\tabcolsep}{0.15cm} % Adjust the column separation to make it narrower
%
\caption{$1 \rightarrow l$, $2 \rightarrow l$, $3 \rightarrow L$, $4 \rightarrow \overline{l}$, $5 \rightarrow \overline{L}$, $6 \rightarrow \overline{L}$}
\end{table}

\begin{table}[H]
\centering
\renewcommand{\arraystretch}{1.4}
\setlength{\tabcolsep}{0.15cm} % Adjust the column separation to make it narrower
%
\caption{$1 \rightarrow \overline{L}$, $2 \rightarrow \overline{L}$, $3 \rightarrow l$, $4 \rightarrow l$, $5 \rightarrow \overline{l}$, $6 \rightarrow L$}
\end{table}

\begin{table}[H]
\centering
\renewcommand{\arraystretch}{1.4}
\setlength{\tabcolsep}{0.15cm} % Adjust the column separation to make it narrower
%
\caption{$1 \rightarrow l$, $2 \rightarrow l$, $3 \rightarrow \overline{l}$, $4 \rightarrow L$, $5 \rightarrow \overline{L}$, $6 \rightarrow \overline{L}$}
\end{table}

\begin{table}[H]
\centering
\renewcommand{\arraystretch}{1.4}
\setlength{\tabcolsep}{0.15cm} % Adjust the column separation to make it narrower
%
\caption{$1 \rightarrow l$, $2 \rightarrow \overline{L}$, $3 \rightarrow \overline{L}$, $4 \rightarrow l$, $5 \rightarrow \overline{l}$, $6 \rightarrow L$}
\end{table}

\begin{table}[H]
\centering
\renewcommand{\arraystretch}{1.4}
\setlength{\tabcolsep}{0.15cm} % Adjust the column separation to make it narrower
%
\caption{$1 \rightarrow l$, $2 \rightarrow \overline{L}$, $3 \rightarrow \overline{l}$, $4 \rightarrow L$, $5 \rightarrow l$, $6 \rightarrow \overline{L}$}
\end{table}

\begin{table}[H]
\centering
\renewcommand{\arraystretch}{1.4}
\setlength{\tabcolsep}{0.15cm} % Adjust the column separation to make it narrower
%
\caption{$1 \rightarrow l$, $2 \rightarrow \overline{L}$, $3 \rightarrow l$, $4 \rightarrow \overline{L}$, $5 \rightarrow \overline{l}$, $6 \rightarrow L$}
\end{table}

\begin{table}[H]
\centering
\renewcommand{\arraystretch}{1.4}
\setlength{\tabcolsep}{0.15cm} % Adjust the column separation to make it narrower
%
\caption{$1 \rightarrow l$, $2 \rightarrow \overline{L}$, $3 \rightarrow L$, $4 \rightarrow \overline{l}$, $5 \rightarrow l$, $6 \rightarrow \overline{L}$}
\end{table}

\begin{table}[H]
\centering
\renewcommand{\arraystretch}{1.4}
\setlength{\tabcolsep}{0.15cm} % Adjust the column separation to make it narrower
%
\caption{$1 \rightarrow l$, $2 \rightarrow l$, $3 \rightarrow \overline{L}$, $4 \rightarrow \overline{L}$, $5 \rightarrow L$, $6 \rightarrow \overline{l}$}
\end{table}

\begin{table}[H]
\centering
\renewcommand{\arraystretch}{1.4}
\setlength{\tabcolsep}{0.15cm} % Adjust the column separation to make it narrower
%
\caption{$1 \rightarrow \overline{L}$, $2 \rightarrow l$, $3 \rightarrow l$, $4 \rightarrow \overline{L}$, $5 \rightarrow L$, $6 \rightarrow \overline{l}$}
\end{table}

\begin{table}[H]
\centering
\renewcommand{\arraystretch}{1.4}
\setlength{\tabcolsep}{0.15cm} % Adjust the column separation to make it narrower
%
\caption{$1 \rightarrow \overline{L}$, $2 \rightarrow l$, $3 \rightarrow \overline{L}$, $4 \rightarrow l$, $5 \rightarrow L$, $6 \rightarrow \overline{l}$}
\end{table}

\begin{table}[H]
\centering
\renewcommand{\arraystretch}{1.4}
\setlength{\tabcolsep}{0.15cm} % Adjust the column separation to make it narrower
%
\caption{$1 \rightarrow \overline{L}$, $2 \rightarrow \overline{L}$, $3 \rightarrow l$, $4 \rightarrow l$, $5 \rightarrow L$, $6 \rightarrow \overline{l}$}
\end{table}

\begin{table}[H]
\centering
\renewcommand{\arraystretch}{1.4}
\setlength{\tabcolsep}{0.15cm} % Adjust the column separation to make it narrower
%
\caption{$1 \rightarrow l$, $2 \rightarrow l$, $3 \rightarrow \overline{L}$, $4 \rightarrow \overline{L}$, $5 \rightarrow \overline{l}$, $6 \rightarrow L$}
\end{table}

\subsubsection{$\bar{L} \, \bar{l} \, \bar{l} \, l \, l \, l \, H$}

\begin{table}[H]
\centering
\renewcommand{\arraystretch}{1.4}
\setlength{\tabcolsep}{0.15cm} % Adjust the column separation to make it narrower
%
\caption{$1 \rightarrow l$, $2 \rightarrow l$, $3 \rightarrow \overline{l}$, $4 \rightarrow \overline{l}$, $5 \rightarrow \overline{L}$, $6 \rightarrow l$}
\end{table}

\begin{table}[H]
\centering
\renewcommand{\arraystretch}{1.4}
\setlength{\tabcolsep}{0.15cm} % Adjust the column separation to make it narrower
%
\caption{$1 \rightarrow l$, $2 \rightarrow \overline{L}$, $3 \rightarrow l$, $4 \rightarrow l$, $5 \rightarrow \overline{l}$, $6 \rightarrow \overline{l}$}
\end{table}

\begin{table}[H]
\centering
\renewcommand{\arraystretch}{1.4}
\setlength{\tabcolsep}{0.15cm} % Adjust the column separation to make it narrower
%
\caption{$1 \rightarrow l$, $2 \rightarrow l$, $3 \rightarrow l$, $4 \rightarrow \overline{L}$, $5 \rightarrow \overline{l}$, $6 \rightarrow \overline{l}$}
\end{table}

\begin{table}[H]
\centering
\renewcommand{\arraystretch}{1.4}
\setlength{\tabcolsep}{0.15cm} % Adjust the column separation to make it narrower
%
\caption{$1 \rightarrow l$, $2 \rightarrow l$, $3 \rightarrow \overline{L}$, $4 \rightarrow l$, $5 \rightarrow \overline{l}$, $6 \rightarrow \overline{l}$}
\end{table}

\begin{table}[H]
\centering
\renewcommand{\arraystretch}{1.4}
\setlength{\tabcolsep}{0.15cm} % Adjust the column separation to make it narrower
%
\caption{$1 \rightarrow \overline{L}$, $2 \rightarrow l$, $3 \rightarrow l$, $4 \rightarrow l$, $5 \rightarrow \overline{l}$, $6 \rightarrow \overline{l}$}
\end{table}

\begin{table}[H]
\centering
\renewcommand{\arraystretch}{1.4}
\setlength{\tabcolsep}{0.15cm} % Adjust the column separation to make it narrower
%
\caption{$1 \rightarrow l$, $2 \rightarrow l$, $3 \rightarrow \overline{l}$, $4 \rightarrow \overline{l}$, $5 \rightarrow l$, $6 \rightarrow \overline{L}$}
\label{tab:seventh_last}
\end{table}

\clearpage

\section{Limits from LEP and TRISTAN}
\label{sec:LEPlimits}

The neutral scalars in some of our UV completions couple to leptons and thus modify leptonic scattering data. For example, the scalar from Sec.~\ref{sec:UVstudy_dim10} will modify $e^+ e^-\to\mu^+\mu^-$ scattering, experimentally studied at LEP. If the scalar mass exceeds the LEP center-of-mass energy of around $\unit[200]{GeV}$, the contribution can be parametrized through a four-lepton contact interaction, exactly the scenario investigated and constrained in LEP's legacy article, Ref.~\cite{ALEPH:2013dgf}, which contains combined measurements from the four LEP experiments ALEPH, DELPHI, L3 and OPAL, using center-of-mass data away from the $Z$ pole, between $\sqrt{s}=\unit[130]{GeV}$ and $\unit[207]{GeV}$, totaling $\unit[3]{fb^{-1}}$ in luminosity.

In particular, LEP~\cite{ALEPH:2013dgf} provides limits on four-fermion contact interactions of the form~\cite{Eichten:1983hw}
\begin{align}
    \L_\text{eff} = \frac{g^2}{(1+\delta_{ef}) \Lambda_\pm^2} \sum_{i,j= L,R} \eta_{ij} \bar{e}_i \gamma_\mu e_i\, \bar{f}_j\gamma^\mu f_j
    \label{eq:Leff}
\end{align}
that continue to be relevant. Here, $g^2/(4\pi)$ is set to $1$ for simplicity, leaving the scale of the contact interaction, $\Lambda_\pm$ as the only parameter. Different chiralities are encoded through $\eta$, which takes on values $\pm 1$ or $0$, a negative sign typically indicating destructive interference with the SM contribution in the process $e^+ e^-\to f^+ f^-$, using the mostly-minus space-time metric. The particular benchmark models shown in Tab.~\ref{tab:etas} are constrained by LEP in Ref.~\cite{ALEPH:2013dgf} for $f=e$, $\mu$, $\tau$, $d$, $u$, and combinations.
We will only be interested in the muon and tau final states here.

\begin{table}[b]
 \begin{center}
 \renewcommand{\arraystretch}{1.3}
 \setlength{\tabcolsep}{3pt} % Increase horizontal spacing
  \begin{tabular}{|c||c|c|c|c|}
   \hline
   Model      & $\eta_{LL}$ & $\eta_{RR}$ & $\eta_{LR}$ & $\eta_{RL}$ \\
   \hline\hline
   LL$^{\pm}$ &   $\pm 1$   &      0      &      0      &      0      \\
   RR$^{\pm}$ &      0      &   $\pm 1$   &      0      &      0      \\
   VV$^{\pm}$ &   $\pm 1$   &   $\pm 1$   &   $\pm 1$   &   $\pm 1$   \\
   AA$^{\pm}$ &   $\pm 1$   &   $\pm 1$   &   $\mp 1$   &   $\mp 1$   \\
   LR$^{\pm}$ &      0      &      0      &   $\pm 1$   &      0      \\
   RL$^{\pm}$ &      0      &      0      &      0      &   $\pm 1$   \\
   V0$^{\pm}$ &   $\pm 1$   &   $\pm 1$   &      0      &      0      \\
   A0$^{\pm}$ &      0      &      0      &  $\pm 1$    &   $\pm 1$   \\
   A1$^{\pm}$ &    $\pm 1$  &   $\mp 1$   &      0      &      0      \\
   \hline
  \end{tabular}
 \end{center}
 \caption[$\eta_{ij}$ choices for different contact interaction models.]
         {$\eta_{ij}$ choices for different contact interaction models.}
 \label{tab:etas}.
\end{table}

The differential cross sections for $e^+ e^-\to \mu^+ \mu^-$ and $\tau^+\tau^-$ through electroweak interactions and the contact interactions of Eq.~\eqref{eq:Leff} take the form~\cite{Kroha:1991mn}
\begin{align}
\begin{split}
    \frac{\dd\sigma}{\dd \cos \theta} &= \frac{\pi\alpha^2 }{2 s} \left\{  \left(\left|\mathcal{A}^{ef}_{LR}(s)\right|^2 +\left|\mathcal{A}^{ef}_{RL}(s)\right|^2 \right) \frac{t^2}{s^2}\right.\\
    &\qquad\quad \left.+ \left(\left|\mathcal{A}^{ef}_{LL}(s)\right|^2 +\left|\mathcal{A}^{ef}_{RR}(s)\right|^2 \right) \frac{u^2}{s^2} \right\} ,
\end{split}
\label{eq:Kroha}
\end{align}
where $s = 4 E_\text{beam}^2$ is the center-of-mass energy, $t = -s (1-\cos\theta)/2$, and $s+t+u=0$ since we treat all fermions as massless.
The individual dimensionless helicity amplitudes are~\cite{Kroha:1991mn}
\begin{align}
    \mathcal{A}^{ef}_{ij}(s) &= Q_e Q_f + c_i^e c_j^f \chi(s) + \eta_{ij} \frac{s}{\alpha \Lambda^2}\,, 
\end{align}
where $\chi(s) = s/(s-M_Z^2 + \ii M_Z \Gamma_Z)$ encodes the $Z$-boson propagators, $Q_f$ the charge of fermion $f$ (e.g.~$Q_e = -1$), and $c_j^f = (T^f_j - s_W^2 Q_f)/(s_W c_W)$ the $Z$-boson coupling to fermion $f$ (e.g.~$c_L^e = (-\tfrac12 + s_W^2)/(s_W c_W)$ \& $c_R^e = (0 + s_W^2)/(s_W c_W)$). Radiative corrections are neglected for simplicity.

The SM limit, which describes the LEP data well~\cite{ALEPH:2013dgf}, arises as $\Lambda\to \infty$, so SM-compatible data generically leads to a lower bound on $\Lambda$, in the ballpark of TeV in the LEP case.
Interestingly, for some models of Tab.~\ref{tab:etas}, the destructive interference case allows for a qualitatively different SM limit with \textit{finite} $\Lambda$. Take the LR${^-}$ case in $e^+e^-\to\mu^+\mu^-$: only $ \mathcal{A}^{e\mu}_{LR}(s)=1 + c_L^e c_R^\mu \chi(s) -s/(\alpha \Lambda^2)$ deviates from the SM, but by solving 
\begin{align}
    1 + c_L^e c_R^\mu \chi(s) -\frac{s}{\alpha \Lambda^2} = -(1 + c_L^e c_R^\mu \chi(s))
    \label{eq:amplitude_flip}
\end{align}
for $\Lambda$ we obtain \textit{exactly} the same differential cross section as in the SM, as we effectively only flip the sign of one amplitude term. At a fixed center-of-mass energy $s$, this scenario is indistinguishable from the SM and opens up an allowed region where $\Lambda$ is not bounded from below but rather by how close it has to be to the value from Eq.~\eqref{eq:amplitude_flip}. In the above example, this value is $\Lambda = \unit[2.2]{TeV}$ for $\sqrt{s} = \unit[207]{GeV}$. This region of parameter space can only be excluded by combining data at different energies. For example, at $\sqrt{s} = \unit[183]{GeV}$, the solution for Eq.~\eqref{eq:amplitude_flip} is at $\Lambda = \unit[2.0]{TeV}$, so it is impossible to satisfy the cancellation over the entire LEP range. However, it just so happens that the data can still be \textit{approximately} satisfied in this cancellation regime, leaving a viable region near this $\Lambda$ even after combining cross-section data from several $s$. This is exactly what happened in the LEP analysis of Ref.~\cite{ALEPH:2013dgf}, explaining the surprisingly low quoted bound of $\Lambda_{\mu\mu}^{LR,-}=\unit[2.2]{TeV}$. After clarifying how this number should be interpreted -- not as a lower limit -- we will show that this region of parameter space can be excluded entirely by combining LEP data with measurements from lower-$s$ experiments.

Ref.~\cite{ALEPH:2013dgf}  provides  total cross sections $\sigma$ and forward--backward asymmetries $A_{FB}$ for $e^+ e^-\to \mu^+ \mu^-$ and $\tau^+\tau^-$ for many center-of-mass bins. 
To fit the theoretical cross section from Eq.~\eqref{eq:Kroha} to data, the parameter $\Lambda$ is replaced by $\epsilon = \Lambda^{-2}$ and a $\chi^2(\epsilon)$ function is constructed. For positive $\eta$, i.e.~in the constructive interference case, $\epsilon$ has to be positive, while in the negative $\eta$ case, $\epsilon$ is can be considered to be negative. The $95\%$ C.L.~limits on $\Lambda_{\pm}$ are then derived through a probability distribution $\propto\exp [- \tfrac12 \chi^2(\epsilon)]$~\cite{DELPHI:2005wxt} by solving
\begin{align}
   \frac{\int_0^{\pm \Lambda_{\pm}^{-2}}\dd \epsilon\, \exp\left[- \tfrac12 \chi^2(\epsilon)\right]}{\int_0^{\pm \infty}\dd \epsilon\, \exp\left[- \tfrac12 \chi^2(\epsilon)\right]} = 0.95 \,.
\end{align}
Despite the crudeness of the used cross section (neglecting radiative corrections) and likelihood function, this method manages to reproduce the $\Lambda$ limits from Ref.~\cite{ALEPH:2013dgf}, collected in Tab.~\ref{tab:limits}, to within $10\%$, thus clearly capturing the overall picture.

\begin{table}[t]
 \begin{center}
 \renewcommand{\arraystretch}{1.1}
\begin{tabular}{|c||rr|rr|rr|}
   \hline
   Model  & \multicolumn{2}{|c|}
            {$\Lambda^{-}_{\mu\mu}~ \,\,\Lambda^{+}_{\mu\mu}$} 
          & \multicolumn{2}{|c|}
            {$\Lambda^{-}_{\tau\tau}~ \,\,\Lambda^{+}_{\tau\tau}$} \\
   \hline
   \hline
   LL &   9.8 &  12.2 &   9.1 &   9.1  \\
   RR &   9.3 &  11.6 &   8.7 &   8.7  \\
   VV &  16.3 &  18.9 &  13.8 &  15.8  \\
   AA &  13.4 &  16.7 &  14.1 &  11.4  \\
   LR &   2.2 \textbf{(8)} &    9.1 &   2.2 \textbf{(5)} &   7.7  \\
   RL &   2.2 \textbf{(8)} &    9.1 &   2.2 \textbf{(5)} &   7.7  \\
   V0 &  13.5 &  16.9 &  12.6 &  12.5  \\
   A0 &  12.1 &  12.6 &   8.9 &  12.1  \\
   A1 &   4.5 &   5.8 &   3.9 &   4.7  \\
   \hline
\end{tabular}
 \end{center}
 \caption{The $95\%$ confidence
         limits on the scale, $\Lambda^{\pm}$, for constructive ($+$)
         and destructive interference ($-$) with the SM, for the
         contact interaction models discussed in the text. Results are
         given for $e^+e^-\to\mu^+\mu^-$ and $e^+e^-\to\tau^+\tau^-$, reproduced from Ref.~\cite{ALEPH:2013dgf}. In bold brackets, we give the lower limit we obtain by combining LEP and TRISTAN~\cite{AMY:1994cjs,VENUS:1997bjf} data, as explained in the text. }
 \label{tab:limits}
\end{table}

\begin{figure}[tb]
    \centering
    \includegraphics[width=0.47\textwidth]{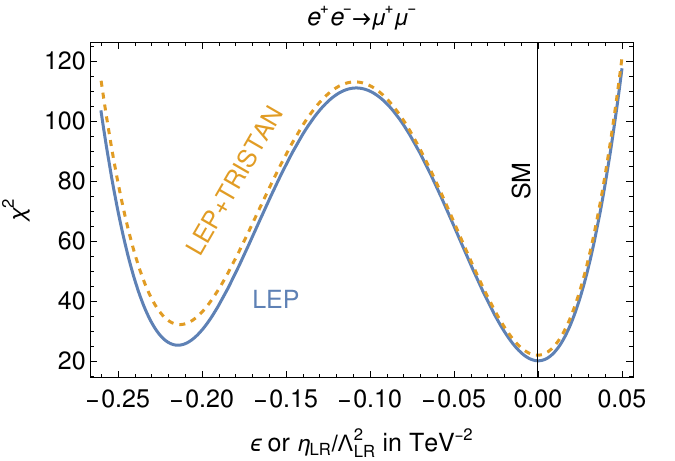}
    \caption{$\chi^2(\epsilon)$ for $\Lambda^{LR}_{\mu\mu}$. The blue curve is for LEP data~\cite{ALEPH:2013dgf} only, the dashed orange one includes low-$\sqrt{s}$ data from TRISTAN~\cite{VENUS:1997bjf}, which lifts the flipped-amplitude local minimum compared the SM one at $\epsilon=0$.
} 
    \label{fig:chimumu}
\end{figure}

In most cases, $\chi^2(\epsilon)$ has two minima, one around $\epsilon=0$, and one for negative $\epsilon$ that corresponds to the sign-flipped amplitude scenario outlined above. The global minimum is always the one near zero, and in most cases the local minimum at negative $\epsilon$ has a far larger $\chi^2$ due to the combination of different $\sqrt{s}$ data and can thus be ignored. However, in the LR case for muons, the two minima only differ by $\Delta\chi^2 \simeq 5.2$, not enough to exclude the sign-flip case outright, as illustrated in Fig.~\ref{fig:chimumu}. Due to the near degeneracy, the above procedure for obtaining $95\%$ CL limits described above yields $\Lambda_{\mu\mu}^{LR,-}\simeq\unit[2.2]{TeV}$. While this matches the number from Ref.~\cite{ALEPH:2013dgf} (reproduced in Tab.~\ref{tab:limits}), it illustrates that the \textit{interpretation} of this number is slightly different from the other cases, seeing as the $\Lambda$ range from $\unit[2.5]{TeV}$ to $\unit[8]{TeV}$ has an extremely low probability (or large $\chi^2$), not captured by the simple lower limit. Instead, we could treat the two minima separately and state that there are two disconnected allowed regions, roughly
\begin{align}
  \unit[2.1]{TeV}\lesssim \Lambda_{\mu\mu}^{LR,-} \lesssim \unit[2.2]{TeV} \,, && \unit[9]{TeV}\lesssim \Lambda_{\mu\mu}^{LR,-} \,,
\end{align}
with the one near $\unit[2]{TeV}$ being disfavored at the $97.8\%$~CL due to the larger $\chi^2$.

The sign-flip scenario survives in the LR$^{-}$ (=RL$^{-}$) case despite fitting to data from $\sqrt{s}=\unit[130]{GeV}$ to $\unit[207]{GeV}$, which could have broken this degeneracy. Indeed, earlier LEP combinations~\cite{Electroweak:2003ram} do not list $\unit[2.2]{TeV}$ in the muon channel, but rather $\unit[7.9]{TeV}$, although the tau limit is still at $\unit[2.2]{TeV}$. 
The energy range covered in LEP is thus seemingly too narrow to exclude the LR$^{-}$ sign-flip scenario with certainty. Luckily, adding data from lower-$\sqrt{s}$ experiments such as TRISTAN~\cite{VENUS:1997bjf} ($\sqrt{s}=\unit[57.8]{GeV}$) can lift the second minimum sufficiently ($\Delta\chi^2 \simeq 10$, see Fig.~\ref{fig:chimumu}) to confidently exclude the flipped-sign case and only leaves the lower bound $\unit[9]{TeV}\lesssim \Lambda_{\mu\mu}^{LR,-}$. Since the $\Lambda$ limits obtained via the above simplified method are systematically larger than the LEP limits, we propose a lower limit of $\unit[8]{TeV}$ on $\Lambda_{\mu\mu}^{LR,-}$, in line with the previous LEP combination~\cite{Electroweak:2003ram}. The limits on other $\mu\mu$ contact interactions are not affected by adding the TRISTAN results.

The tauon case is similar, here we can reproduce all LEP limits to better than $10\%$ except $\Lambda_{\tau\tau}^{LR,+}$, which comes out $20\%$ larger than LEP's. In any case, the LR$^{-}$ case once again    features two minima, this time even more degenerate, $\Delta\chi^2 \simeq 3.9$. Adding TRISTAN data~\cite{AMY:1994cjs} again lifts the second minimum sufficiently to exclude the sign-flip case and provides a conservative lower limit on $\Lambda_{\tau\tau}^{LR,-}$ of $\unit[5]{TeV}$. 

By combining LEP and TRISTAN data, all amplitude sign-flip scenarios can be excluded, with the exception of A1$^{-}$. Here, the sign-flip value for $\Lambda$ is around $\unit[6]{TeV}$, \textit{above} the quoted limits. The data is not able to resolve the two different minima, resulting in one broad minimum that gives rise to weak limits, as can be seen in Tab.~\ref{tab:limits}. Improving this bound requires experiments with larger $s$.

Having re-derived the four-lepton contact-interaction limits from LEP, we can go beyond it and fit our neutral-scalar scenario to the data. We calculate the cross section and forward--backward asymmetry for $e^+e^-\to\mu^+\mu^-$ in our model at tree level and fit to the same data as above, obtaining a $\chi^2$ function as a function of the Yukawa coupling and scalar mass. After including TRISTAN data to the LEP set, this gives an upper limit on the Yukawa for a given mass, shown in Fig.~\ref{fig:mu-e-scalar_limits}. For large masses, this matches the contact-interaction limit from $\Lambda_{\mu\mu}^{LR,-}$, slightly different since our limits are now derived for two parameters rather than one.

%%%%%%%%%%%%%%%%%%%%%%%%%

\bibliographystyle{utcaps_mod}
\bibliography{BIB.bib}

\end{document}